\documentclass[rmp,twocolumn,floatfix,superscriptaddress,showpacs]{revtex4-1}

\usepackage{lineno}

\usepackage{fancyhdr}

\usepackage{graphicx,dcolumn,bm}
\usepackage{rotating}
\usepackage{amsmath}
\usepackage{amssymb}
\usepackage{array}
\usepackage{wrapfig}
\usepackage{units}
\usepackage{color}
\usepackage{cancel}
\usepackage{enumerate}

\newcommand{\be}{\begin{equation}}
\newcommand{\ee}{\end{equation}}
\newcommand{\bee}{\begin{equation*}}
\newcommand{\eee}{\end{equation*}}
\newcommand{\bea}{\begin{eqnarray}}
\newcommand{\eea}{\end{eqnarray}}
\newcommand{\bean}{\begin{eqnarray*}}
\newcommand{\eean}{\end{eqnarray*}}

\usepackage{color}

\newcommand{\ba}{\begin{eqnarray}}
\newcommand{\ea}{\end{eqnarray}}

\newcommand{\ket}[1]{\left\lvert #1\right\rangle}
\newcommand{\bra}[1]{\left\langle #1\right\rvert}

\newcommand{\diracslash}[1]{\not\!\! #1}

\newcommand{\baa}{\begin{array}}      
\newcommand{\eaa}{\end{array}}
\newcommand{\bit}{\begin{itemize}}    
\newcommand{\eit}{\end{itemize}}
\newcommand{\ben}{\begin{enumerate}}  
\newcommand{\een}{\end{enumerate}}
\newcommand{\bce}{\begin{center}}     
\newcommand{\ece}{\end{center}}
\newcommand{\bfl}{\begin{flushright}} 
\newcommand{\efl}{\end{flushright}}
\newcommand{\btb}{\begin{tabular}}    
\newcommand{\etb}{\end{tabular}}

\newcommand{\vp}{\varphi}
\newcommand{\tvp}{\widetilde{\varphi}}

\def\wt{\widetilde}
\def\gpbz{{\bar g}_\pi^{(0)}}
\def\gpiz{{\bar g}_\pi^{(0)}}
\def\gpio{{\bar g}_\pi^{(1)}}
\def\gpbo{{\bar g}_\pi^{(1)}}
\def\gpbt{{\bar g}_\pi^{(2)}}
\def\gpbi{{\bar g}_\pi^{(i)}}

\newcommand{\ecm}{\,{\it e}~{\rm cm}}

\begin{document}

\title{Electric Dipole Moments of Atoms, Molecules, Nuclei and Particles}

 \author{T.E. Chupp}
 \affiliation{Department of Physics, University of Michigan, Ann Arbor, Michigan 48109 USA}
 \author{P. Fierlinger}
 \affiliation{Physik Department and Excellence-Cluster ``Universe'', Technische Universit{\" a}t M{\" u}nchen, 85748 Garching bei M{\" u}nchen, Germany}
 \author{M.J. Ramsey-Musolf}
 \affiliation{Amherst Center for Fundamental Interactions and Department of Physics, University of Massachusetts-Amherst, Amherst, Massachusetts 01003 USA}
 \author{J.T. Singh}
 \affiliation{National Superconducting Cyclotron Laboratory and Department of Physics and Astronomy, Michigan State University
East Lansing, Michigan 48824 USA}


\begin{abstract}
A permanent electric dipole moment (EDM) of a particle or system is a separation of charge along its angular-momentum axis and is a direct signal of T-violation and, assuming CPT symmetry, CP violation. For over sixty years EDMs have been studied, first as a signal of a parity-symmetry violation and then as a signal of CP violation that would clarify its role in nature and in theory. Contemporary motivations include the role that CP violation plays in explaining the cosmological matter-antimatter asymmetry and the search for new physics. Experiments on a variety of systems have become ever-more sensitive, but provide only upper limits on EDMs, and theory at several scales is crucial to interpret these limits. Nuclear theory provides connections from Standard-Model and Beyond-Standard-Model physics to the observable EDMs, and atomic and molecular theory reveal how CP-violation is manifest in these systems. EDM results in hadronic systems require that the Standard Model QCD parameter of $\bar\theta$ must be exceptionally small, which could be explained by the existence of axions - also a candidate dark-matter particle. Theoretical results on electroweak baryogenesis show that new physics is needed to explain the dominance of matter in the universe. Experimental and theoretical efforts continue to expand with new ideas and new questions, and this review provides a broad overview of theoretical motivations and interpretations as well as details about experimental techniques, experiments, and prospects. The intent is to provide specifics and context as this exciting field moves forward.
\end{abstract}

\hskip -3.8 truein 
ACFI-T17-18
\hskip 3.3 truein
To be published in Reviews of Modern Physics


 \maketitle
  
 \date{\today.  To be submitted to RMP}
 \tableofcontents

\section{Introduction}
\label{sec:Introduction}

The measurement and interpretation of permanent Electric Dipole Moments or EDMs of  particles and quantum systems have been a unique window into the nature of elementary particle interactions from the very first proposal of~\textcite{rf:Purcell1950} to search for a neutron EDM as a signal of parity-symmetry (P) violation.  The neutron EDM was not observed then or since, and we now recognize, as pointed out by~\textcite{LY57a} and \textcite{Lan57c,Lan57b-ru,Lan57b}, that EDMs  also violate time-reversal symmetry (T).  An EDM is  a direct signal of T-violation,
and CPT symmetry (C is charge conjugation), required of any relativistic field theory~\cite{Tureanu:2013psa} therefore implies that observation of a non-zero P-odd/T-odd EDM is also a signal of CP violation. EDMs have become a major focus of contemporary research for several interconnected reasons: 
\begin{enumerate}[i.]
\item EDMs provide a direct experimental probe of CP violation, a feature of the Standard Model (SM) and Beyond-Standard-Model (BSM) physics;
\item the P-violating and T-violating EDM signal distinguishes the much weaker CP-violating interactions from the dominant strong and electromagnetic interactions; 
\item  CP violation is a required component of Sakharov's recipe~\cite{SakharovBaryogenesis-ru,SakharovBaryogenesis} for the baryon asymmetry, the fact that there is more matter than antimatter in the universe; however SM-CP violation cannot produce the observed asymmetry, and new CP-violating interactions are required.
\end{enumerate}

The EDM of a  system $\vec d$ must be parallel (or antiparallel) to the average angular momentum of the system $\hbar\langle\vec J\rangle$ . Thus, relative to the center of mass ($\vec r=0$):
\begin{equation}
\label{eq:EDMdef}
\vec d =\int\ \vec r\rho_Q d^3r = d \frac{\langle\vec J\rangle}{J}, 
\end{equation}
where $\rho_Q$ is the electric-charge distribution. 
This is analagous to the magnetic dipole moment~\cite{RamseyMolBeamsBookp77}
\begin{equation}
\label{eq:magmomdef}
\vec\mu =\frac{1}{2} \int\ \vec r\times\!\vec J_Q d^3r = \mu  \frac{\langle\vec J\rangle}{J},
\end{equation}
where $\vec J_Q$ is the current density. For a neutral system, {\it e.g.} the neutron, $\vec d$ can be considered a separation of equal charges; however for a  system with charge $Q\ne 0$, $\vec d/Q=\vec r_Q$ is the center of charge with respect to the  center of mass.

The interaction of a fermion with magnetic moment $\mu$ and EDM $d$ with  electric and magnetic fields 
can be written
\begin{equation}
{\cal L_{EM}}=-\frac{\mu}{2}\bar\Psi \sigma^{\mu\nu}F_{\mu\nu}\Psi -i\frac{d}{2}\bar\Psi \sigma^{\mu\nu}\gamma^5 F_{\mu\nu}\Psi,
\label{eq:EMLagrangian}
\end{equation}
where $\Psi$ is the fermion field, and  $F_{\mu\nu}=\partial_\mu A_\nu-\partial_\nu A_\mu$ is the electromagnetic field tensor with $A_\mu$ the four-vector electromagnetic potential.
The second term of Eqn.~\ref{eq:EMLagrangian}, first written down in this way by~\textcite{rf:Salpeter1958}\footnote{In his paper ``The Quantum Theory of the Electron'', \textcite{Dirac} revealed the purely imaginary coupling of the electric field to an electric moment. This was considered, at the time, unphysical because it was derived from a real Hamiltonian.} in analogy to the anomalous magnetic moment term, reveals parity and time-reversal violation in the Dirac matrix $\gamma^5$ and the imaginary number $i$, respectively. The corresponding non-relativistic Hamiltonian for a quantum system is
\begin{equation}
H=-(\vec\mu\cdot \vec B+\vec d\cdot\vec E) = -(\mu\vec J\!\cdot\!\vec B+d\vec J\!\cdot\!\vec E)/J.
\label{eq:EDMHamiltonian}
\end{equation}
The magnetic field $\vec B$ and the angular momentum operator $\vec J$ are both even under P but odd under T, while the electric field $\vec E$ is odd under P but even under T. The second term, proportional to $\vec J\cdot\!\vec E$,
is thus P-odd and T-odd and a direct signal of CP violation assuming CPT invariance.

A common approach to measuring an EDM is to apply a strong electric field and a very well controlled and characterized magnetic field and to measure the shift in the energy, or more  commonly the frequency, of the splitting between magnetic sub-levels when $\vec E$  is changed. For a system with total angular momentum $\hbar J$,  the EDM frequency shift for two adjacent levels ($|\Delta m_J|=1$) is
\begin{equation}
|\Delta \omega|=\frac{|dE|}{\hbar J}.
\label{eq:EDMFreqEquation1}
\end{equation}
The precision of a single frequency measurement depends on the interrogation time $\tau$ and the signal-to-noise ratio (SNR) for the measurement. The SNR depends on the specifics of the 
technique.
For a count-rate-limited experiment with $N$ particles measured or interrogated in a single measurement SNR\,$\propto$\, $\sqrt{N}$, and the statistical uncertainty of a single frequency measurement is given by $\sigma_\omega = 1/( \tau \sqrt{N})$. 
In phase-noise-limited experiments, for example those using a SQUID magnetometer, the statistical uncertainty of a single frequency measurement for constant signal and SNR is given 
by $\sigma_\omega = \sqrt{12}/( \tau [{\rm SNR}])$~\cite{rf:ChuppMaser1,clps}, where SNR generally increases as $\sqrt{\tau}$.
The EDM sensitivity for a {\bf pair} of frequency measurements with opposite electric field each lasting a time $\tau$ therefore scales as
\begin{eqnarray}
\sigma_d&\gtrsim&\frac{\hbar J}{E}\frac{1}{\sqrt{2N}}\tau^{-1}\quad\quad\ \ {\rm (counting)},\nonumber\\
\sigma_d&\gtrsim&\frac{\hbar J}{E}\sqrt{\frac{3}{\pi}}\frac{v_n}{V_0}\tau^{-3/2}\quad {\rm (phase\ noise)}.
\label{eq:EDMSigmaEquation1}
\end{eqnarray}
Here $V_0$ is the signal size and $v_n \sqrt{B}$ is the noise in a bandwidth $B=1/(2\pi\tau)$, with $B$ in Hz.

The experimental challenges are to have the largest possible electric field magnitude $E$, the longest possible $\tau$ and the highest possible  $N$ or $V_0/v_n$. Additionally an ideally small, stable, and well characterized applied magnetic field is required to suppress frequency fluctuations due to changes in the magnetic moment interaction $\vec\mu\cdot\vec B$. Experimenters also strive to find systems in which the EDM is in some way enhanced, basically due to a large intrinsic (P-even, T-even) electric dipole moment and/or increased electric polarizability of the system, which is the case for a molecule or an atomic nucleus with octupole collectivity.
To date all EDM searches (see Table~\ref{tb:EDMResults}), including the neutron, atoms (Cs, Tl, Xe, Hg, and Ra) and molecules (TlF, YbF, ThO, and HfF$^+$),
have results consistent with zero but also consistent with the Standard Model.

\begin{figure*}[tb]
\includegraphics[width = 7 truein]{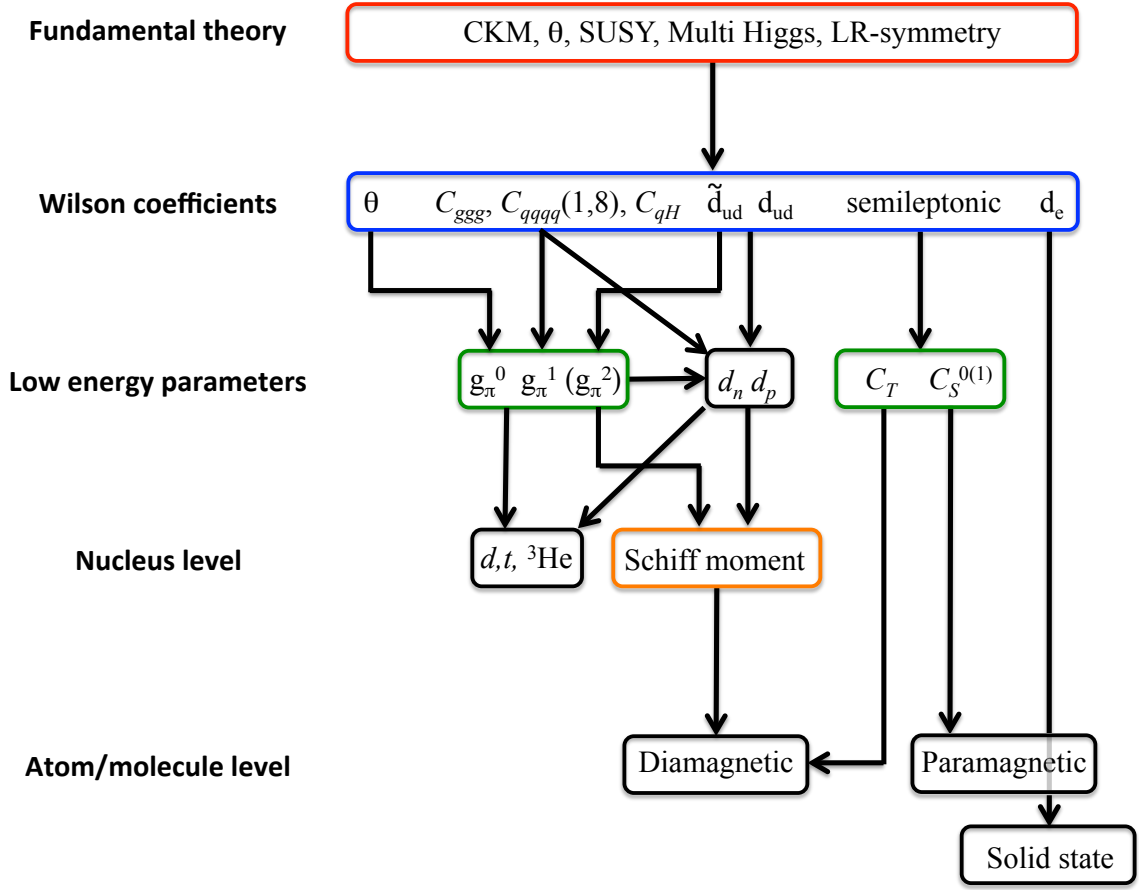}
\caption{\label{fg:EDMSubwayMap} (Color online) Illustration of the connections from a fundamental theory at a high energy scale to an EDM in a measurable low-energy system. The dashed boxes indicate levels dominated by theory, and the solid boxes identify systems that are the object of current and future experiments. The fundamental CP-violating Lagrangian at the top, a combination of SM and BSM physics, is reduced to the set of effective-field-theory Wilson coefficients that characterize interactions at the electroweak energy scale of $\approx 300$ GeV, the vacuum-expectation value of the Higgs. The  set of low-energy parameters defined in Sec.~\ref{sec:Theory} enter calculations that connect the electroweak-scale Wilson coefficients directly to electrons and nuclei. Finally atomic, molecular and condensed-matter structure calculations connect the low-energy parameters to the observables  in experimentally accessible systems.}
 \end{figure*}

Figure~\ref{fg:EDMSubwayMap} 
shows the connections from fundamental theory, including  Standard Model and Beyond-Standard-Model physics, through a series of theory levels at different energy scales to the experimentally accessible P-odd/T-odd observables in a variety of systems.
SM  CP violation arises from a complex phase in the Cabibbo-Kobayashi-Maskawa (CKM) matrix parameterizing the weak interaction~\cite{Kobayashi:1973fv} and in the gluon
$G{\widetilde G}$ contribution to the  strong interaction, which is proportional to the   parameter ${\bar\theta}$~\cite{'tHooft:1976up,Jackiw:1976pf,Callan:1976je}. The CKM contribution to any observable EDM is many orders of magnitude smaller than current upper limits, providing a window of opportunity for discovering EDMs that arise from a non-zero $\bar\theta$ or  BSM physics.
 In contrast to CKM CP-violation, contributions of BSM physics need not be suppressed unless the CP-violating parameters themselves are small, or the mass scales are high. 
New BSM interactions are also required for baryogenesis to account for the cosmic matter-antimatter asymmetry.  EDMs provide a particularly important connection to baryogenesis if the CP-violation energy scale is not too high compared to the scale of electroweak symmetry-breaking, and if the responsible P-odd/T-odd interactions are flavor diagonal~\cite{Morrissey:2012db}.
 
This review is intended as a broad summary of how EDM experiment and theory have reached this point and how it will progress. To do so the motivations and impact of EDM measurements along with a context for interpreting the results in terms of a set of P-odd/T-odd low-energy parameters are presented in Sec.~\ref{sec:Theory}. State-of-the art experimental techniques and improvements that will drive progress are presented in some detail in Sec.~\ref{sec:Techniques}, followed by a review of the current status of all experiments and prospects for new and improved approaches. The interpretation of these experiments and the impact of improvements in the context of the low energy theory parameters are presented in Sec~\ref{sec:Interpretation}. The conclusions emphasize what will be necessary from both theory and experiment for continued progress and, perhaps, the discovery of an EDM.

We also draw attention to a long list of  important and classic reviews  that include or are fully devoted to EDMs and provide a number of different perspectives as well as the history of the motivations and context:~\textcite{Garwin1959},
 \textcite{rf:Shapiro1968,rf:Shapiro1968-ru}, \textcite{Sandars1975,Sandars1993},  \textcite{RamseyEDMReview,doi:10.1146/annurev.ns.0.120190.000245}, \textcite{BM89},  \textcite{Bernreuther:1990jx,Bernreuther:1990jx-erratum}, \textcite{1402-4896-1993-T46-012}, \textcite{rf:GolubLamoreaux},   \textcite{KL1997}, \textcite{COMMINS19991}, \textcite{doi:10.1080/00107510010027781}, \textcite{Ginges:2003qt}, \textcite{Pospelov:2005pr}, \textcite{doi:10.1143/JPSJ.76.111010}, \textcite{LDM-2009},  \textcite{CHUPP2010129}, \textcite{Dubbers:2011ns}, \textcite{Engel:2013lsa},  \textcite{doi:10.1146/annurev-nucl-102014-022331}, and \textcite{Yamanaka:2017mef}.

\subsection{Experimental landscape}
The neutron was the objective of the early direct EDM measurements of Smith, Purcell and Ramsey due to the reasoning that it was a neutral hadronic (weakly interacting) system and would not be accelerated from the measurement region by a large static electric field~\cite{rf:Purcell1950,rf:Smith1957}. The early neutron-EDM measurements~\cite{Miller1967nEDM,rf:Baird1969,rf:Cohen1969} culminating in the measurement reported by~\textcite{rf:Dress1977}, used molecular-beam techniques developed to measure the neutron magnetic moment. This limited the observation time for a neutron transiting a meter-scale apparatus to milliseconds, resulting in linewidths of hundreds of Hz or more. The beam approach also had significant limitations due to a number of systematic effects, including the interaction of the neutron magnetic moment with the motional magnetic field ($\vec E\times\vec v/c^2$) and leakage currents, both of which shifted the energy when the electric field was changed. By 1980, advances in ultra-cold neutron (UCN) production at Leningrad Nuclear Physics Institute (LNPI)~\cite{Altarev1980,Altarev1981b} and at the Institute Laue-Langevin (ILL) in Grenoble, France~\cite{Smith1990}, enabled the first EDM measurements with neutrons stored in a ``cell.'' The much smaller UCN speeds and the nearly zero average velocity mitigated the motional effects, leading to a series of increasingly precise neutron EDM measurements, which are discussed in section~\ref{sec:NeutronEDM}. As the rate of UCN production improved, leading to smaller statistical errors, the control of the magnetic field required advances in magnetic shielding and magnetometry discussed in Sec~\ref{sec:Techniques}. Comagnetometry (the use of a second species  less sensitive to CP-violation but with a similar magnetic moment in the same measurement volume and at the same time) mitigated magnetic-field instability and a number of systematic effects. The neutron-EDM experiments are rate/statistics limited with typically only a few thousand UCN per measurement cycle, and advances require new UCN sources, which are discussed in detail in Sec~\ref{sec:UCNSources}, and corresponding improvements to magnetic shielding, magnetometry and understanding of systematic effects.


The earliest limits  on proton and electron EDMs were established by studies of corrections to the Lamb shift in hydrogen respectively by \textcite{Sternheimer1959} and by~\textcite{rf:Salpeter1958} and~\textcite{Feinberg1958}. Limits on the proton EDM were also set by analyzing the out-of-plane component of the proton spin polarization in an scattering asymmetry experiment  from a carbon target.~\cite{Rose:1960cla}, and  electron EDM limits were derived from frequency shifts in electron paramagnetic resonance
\cite{RB63},  anomalous magnetic-moment ($g-2$) measurements~\cite{Nelson1959,WC63}, and scattering measurements with spin-zero targets~\cite{PhysRev.114.1530} of helium~\cite{rf:Goldemberg1963,BURLESON196068,AVAKOV1959685} and carbon~\cite{PhysRev.140.B1605}. An early limit on the EDM of the muon was  derived from from analyzing the vertical component of the muon spin polarization the Nevis Cyclotron and fringe
fields, by the measuring asymmetry of the muon decay electrons.~\cite{BG60muon,PhysRevLett.1.144,PhysRevLett.1.144-err,Charpak1961}.

Starting in the 1960's, experimenters turned their attention to stable atoms and molecules beams in early beams experiments pioneered by~\textcite{rf:Sandars1964}.  It was recognized that these systems provided a rich set of possible contributions to the P-odd/T-odd observables, but the charged constituents, the electron and nucleus, are significantly shielded from the large external field by the polarization of the atom. This is embodied in Schiff's theorem~\cite{rf:Schiff1963}, which 
states that for a bound system of point-like charged particles the net force and the net electric field at the position of each charged particle are exactly zero. The shielding is not perfect for a nucleus of finite size and in the case of unpaired  electrons (paramagnetic systems) due to relativistic effects. 
In fact for paramagnetic atoms there is an effective enhancement of the sensitivity to an electron EDM that is approximately $10Z^3\alpha^2$ as explained by~\textcite{rf:Sandars1965,rf:Sandars1966,rf:Sandars1968I,rf:Sandars1968II},~\textcite{Ignatovich:1969tv,
Ignatovich:1969tv-ru}, and~\textcite{doi:10.1119/1.2710486}. Moreover, an atomic EDM can arise due to T and P violation in the electron-nucleus interaction that may have a scalar or tensor nature,  and these effects also increase with $Z$. 

Paramagnetic systems with one or more unpaired electrons (Cs, Tl, YbF, ThO, HfF$^+$ {\it etc.}) are most sensitive to both the electron EDM $d_e$ and the nuclear spin-independent component of the electron-nucleus coupling ($C_S$), which are likely to be several orders of magnitude stronger than tensor and pseudoscalar contributions, given comparable strength of the intrinsic couplings~\cite{Ginges:2003qt}. 
Diamagnetic systems, including $^{129}$Xe, $^{199}$Hg and $^{225}$Ra atoms, and the molecule TlF are most sensitive to purely hadronic CP-violating sources that couple through the Schiff moment $\vec S$, the $r^2$-weighted electric-dipole charge distribution for a nucleus with $Z$ protons,
\begin{equation}
\vec S =\frac{1}{10}\int\ r^2\vec r\rho_Q d^3r-\frac{1}{6Z} \int r^2d^3r \int\vec r\rho_Q d^3r.
\label{eq:SchiffDef}
\end{equation}
The EDM of the nucleus $\int\vec r\rho_Q d^3r=\vec d_N$ is unobservable in a neutral atom and the second term is therefore subtracted from $\vec S$. \textcite{flambaum02} have shown that an effective model of the Schiff moment is a constant electric field with the nucleus directed along the nuclear spin, which is probed by the atomic or molecular electrons through the interaction
\begin{equation}
H=-4\pi\vec\nabla\rho_e(0)\cdot\vec S,
\end{equation}
where $\vec\nabla\rho_e(0)$ is the gradient of the electron density at the nucleus.
As the atomic electrons penetrate the nucleus, the Schiff-moment electric force moves the electron cloud with respect to the atom's center of mass and induces an EDM along the spin.\footnote{We note that Schiff's Theorem has been recently reevaluated in work showing that these formulas may be unjustified approximations~\cite{rf:Liu2007}; however, there is disagreement on the validity of this reformulation \cite{rf:Senkov2008}.
}
 In addition, the EDM of a diamagnetic atom or molecule can arise due to the tensor component of the electron-nucleus coupling $C_T$ for atoms and molecules. The electron EDM and $C_S$ contribute to the EDM of diamagnetic atoms in higher order.  The magnetic quadrupole moment, a P-odd and T-odd distribution of currents in the nucleus is not shielded in the same way as electric moments and   induces an atomic EDM by coupling to an unpaired electron~\cite{Flambaum:1984fb,Flambaum:1984fb-ru}. The magnetic quadrupole moment requires a paramagnetic atom with nuclear spin $I>1/2$, and cesium is the only experimental system so far that meets these requirements. \textcite{Murthy:1989zz} have presented an analysis of their experiment on cesium that extracts the magnetic quadrupole moment.

EDM searches are not confined to neutral systems. Charged particles and ions can be contained in storage rings or with time-dependent fields. For example, the paramagnetic molecular ion HfF$^+$ was stored with a rotating electric field~\cite{Cairncross:2017fip}, and the EDM of the muon was measured in conjunction with the $g-2$, magnetic-moment anomaly measurements at Brookhaven~\cite{Bennett:2006fi}. In the muon experiment, spin-precession due to the EDM coupling to the motional electric field ($\vec v\times \vec B$) was measured, and an upper limit on $d_\mu$ was reported~\cite{Bennett:2008dy}. Though not a dedicated EDM measurement, the technique has demonstrated the possibility of a significantly improved measurement, which is motivated by theoretical suggestions that  lepton EDMs may scale with a power of the lepton mass~\cite{Babu:2000cz}. Storage-ring EDM searches have also been proposed for light nuclei, {\it i.e.} the proton, deuteron and helion ($^3$He$^{++}$)~\cite{Farley04,Khriplovich98,Rathmann:2013rqa}. The electron EDM can also be measured in special ferro-electric and paramagnetic solid-state systems with quasi-free electron spins that can be subjected to applied electric and magnetic fields~\cite{Eckel:2012aw}. We also note that the EDM of the $\Lambda$ hyperon was measured in a spin-precession measurement~\cite{rf:Pondrom1981}, and that limits on the $\tau$ lepton EDM~\cite{Inami:2002ah} and on neutrino EDMs have been derived~\cite{COMMINS19991, doi:10.1143/JPSJ.76.111010}.

A compilation of experimental results is presented in Table~\ref{tb:EDMResults}, which separates paramagnetic (electron-spin dependent) systems from diamagnetic (nuclear and nucleon spin-dependent) systems. 
In order to cast all results consistently, we have expressed the upper limits (u.l.) at 95\% confidence levels.\footnote{The upper limit $l$ is defined by the $\int_{-l}^l P^\prime(x)dx=0.95$ for a normalized Gaussian probability distribution $P^\prime(x)$ with central value and $\sigma$ given by the total error given in Table~\ref{tb:EDMResults}}

\begin{table}
\centering
\begin{tabular}{|c|l|cr|c|}
\hline\hline
  & {Result}&\multicolumn{2}{|c|} {95\% u.l.} & ref. \\
\hline
\multicolumn{5}{|c|}{Paramagnetic systems}\\
\hline
Xe$^m$ & $d_A=(\ \ 0.7\pm 1.4)\times 10^{-22}$ & $3.1\times 10^{-22}$  & \ecm &  $a$\\
\hline
Cs  &  $d_A=(-1.8\pm6.9)\times 10^{-24}$  & $1.4\times 10^{-23}$& \ecm & $b$ \\
& $d_e=(-1.5\pm 5.7)\times 10^{-26}$    & $1.2\times 10^{-25}$ &\ecm &  \\
&$C_S =  {(2.5\pm9.8)}\times 10^{-6}$ & $2\times 10^{-5}$& &\\
&$Q_m = {(3\pm13)}\times 10^{-8} \  $  & $2.6\times 10^{-7}$&$\mu_N R_\mathrm{Cs}$ &\\
\hline
Tl  &$d_A=(-4.0\pm 4.3)\times 10^{-25}$   & $1.1\times 10^{-24}$ &\ecm &$c$ \\
&   $d_e=(\quad 6.9\pm 7.4)\times 10^{-28}$  & $1.9\times 10^{-27}$& \ecm &  \\
 \hline
YbF &  $d_e=(-2.4\pm 5.9)\times 10^{-28}$     & $1.2\times 10^{-27}$ & \ecm& $d$ \\
\hline
 ThO    &  $d_e=(-2.1\pm 4.5)\times 10^{-29}$   &$9.7\times 10^{-29}$ &\ecm & $e$  \\
     &  $C_S=(-1.3\pm 3.0)\times 10^{-9}$ &$6.4\times 10^{-9}$ & &  \\  
\hline
HfF$^+$     &  $d_e=(0.9\pm 7.9)\times 10^{-29}$  &$1.6\times 10^{-28}$ &\ecm  & $f$  \\
\hline
\multicolumn{ 5}{|c|}{Diamagnetic systems}\\
\hline 
 $^{199}$Hg & $d_A=(2.2\pm 3.1)\times 10^{-30}$  &  $7.4\times 10^{-30}$  & \ecm&$g$\\
\hline
$^{129}$Xe &   $d_A=(0.7\pm 3.3)\times 10^{-27}$  & $6.6\times 10^{-27}$ & \ecm&$h$\\
\hline
$^{225}$Ra &   $d_A=(4\pm 6)\times 10^{-24}$ & $1.4\times 10^{-23}$ &\ecm&$i$\\
\hline
TlF &  $d_{\rm\ }=(-1.7\pm 2.9)\times 10^{-23}$   & $6.5\times 10^{-23}$ & \ecm&$j$\\
\hline
n &  $d_n=(-0.21\pm1.82)\times 10^{-26}$     & $3.6\times 10^{-26}$  &\ecm&$k$\\
\hline\multicolumn{ 5}{|c|}{Particle systems}\\
\hline 
$\mu$ & $d_\mu=(0.0\pm 0.9)\times 10^{-19}$  &$1.8\times 10^{-19}$  & \ecm&$l$\\
\hline 
$\tau$ & $Re(d_\tau)=(1.15\pm 1.70)\times 10^{-17}$  &$3.9\times 10^{-17}$  & \ecm&$m$\\
\hline $\Lambda$ & $d_{\Lambda}=(-3.0\pm 7.4)\times 10^{-17}$  & $1.6\times 10^{-16}$& \ecm& $n$\\
\hline
\hline 
\end{tabular}
\caption{Systems with EDM results and the most recent results as presented by the authors. When $d_e$ is presented by the authors, the assumption is $C_S=0$, and for ThO, the $C_S$ result assumes $d_e=0$. $Q_m$ is the magnetic quadrupole moment, which requires a paramagnetic atom with nuclear spin $I>1/2$. ($\mu_N$ and $R_{\rm Cs}$ are the nuclear magneton and the nuclear radius of $^{133}$Cs, respectively.)
We have combined statistical and systematic errors in quadrature for cases where they are separately reported by the experimenters. References;
$a$~\cite{rf:Player1970}; $b$ \cite{Murthy:1989zz}; $c$  \cite{Regan:2002ta}; $d$ \cite{Hudson:2011zz}; $e$  \cite{Baron:2013eja}; $f$  \cite{Cairncross:2017fip}; $g$  \cite{Graner:2016ses-erratum}; $h$ \cite{rf:Rosenberry2001}; $i$ \cite{rf:Parker2015}; $j$ \cite{rf:Cho1991}; $k$ \cite{Afach:2015sja}; $l$ \cite{Bennett:2008dy}; $m$~\cite{Inami:2002ah}; $n$ \cite{rf:Pondrom1981}. 
 \label{tb:EDMResults}
}
\end{table}

\subsection{Theoretical interpretation}

\label{sec:TheoreticalInterpretation}

\begin{figure}[tb]
\vskip -0.75 truein
\centerline{\includegraphics[width=4.25 truein]{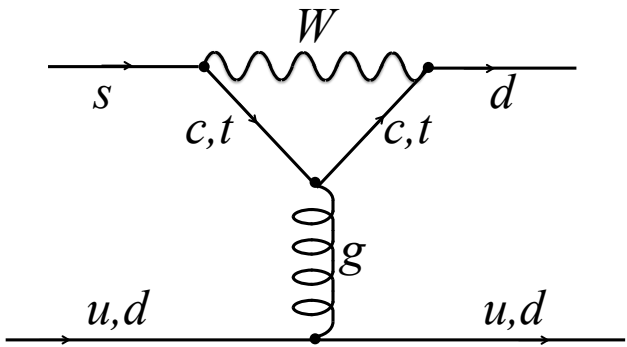}}
\vskip -0.75 truein
\caption{\label{fg:NeutronEDMVertex}  Penguin diagram  giving the SM-CKM $\Delta S=1$ CP-violating effective interaction. Adapted from \textcite{Pospelov:2005pr}.}
\end{figure}

The results on EDMs presented in Table~\ref{tb:EDMResults} have significant theoretical impact in several contexts by constraining explicit parameters of  SM and BSM physics. The Standard Model has two explicit CP-violating parameters: the phase in the CKM matrix, and the coefficient $\bar\theta$ 
 in the SM strong interaction Lagrangian.  
EDMs arising from the CKM-matrix vanish up to three loops for the electron~\cite{Bernreuther:1990jx} and up to two loops for quarks~\cite{Shabalin:1978rs,Shabalin:1978rs-ru,Shabalin:1982sg,Shabalin:1982sg-ru}.  The leading SM contributions to the neutron EDM, however, arise from a combination of hadronic one-loop and 
resonance contributions,  each a combination of two $\Delta S=1$ hadronic interactions (one CP violating and one CP-conserving). The CP-violating $\Delta S=1$ vertex is itself a one-loop effect, arising from the QCD  \lq\lq Penguin" process (See FIG.~\ref{fg:NeutronEDMVertex}). 
The estimate of the corresponding neutron EDM is $(1-6)\times 10^{-32}$ \ecm~\cite{Seng:2014lea}, where the range reflects the present hadronic uncertainties. For both the electron and the neutron, the SM CKM contribution lies several orders of magnitude below the sensitivities of recent and next-generation EDM searches. The Penguin process generated by the exchange of a kaon between two nucleons induces CP-violating effects in nuclei; however \textcite{Donoghue:1987dd} and \textcite{Yamanaka:2015ncb} show that this contribution is also many orders of magnitude below current experimental sensitivity for diamagnetic atom EDMs.
EDMs of the neutron and atoms also uniquely constrain the SM strong-interaction parameter $\bar\theta$ which sets the scale of strong CP violation as discussed in Sec.~\ref{sec:Theory}). 


BSM theories generally provide new degrees of freedom and complex CP-violating couplings that often induce EDMs at the one-loop level. The most widely-considered BSM scenarios for which implications have been analyzed include supersymmetry~\cite{Pospelov:2005pr,RamseyMusolf:2006vr}, the two-Higgs model~\cite{Inoue:2014nva}, and left-right symmetric models~\cite{Pati:1974yy,Pati:1974yy-erratum,Mohapatra:1974hk,Senjanovic:1975rk}.


A complementary, model-independent framework for EDM interpretation relies on effective field theory (EFT), presented in detail Sec.~\ref{sec:EFTParameters}). 
The EFT approach assumes that the BSM particles are sufficiently heavy that their effects can be compiled into  a  set of residual weak-scale, non-renormalizable operators involving only SM fields. The corresponding operators are dimension six  
 and  effectively depend on $(v/\Lambda)^2$, where 
$v=246$ GeV is the Higgs vaccum-expectation-value and $\Lambda$ is the energy scale of the new physics. The strength of each operator's contribution is characterized by a corresponding Wilson coefficient.  In addition to $\bar\theta$ there are the following 12 dimension-six BSM Wilson coefficients representing the intrinsic electron EDM, up-quark and down-quark EDMs and CEDMs, one CP-violating three gluon operator, five four-fermion operators, and one quark-Higgs boson interaction operator. Experimental EDM results constrain the Wilson coefficients, while a given BSM theory provides predictions for the Wilson coefficients in terms of the underlying model parameters. 

Interactions involving light quarks and gluons are, of course, not directly accessible to experiment. Consequently, it is useful to consider their manifestation in a low-energy effective theory (below the hadronic scale $\Lambda_\mathrm{had}\sim 1$ GeV) involving electrons, photons, pions, and nucleons. Hadronic matrix elements of the quark and gluon EFT operators then yield the hadronic operator coefficients. At lowest non-trivial order, one obtains the electron EDM ($d_e$); scalar, pseudoscalar, and tensor electron-nucleon interactions ($C_S$, $C_P$, and $C_T$, respectively\footnote{Each interaction has an isoscalar and isovector component, which we have suppressed here for notational simplicity.}; short-range neutron and proton EDMs ($\bar d_n^{sr}$ and $\bar d_p^{sr}$);  isoscalar, isovector, and isotensor pion-nucleon couplings ($\gpbi$, $i=0,1,2$); and a set of four-nucleon operators. 
In the context of this hadronic-scale EFT, it is appropriate to express the combination of contributions to a measured atomic  EDM for paramagnetic systems, diamagnetic systems, and nucleons as
\begin{equation}
d_i=\sum_{j} \alpha_{ij} C_j,
\label{eq:d_i}
\end{equation}
where $i$ labels the system, and $j$ labels the specific low-energy parameter ({\em e.g.}, $d_e$, $C_S$, {\em etc.}). 
The $\alpha_{ij}=\partial d_i/\partial C_j$ are provided by theoretical calculations at various scales from atomic to nuclear to short-range and are presented in Sec.~\ref{sec:Theory}. Note that the coefficients $\alpha_{ij}$ have various labels in the literature for notation developed for the different experimental systems. 

One approach to interpreting the experimental limits assumes that the EDM in a specific system arises from only one source -- the \lq\lq sole-source" approach. In the sole-source approach,  the constraint on each parameter is derived assuming that all other contributions are negligible and so one experimental result may appear to set limits on a large number of individual CP-violating parameters. An alternative approach  -- the global analysis  presented in~\textcite{Chupp:2014gka} and Sec.~\ref{sec:GlobalAnalysis} -- assumes simultaneous non-zero values of the dominant parameters globally constrained by the experimental results. In the global analysis, paramagnetic systems are used to set limits on the electron EDM $d_e$ and the nuclear spin-independent electron-nucleus coupling $C_S$.  Diamagnetic systems set limits on  four dominant parameters: two pion-nucleon couplings ($\gpbz,\gpbo$), a specific isospin combination of nuclear spin-dependent couplings, and the  \lq\lq short distance" contribution to the neutron EDM, ${\bar d}^{sr}_n$. There is, unfortunately, significant variation and uncertainty in the $\alpha_{ij}$, in particular for the nuclear and hadronic calculations, which soften the constraints on the low-energy parameters.

\subsection{Reach and complementarity}
\label{sec:Reach}

EDMs arising from BSM CP violation depend on a combination of factors, including new CP-violating phases, $\phi_{\rm CPV}$, the mass scale $\Lambda$ associated with the new particles, and the underlying dynamics. In general, an elementary fermion EDM can be expressed as
\begin{equation}
d \approx (10^{-16}  \ecm~)  \left(\frac{v}{\Lambda}\right)^2\ (\sin\phi_{\rm CPV}) (y_f\, F) \ \ \ ,
\label{eq:edmbsm}
\end{equation}
where $v=246$ GeV is the Higgs vacuum expectation value, $y_f$ is a Yukawa coupling typically associated with the SM fermions in the system of interest, and $F$ accounts for the dynamics, which may be perturbative or non-perturbative and will differ depending on the system. 

For an electron EDM that arises through perturbative dynamics at the  one-loop level, $F\sim g^2/(16\pi^2)$ where $g$ is the BSM coupling strength. The present electron EDM upper limit of $\approx 1\times 10^{-28}$ \ecm~ 
implies that $\Lambda\gtrapprox 1-2$ TeV for $g$ of order the SM SU(2$)_L$ gauge coupling strength, and $\sin\phi_{\rm CPV}\approx 1$. This energy scale for $\Lambda$, which is comparable to the reach of the neutron and diamagnetic atom EDM limits, rivals the BSM physics reach of the LHC.  
It is important to note that exceptions to these na\"ive estimates of mass scale sensitivity can occur. For example at the level of the underlying elementary particle physics, an EDM may be enhanced by contributions of heavy fermion intermediate states, {\em e.g.}, the top quark, leading to the presence of a larger Yukawa coupling in Eqn.~(\ref{eq:edmbsm}). 
In paramagnetic atoms and molecules an EDM
 may also be generated by a  nuclear-spin independent scalar T-odd/P-odd electron-quark interaction at tree-level, which generally scales with the number of nucleons. 
In this case, the resulting mass reach for current experimental sensitivies is as high as $\sim 13,000$ TeV, as discussed in Section~\ref{sec:Interpretation}. 
Models that generate EDMs at two-loop or higher-loop order allow for lighter BSM particles with CP-violating interactions.

For the Schiff moment, the additional power of $r^2$ in Eqn.~(\ref{eq:SchiffDef})
 implies that for a given underlying source of CP-violation, the contribution to a diamagnetic atomic or molecular EDM is suppressed compared to that of the neutron by $(R_N/R_A)^2$, where $R_N$ and $R_A$ are the nuclear and atomic radii, respectively. As a concrete illustration the bound on $\bar{\theta}$ arising from the $^{199}$Hg EDM limit is comparable to the bound from $d_n$, even though the respective EDM limits differ by nearly four orders of magnitude (see Table~\ref{tb:EDMResults}).

An example of unique constraints set by EDM searches is found in the strong CP contribution to the neutron EDM given by~\textcite{Pospelov:1999ha,Crewther:1979pi,rf:Crewther1979-erratum} and \textcite{Shindler:2015aqa}
\begin{equation}
d_n \approx (10^{-16} e-{\rm cm})\ {\bar\theta}\ \ \ .
\label{eq:dnthetabar}
\end{equation}
The parameter $\bar\theta$ is na\"ively expected to be of order unity. However assuming this is the only contribution to the neutron EDM, the current upper bound, from Table~\ref{tb:EDMResults} implies
$|{\bar\theta}|\lessapprox 10^{-10}$. The corresponding bound obtained from the $^{199}$Hg EDM limit is comparable. This severe constraint on $\bar\theta$ has 
motivated a variety of  theoretical explanations. The most widely considered explanation is the existence of a spontaneously-broken Peccei-Quinn symmetry~\cite{Peccei:1977ur,Peccei:1977hh} and an associated particle - the axion~
\cite{Weinberg:1977ma,Wilczek:1977pj}. The axion is also  a candidate for the observed relic density of cold dark matter. The axion proposal has been very compelling and has spawned a number of experimental endeavors summarized, for example by~\textcite{doi:10.1146/annurev-nucl-102014-022120}.


\begin{figure*}[tb]
\vskip -1 truein
\includegraphics[width=7 truein]{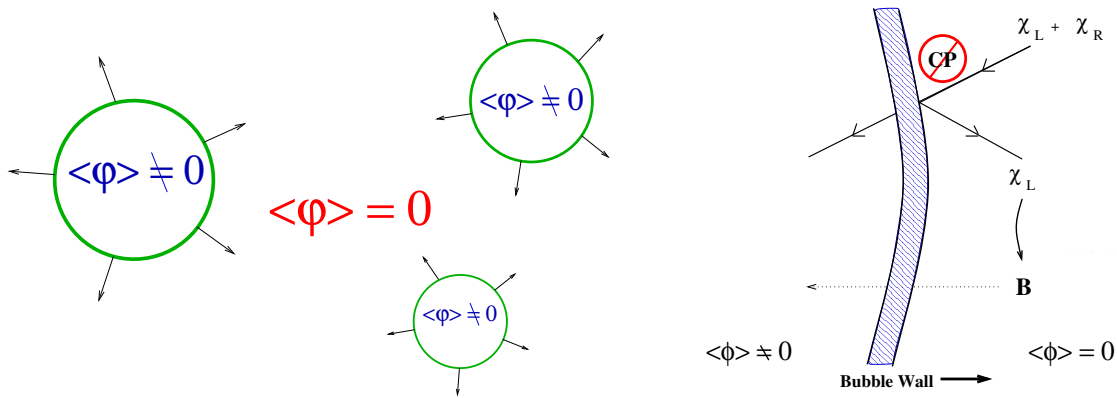}
\vskip -1.5 truein
\caption{\label{fg:EWBGBubbles}  (Color online) Left side: bubble nucleation during first order electroweak phase transition. Right side: CP- and C-violating interactions in the $\langle\phi_H\rangle = 0$ background near the bubble walls that produce baryons. Figure \copyright~ IOP Publishing Ltd and Deutsche Physikalische Gesellschaft;  reproduced from~\textcite{Morrissey:2012db}  by permission of IOP Publishing. }
\vskip .5 truein
\end{figure*}

\subsection{EDMs and baryogenesis}

Baryogengesis, the generation of a net asymmetry of matter over antimatter in the early universe, requires three components as first explained by~\textcite{SakharovBaryogenesis-ru,SakharovBaryogenesis}: (1) violation of baryon number $B$; (2) departure from thermodynamic equilibrium (assuming CPT invariance); and 3) both C-violating and CP-violating  processes. A number of baryogenesis scenarios that satisfy these requirements have been proposed, each typically focusing on a certain era in cosmic history and corresponding energy scale. Among the most widely considered and experimentally testable scenarios is electroweak baryogenesis. For  recent reviews of electroweak baryogenesis, see \textcite{Morrissey:2012db}; and see~\textcite{Riotto:1999yt} and~\textcite{Dine:2003ax} for more general baryogenesis reviews. 

In the electroweak baryogenesis (EWBG) scenario, the universe proceeds from initial conditions having no net baryon number and vanishing Higgs field expectation value $\langle\phi_H\rangle=0$, implying that the Standard-Model SU(2)$_L\times$U(1)$_Y$ electroweak symmetry has not yet been broken by the Higgs mechanism (giving mass to the $W^\pm$ and $Z$ bosons). As the plasma  cools below the electroweak scale of $\approx$ 100 GeV, $\langle\phi_H\rangle$ becomes non-zero, breaking electroweak symmetry. EWBG requires that the transition be a first order phase transition, proceeding via the nucleation of bubbles of broken symmetry with $\langle\phi_H\rangle \ne 0$ as suggested in the left side of FIG.~\ref{fg:EWBGBubbles}. These bubbles expand and fuse into a single  phase with $\langle\phi_H\rangle \ne 0$. 
CP-violating and C-violating interactions in the $\langle\phi_H\rangle = 0$ background near the bubble walls produce $B\ne0$ in processes that convert baryons to antileptons or antibaryons to leptons illustrated in the  right side of FIG.~\ref{fg:EWBGBubbles}. 
These processes are referred to  as sphaleron transitions that arise from a configuration of the Standard-Model fields in the $\langle\phi_H\rangle=0$ phase that corresponds to a saddle point of the electroweak effective action~\cite{Klinkhamer:1984di}. As the bubbles expand and merge, they sweep up the $B\ne 0$ regions as they eventually coalesce into the universe that persists to the present epoch. However if sphaleron transitions persist inside the bubbles the baryon number would not be preserved.  Thus, the first order transition must be sufficiently \lq\lq strong"  so as to quench the sphaleron transitions inside the broken phase.  

In principle, the SM with CP-violation from the CKM matrix provides all of the ingredients for this scenario; however the phase transition cannot be first order for a Higgs mass greater than $\sim$70 GeV. Given the observed Higgs mass  $m_H=125.09\pm 0.24$ GeV~\cite{rf:HiggsMassdoi10.1103/PhysRevLett.114.191803}, a first order phase transition cannot have occurred in a purely SM universe. Even if the value of $m_H$ were small enough to accommodate a first order electroweak phase transition,  the effects of CKM CP-violation are too feeble to have resulted in the observed matter-antimatter asymmetry. Thus, electroweak baryogenesis requires BSM physics for two reasons: generation of a strong, first order electroweak phase transition and production of sufficiently large CP-violating asymmetries during the transition. New particle searches at colliders may discover new interactions responsible for a first-order phase transition~\cite{Assamagan:2016azc,Contino:2016spe}, but it is EDMs that provide the most powerful probe of the new CP-violating interactions.

\begin{figure*}[htb]
\includegraphics[width=3.25 truein]{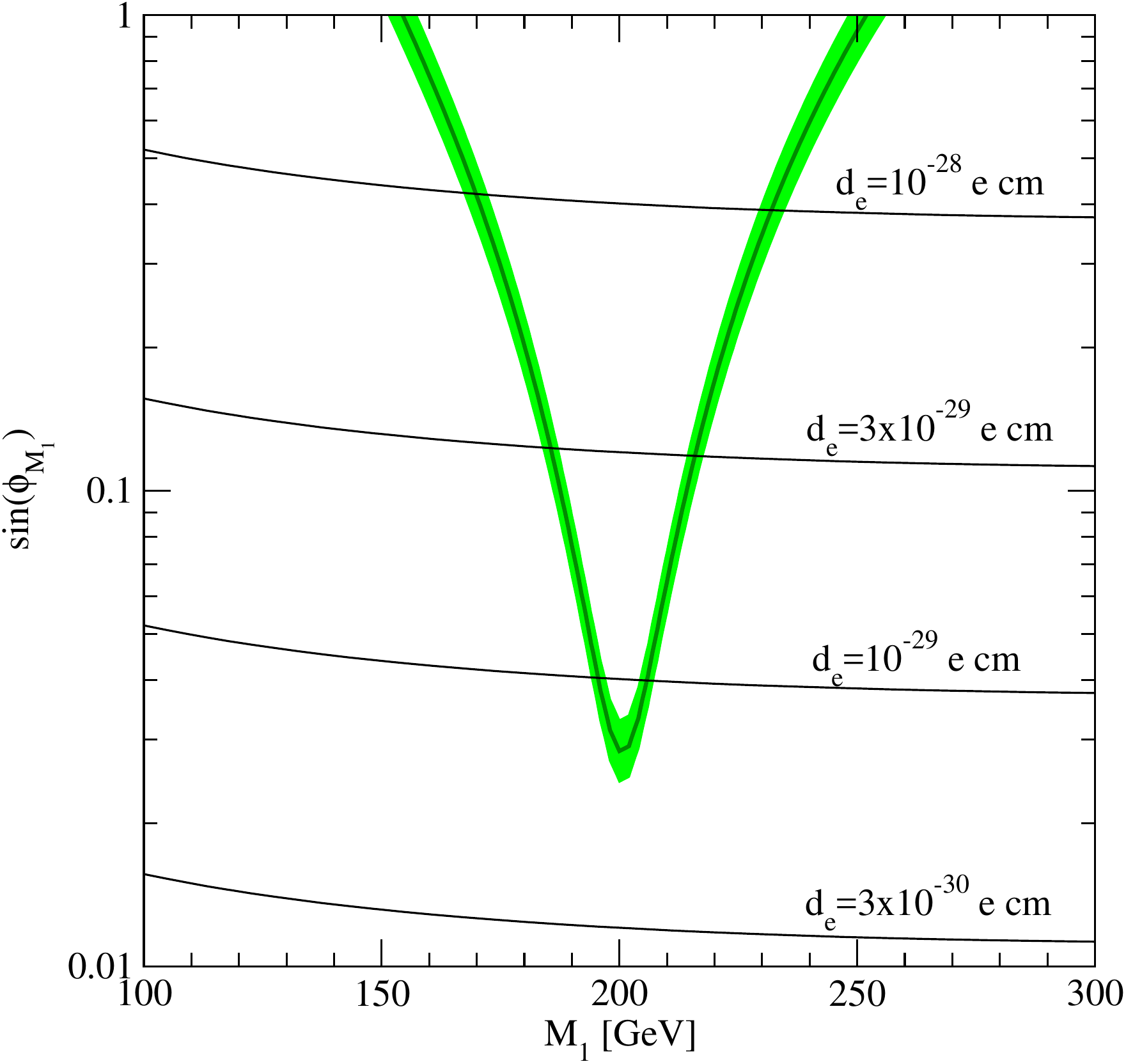}\quad\includegraphics[width=3.25 truein]{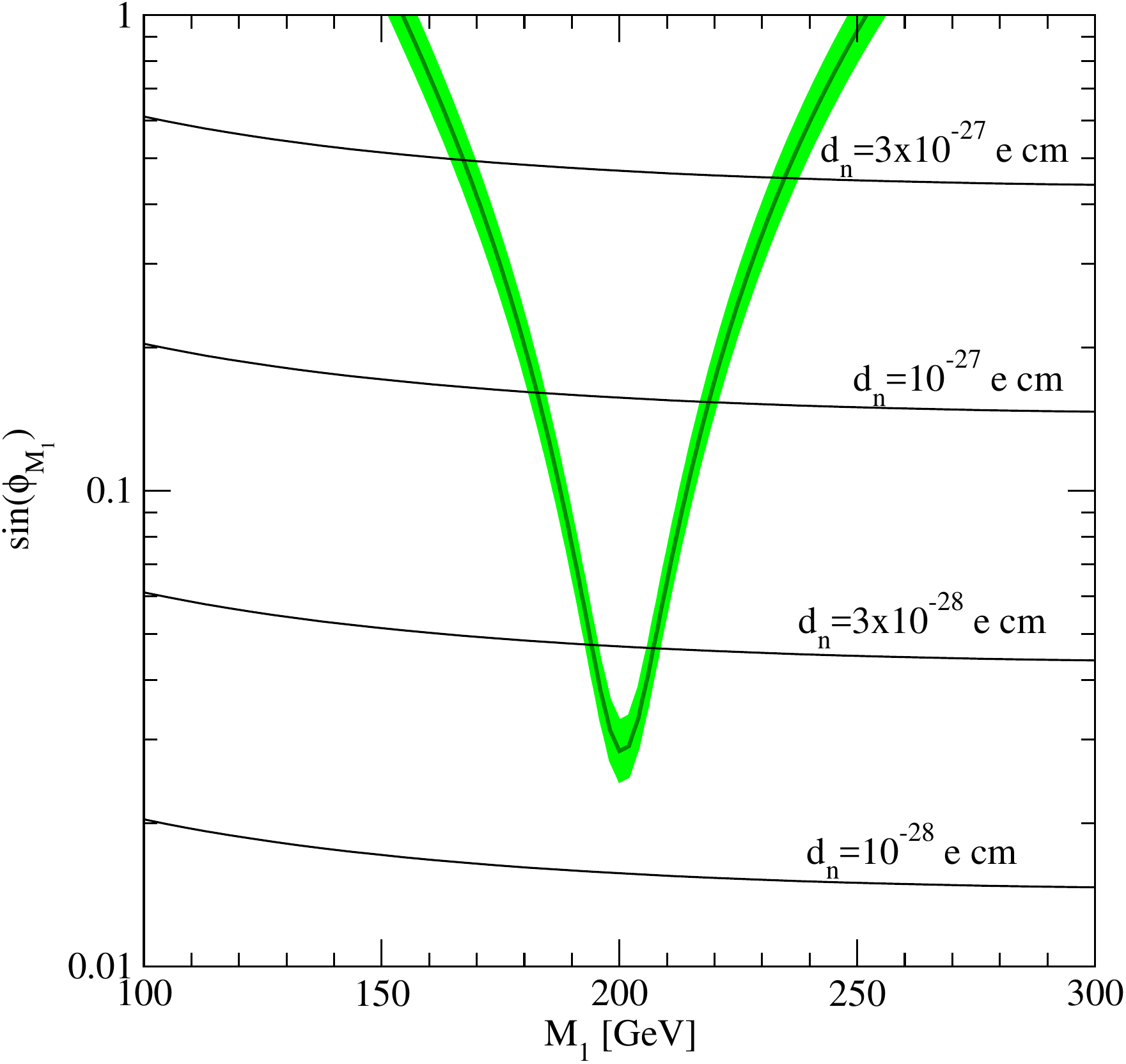}
\caption{\label{fg:MSSMdedn}  (Color online) Sensitivity of the electron EDM (left panel) and neutron EDM (right panel) to the baryon asymmetry in the MSSM.  The horizontal axes give the bino soft mass parameter, $M_1$; the vertical axes give the sine of the relative phase of $M_1$, the supersymmetric $\mu$ parameter, and the soft Higgs mass parameter $b$. The green bands indicate the values of these parameters needed to obtain the observed baryon asymmetry. Nearly horizontal lines give contours of constant EDMs. Figure originally published in \textcite{Li:2008ez}.  }
\end{figure*}

Electroweak baryogenesis  provides an additional constraint on the BSM mass scale $\Lambda$ and on CP-violating phase(s) that set the scale of EDMs. Eqn.~\ref{eq:edmbsm} shows that experimental limits on EDMs constrain the ratio $\sin\phi_{\rm CPV}/\Lambda^2$ but do not separately constrain $\Lambda$  and $\sin\phi_{\rm CPV}$. However the requirements for electroweak baryogenesis do provide complementary constraints on the mass scale and CP-violating phases. We illustrate this in FIG.~\ref{fg:MSSMdedn} from~\textcite{Li:2008ez}, which shows constraints on parameters of the minimal supersymmetric Standard Model (MSSM) needed to generate the observed matter-antimatter asymmetry and the corresponding EDMs that would arise. The horizontal and vertical axes give the soft SUSY-breaking {\it bino} mass parameter $M_1$ and the CP-violating \lq\lq  bino phase" $\sin(\Phi_M)=\sin\mathrm{Arg}(\mu M_1b^\ast)$. The green band shows the relationship between these parameters needed to produced the matter-antimatter asymmetry, while the nearly horizontal lines indicate values of the electron (left panel) and neutron (right panel) EDMs. The EDMs have been computed in the limit of heavy {\it sfermions}, which is consistent with LHC results, so that EDMs arise from two-loop graphs containing the electroweak {\it gauginos}. Note that the present $d_e$ limit roughly excludes the region   $d_e>10^{-28}$ \ecm, while the current neutron-EDM bound does not yet constrain the indicated parameter space. The next generation electron and neutron EDM experiments are expected to probe below $d_e<10^{-29}$ \ecm~ and $d_n<10^{-27}$ \ecm.

\section{Theoretical background}
\label{sec:Theory}

\subsection{CP/T Violation}


Parity (P), Time Reversal (T) and Charge Conjugations (C) are the discrete symmetry transformations of quantum mechanics and quantum field theory. Experiment shows that the strong and electromagnetic interactions are symmetric under C, P and T separately and under CP and T. The weak interaction, which  involves only left-handed neutrinos and right-handed antineutrinos, is maximally antisymmetric under P and C . CP is also violated in weak decays of kaons and b-mesons. Symmetry under the combined transformations of C, P, and T (or CPT) is consistent with experiment and is also required for any Lorentz-invariant quantum field theory as embodied in the CPT theorem~\cite{Tureanu:2013psa,rf:LudersCPTTheorem,Jost:1957zz}.

Parity, a unitary transformation described by $(t,\vec r)\rightarrow (t_P,\vec r_P)=(t,-\vec r)$, reverses the handedness of the coordinate system, {\it i.e.} $\hat x\times\hat y=\hat z$ while $\hat x_P\times \hat y_P=-\hat z_P$. Particles have intrinsic parity, that is the field describing a particle acquires a phase of $\pm1$ under the parity transformation. For fermions, particles and antiparticles have opposite intrinsic parity. 

The time-reversal transformation is described by $(t,\vec r)\rightarrow (t_T,\vec r_T)=(-t,\vec r)$, but when this is applied to wave functions or fields not only is motion reversed (e.g. $\vec p\rightarrow \vec p$ and $\vec J\rightarrow -\vec J$) but the imaginary phase is reversed as well, {\it i.e.} time reversal includes complex conjugation. Moreover for scattering and decay processes, the initial and final state are reversed, which is a complication for interpreting any experiment subject to final sate corrections such as detailed balance or decay correlation measurements~\cite{rf:FSI}.
Thus the {\it anti-unitary} time-reversal transformation involves motion reversal, complex conjugation and reversal of initial and final states. For an EDM, however, the initial and final states are the same, so there is no complication from final-state effects and a definitive observation of an EDM is a direct signature of T-violation and, invoking the CPT theorem, of CP-violation.

Charge conjugation transforms particles into antiparticles, reversing charge without reversing the handedness or spin. It is interesting to note, therefore, that CPT symmetry requires the EDM   of a particle  and the EDM of its antiparticle  be equal in magnitude and opposite in sign:
\begin{equation}
d\langle\vec J\rangle\xrightarrow{C}\bar d\langle\vec J\rangle\xrightarrow{P}+\bar d\langle\vec J\rangle\xrightarrow{T}\bar d\langle-\vec J\rangle=-\bar d\langle\vec J\rangle.
\end{equation}
Similarly, for the magnetic moments:
\begin{equation}
\mu\langle\vec J\rangle\xrightarrow{C}\bar\mu\langle\vec J\rangle\xrightarrow{P}+\bar\mu\langle\vec J\rangle\xrightarrow{T}\bar\mu\langle-\vec J\rangle=-\bar\mu\langle\vec J\rangle.
\end{equation}

\subsection{General framework}


As indicated by FIG.~\ref{fg:EDMSubwayMap}, EDMs in experimentally accesible systems arise from CP-violation at a fundamental level that is manifest at several energy or length scales. The Lagrangian for a fundamental theory incorporating SM CKM and  $\bar\theta$ and contributions together with BSM physics can be written
\begin{equation}
\label{eq:LCPV1}
\mathcal{L}_\mathrm{\rm CPV} = \mathcal{L}_\mathrm{CKM}+\mathcal{L}_{\bar\theta}
+\mathcal{L}_\mathrm{BSM}. 
\end{equation}
The general framework that connects this to experiment, Effective Field theory (EFT),  absorbs higher-energy processes into a set of operators that contribute at a scale $\Lambda$ resulting in a set of weak scale, non-renormalizable operators involving only SM fields. The corresponding amplitudes scale as $(v/\Lambda)^{d-4}$, where $d$ is the operator's canonical dimension and $v=246$ GeV is the Higgs vacuum expectation value. 

The $\bar\theta$ term in $\mathcal{L}_\mathrm{\rm CPV}$ enters at EFT dimension four, while  CKM-generated fermion EDMs 
are dimension five, but elecro-weak $SU(2)\times U(1)$ gauge invariance requires coupling through the Higgs field making these effectively dimension six. 
BSM physics enters at dimension six and higher
 {\it i.e},
\begin{equation}
\mathcal{L}_\mathrm{BSM}\rightarrow \mathcal{L}_\mathrm{\rm CPV}^\mathrm{eff} = \sum_{k,d}\ \alpha_k^{(d)} \left(\frac{1}{\Lambda}\right)^{d-4} \mathcal{O}_k^{(d)},
\label{eq:bsmeff}
\end{equation}
where $\alpha_k^{(d)}$ is the Wilson coefficient for each operator $\mathcal{O}_k^{(d)}$,  $k$ denotes all operators for a given $d$ that are invariant under both $SU(2)$ and $U(1)$, and the operators contain only SM fields. However when considering only first generation SM fermions and SM bosons, it is sufficient to consider only $d=6$.
At this order, the relevant set of operators, {\it i.e.} the \lq\lq CP-violating sources"  listed in Table~\ref{tb:ThirteenParameters},
 include the fermion SU(2)$_L$ and U(1)$_Y$ electroweak dipole operators and the SU(3$)_C$ chromo-electric-dipole operators; a set of four fermion semi-leptonic and non-leptonic operators; a CP-violating three-gluon operator; and a CP-violating fermion-Higgs operator.  After electroweak symmetry-breaking, the  dipole operators induce the elementary fermion EDMs and Chromo-EDMs (CEDMs) as well as analogous fermion couplings to the massive electroweak gauge bosons that are not directly relevant to the experimental observables discussed in this review. The fermion-Higgs operator induces a four-quark CP-violating operator whose transformation properties are distinct from the other four-quark operators listed in Table~\ref{tb:ThirteenParameters}.


\begin{table}[t]
\centering \renewcommand{\arraystretch}{1.5}
\begin{tabular}{||c|c|c||}
\hline\hline
$\mathcal{O}_{fW}$ &$(\bar F \sigma^{\mu\nu} f_R) \tau^I \Phi\, W_{\mu\nu}^I$  & fermion SU(2)$_L$  EDM \\
$\mathcal{O}_{fB}$ & $(\bar F \sigma^{\mu\nu} f_R) \Phi\, B_{\mu\nu}$ & fermion U(1)$_Y$  EDM \\
$\mathcal{O}_{uG}$ & $(\bar Q \sigma^{\mu\nu} T^A u_R) \tvp\, G_{\mu\nu}^A$ & u-quark Chromo EDM\\
$\mathcal{O}_{dG}$ & $(\bar Q \sigma^{\mu\nu} T^A d_R) \vp\, G_{\mu\nu}^A$& d-quark Chromo EDM\\
\hline
$Q_{ledq}$ & $(\bar L^j e_R)(\bar d_R Q^j)$ & CP-violating semi-leptonic \\
$Q_{lequ}^{(1)}$ & $(\bar L^j e_R) \epsilon_{jk} (\bar Q^k u_R)$ & \\
$Q_{lequ}^{(3)}$ & $(\bar L^j \sigma_{\mu\nu} e_R) \epsilon_{jk} (\bar Q^k
\sigma^{\mu\nu} u_R)$ &  \\
\hline
$\mathcal{O}_{\wt G}$ & $f^{ABC} \wt G_\mu^{A\nu} G_\nu^{B\rho} G_\rho^{C\mu} $ & CP-violating 3 gluon \\
$Q_{quqd}^{(1)}$ & $(\bar Q^j u_R) \epsilon_{jk} (\bar Q^k d_R)$ & CP-violating four quark\\
$Q_{quqd}^{(8)}$ & $(\bar Q^j T^A u_R) \epsilon_{jk} (\bar Q^k T^A d_R)$ & \\
\hline
$Q_{\varphi ud}$ & $i\left({\tilde\varphi}^\dag D_\mu \varphi\right) {\bar u}_R\gamma^\mu d_R$ & CP-violating  quark-Higgs\\
\hline\hline
\end{tabular}
\caption{Dimension-six P-odd/T-odd operators that induce atomic, hadronic, and nuclear EDMs. Here $\varphi$ is the SM Higgs doublet, $\tvp=i\tau_2\vp^\ast$, and $\Phi=\vp$ ($\tvp$) for $I_3^f< 0$ ($I_3^f>0$). The notation is adapted from~\textcite{Engel:2013lsa}.
\label{tb:ThirteenParameters}}
\end{table}


The second term of the electromagnetic Lagrangian in Eq~\ref{eq:EMLagrangian} describes the EDM interaction for an elementary fermion $f$, which couples left-handed to right-handed fermions. Letting the Wilson coefficient $\alpha_{fV_k}^{(6)} = g_k C_{fV_k}$, where $k=B,\ W,\ G$ labels the Standard-Model  electroweak ($B$ and $W$) and gluon ($G$) gauge fields 
\bea
{\cal L_\mathrm{EDM}}&=&-i\frac{d_f}{2}\bar\Psi \sigma^{\mu\nu}\gamma^5 F_{\mu\nu}\Psi\nonumber\\
&=&\frac{1}{\Lambda^2}(g_B C_{f B}\mathcal{O}_{fB}+2I_3g_W C_{f W}\mathcal{O}_{fW}),\nonumber\\
\eea
where $d_f$ is the  fermion EDM, which couples to the EM field $A^\mu=B^\mu\cos\theta_W+W_3^\mu\sin\theta_W$ ($\theta_W$ is the SM Weinberg angle), is
 \bea
\label{eq:prdef}
\nonumber
d_f  &=&   - \frac{\sqrt{2} e}{v}\ \left(\frac{ v}{\Lambda}\right)^2\ 
(\mathrm{Im}\ C_{f B}+ 2I_3^f \; \mathrm{Im}\ C_{f W})\\
& = & -(1.13\times 10^{-13}\ e\ \mathrm{fm})\  \left(\frac{ v}{\Lambda}\right)^2\  \mathrm{Im} C_{f\gamma}.
\eea
Here 
\be
\mathrm{Im} C_{f\gamma} =\mathrm{Im}\ C_{f B}+ 2I_3^f \; \mathrm{Im}\ C_{f W}
\ee
and $I_3^f$ is the third component of weak isospin for fermion $f$.  
The CP-violating quark-gluon interaction can be written in terms of an analogous chromo-EDM (CEDM) $\tilde d_q$: 
\be
\label{eq:cedmdef}
\mathcal{L}_\mathrm{CEDM} = -i\sum_q\ \frac{g_3 {\tilde d}_q}{2}\ 
{\bar q} \sigma^{\mu\nu} T^A\gamma_5 q\ G_{\mu\nu}^A \ ,
\ee
where $T^A$ ($A=1, \ldots, 8$) are the generators of the QCD color group. Note that $\tilde d_q$ has dimensions of length, because it couples to gluon fields, not EM fields.


Due to electroweak gauge invariance, the coefficients of the operators that generate EDMs and CEDMs   ($Q_{q \wt G}$, $Q_{f \wt W}$, $Q_{f \wt B}$) contain explicit factors of the Higgs 
field with Yukawa couplings $Y_f=\sqrt{2} m_f/v$. We can write
$\mathrm{Im}\ C_{{f\gamma }}  \equiv  Y_f\, {\delta}_f $, {\it etc.} so that
\be
 d_f =  -(1.13 \times 10^{-3}\ e\, \mathrm{fm})  \, 
\left(\frac{v}{\Lambda}\right)^2\, Y_f\,  {\delta}_f \  ,
\label{eq:nda1prime}
\ee
and
\be
{\tilde d}_q = -(1.13 \times 10^{-3}\ \mathrm{fm})\,  
\left(\frac{v}{\Lambda}\right)^2\, Y_q\,  {\tilde\delta}_q.
\label{eq:nda1}
\ee
In general, we expect $\delta_q\sim \delta_\ell$, thus the up and down-quark EDMs would be comparable, but
 light quark EDM $d_q$ would be roughly an order of magnitude larger than the  electron EDM. As noted earlier, exceptions to this expectation can arise when the Higgs couples to heavy degrees of freedom in the loop graphs that generate quark EDMs and CEDMs. 

Considering only first generation fermions, there are fifteen independent weak-scale coefficients. Translating the electroweak dipole operators into the elementary fermion EDMs and neglecting couplings to massive gauge bosons leads to a set of twelve $d=6$ CP-violating sources -- in addition to the $\bar{\theta}$ parameter -- that induce atomic, hadronic and nuclear EDMs.

\subsection{Low-energy parameters}
\label{sec:LowEnergyParameters}
\label{sub:lowE}

As indicated in FIG.~\ref{fg:EDMSubwayMap}, the Wilson coefficients are connected to the experimental observables at the hadronic scale, $\Lambda_\mathrm{had}\sim 1$ GeV, through a set of low-energy parameters involving pions, and nucleons in place of quarks and gluons as well as photons and electrons. 
Considering first purely hadronic interactions, the starting point is a T-odd/P-odd (or TVPV) effective, non-relativistic Lagrangian containing pions and nucleons~\cite{Engel:2013lsa}:
\bea
\label{eq:pinn}
\mathcal{L}_{\pi NN}^\mathrm{TVPV}\!\!\! &=&\!\! -2{\bar N} \left(\bar{d}_0+\bar{d}_1\tau_3\right)S_\mu N v_\nu F^{\mu\nu}\nonumber\\
\nonumber
&+&\!\!\! {\bar N}\left[\gpbz{\vec\tau}\cdot{\vec\pi} +\gpbo \pi^0 + \gpbt\, \left(3\tau_3\pi^0-{\vec\tau}\cdot{\vec\pi}\right)\right] N,\nonumber\\
\eea
where $S_\mu$ is the spin of a nucleon $N$ having velocity $v_\nu$, $\vec\tau$ is the isospin operator, and $\vec\pi={\pi^+,\pi^0,\pi^-}$ represents the pion field. Four-nucleon interactions are currently being studied and are not considered in this discussion.
Combinations $\bar d_0+\bar d_1\tau_3=\bar d_0\mp \bar d_1$ correspond to the short-range contributions to the neutron and proton EDMs.
The quark EDMs contribution to the $\bar{d}_{0,1}$ while the quark CEDMs, the three-gluon operator, and the CP-violating four-quark operators (including the operator induced by $Q_{\varphi ud}$) will contribute to both $\bar{d}_{0,1}$ and $\bar g_\pi^{(0),(1),(2)}$. Generally, the sensitivity of the isotensor coupling $\gpbt$ is significantly suppressed compared to that of $\gpbz$ and $\gpbo$.  
The T-odd/P-odd pion-nucleon interactions parameterized by the couplings $\gpbi$, contribute to  nucleon EDMs as well as to  nucleon-nucleon interactions  that generate the Schiff moment.

The semi-leptonic operators $\mathcal{O}_{\ell e dq}$ and $\mathcal{O}_{\ell equ}^{(1,3)}$ induce effective nucleon spin-independent (NSID)  and nuclear spin-dependent electron-nucleon interactions, described by the scalar ($S$) and tensor ($T$) interactions:
\bea
\label{eq:NSID}
\mathcal{L}_{S} & = & -\frac{G_F}{\sqrt{2}}
{\bar e}i\gamma_5 e\ {\bar N} \left[ C_S^{(0)} +C_S^{(1)}\tau_3\right] N
\eea
\bea
\label{eq:NSD}
\mathcal{L}_\mathrm{T} & = & \frac{8 G_F}{\sqrt{2}}
{\bar e} \sigma^{\mu\nu} e\ v_\nu
{\bar N} \left[ C_T^{(0)} +C_T^{(1)}\tau_3\right] S_\mu N
+\cdots \ ,\nonumber\\
\eea
where the Dirac matrices act on the electron wave function, $G_F$ is the Fermi constant, and $N$ is a nucleon spinor; the sum over all nucleons is implied,  and where the $+\cdots$ indicate sub-leading contributions arising from the electron-scalar--nucleon-pseudoscalar interaction. 
 
The coefficients $C_{S,T}^{(0,1)}$ can be expressed in terms of the underlying semileptonic operator coefficients and the nucleon scalar and tensor form factors:
\begin{eqnarray}
\nonumber
C_S^{(0)}  &=& -g_S^{(0)}\, \left(\frac{v}{\Lambda}\right)^2\,  
\mathrm{Im}\ C_{eq}^{(-)}\\
\nonumber
C_S^{(1)}  &=&  g_S^{(1)}\, \left(\frac{v}{\Lambda}\right)^2\,  
\mathrm{Im}\ C_{eq}^{(+)}  \\
\nonumber
C_T^{(0)} & = & -g_T^{(0)}\, \left(\frac{v}{\Lambda}\right)^2\,  
\mathrm{Im}\ C_{\ell e qu}^{(3)}\\
C_T^{(1)} & = & -g_T^{(1)}\, \left(\frac{v}{\Lambda}\right)^2\,  
\mathrm{Im}\ C_{\ell e qu}^{(3)},
\label{eq:CSi}
\end{eqnarray}
where 
\be
\label{eq:Ceqdef}
C_{eq}^{(\pm)}= C_{\ell e dq} \pm C_{\ell e q u}^{(1)} \ \ \ .
\ee 
The isoscalar and isovector  form factors $g_\Gamma^{(0,1)}$ are given by
\bea
\label{eq:ffdef}
\frac{1}{2} \bra{N} \left[{\bar u} \Gamma u + {\bar d}\Gamma d\right]\ket{N} 
&\equiv& g_\Gamma^{(0)} {\bar \psi_N} \Gamma \psi_N\ ,\\
\frac{1}{2} \bra{N} \left[{\bar u} \Gamma u - {\bar d}\Gamma d\right]\ket{N} 
&\equiv& g_\Gamma^{(1)} {\bar \psi_N} \Gamma \tau_3 \psi_N\ ,
\eea
where $\Gamma = 1$ for $S$  and $\sigma_{\mu\nu}$ for $T$~\cite{Engel:2013lsa}.

\subsection{EDMs in the Standard Model}

CP violation in the CKM matrix leads to non-vanishing coefficients of the $d=6$ CP-violating sources at the multi-loop level. The primary theoretical interest has been the elementary fermion EDMs. 
The CKM Lagrangian for mixing of left-handed down-type quarks and up-type quarks is
\begin{equation}
\label{eq:ccint}
\mathcal{L}_\mathrm{CKM} = -\frac{ig_2}{\sqrt{2}}\sum_{p,q} V^{pq} {\bar
U}_L^p \diracslash{W}^+ D_L^q +\mathrm{h.c.}\ .
\end{equation}
Here $g_2$ is the weak coupling constant,  $\diracslash{W}^+=\gamma^\mu W_\mu^{+}$  is the charged $W$-boson coupling,
 $U_L^p=u,c,t$ and $D_L^p=d,s,b$ are a generation-$p$ left-handed up-type
and down-type quark fields, and $V^{pq}$ denotes the element of the CKM matrix.
The constraints from unitarity and quark-field rephasing for the three quark generations allow four free parameters: three magnitudes and a CP-violating phase.  Writing
\begin{eqnarray}
V_\mathrm{CKM}&= &
\begin{bmatrix}
V^{ud} & V^{us} &V^{ub}\\
V^{cd} & V^{cs} & V^{cb}\\
V^{td} & V^{ts} & V^{tb}\\
\end{bmatrix}
\end{eqnarray}
the CP violating effects are proportional to the Jarlskog invariant 
\be
\bar\delta=\mathrm{Im}(V_{us}V_{cs}^*V_{cb}V_{ub}^*).
\ee
 A global analysis of experimental determinations of CP-violating observables in the neutral kaon and $B$-meson systems gives $\bar\delta\approx 5\times 10^{-5}$ 
~\cite{rf:PDG2014,CKMFitter}. 


The electron EDM arises at four-loop level and has been estimated  by~\textcite{rf:NgNg1996} to be 
 \be
 d_e^{\rm CKM}\approx \frac{eG_F}{\pi^2} \left (\frac{\alpha_{EM}}{2\pi}\right )^3 m_e \bar\delta\approx 10^{-38}\ {\rm \ecm}.
 \ee
 For the neutron, the contribution of the valence $u$- and $d$-quarks has been computed by~\textcite{Czarnecki:1997bu} to be 
 \ba
d_d&\approx& \frac{m_dm_c^2\alpha_S G_F^2 \bar\delta}{108\pi^5} \times f\left (\ln\frac {m_b^2}{m_c^2},\ln\frac{m_W^2}{m_b^2}\right )\nonumber\\
&\approx& -0.7\times 10^{-34} \text{\ecm~}
\ea
\ba
d_u&\approx& \frac{m_um_s^2\alpha_S G_F^2 \bar\delta}{216\pi^5}\times f\left (\ln\frac {m_b^2}{m_s^2},\ln\frac{m_c^2}{m_s^2},\ln\frac{m_b^2}{m_c^2},\ln\frac{m_W^2}{m_b^2} \right )\nonumber\\
 &\approx& -0.15\times 10^{-34} \text{\ecm},
\ea
where the $f's$ are functions of the natural logarithms of the mass ratios $\frac {m_b^2}{m_c^2}$, {\it etc}.
The valence-quark contribution to the neutron EDM is
\be
 d_n^{\rm CKM}=\frac{4}{3} d_d-\frac{1}{3}d_u
 \approx -0.9\times 10^{-34}\ \text{\rm \ecm}.
 \ee

\begin{figure}
\vskip -0.5 truein
\centerline{\includegraphics[width = 4.25 truein]{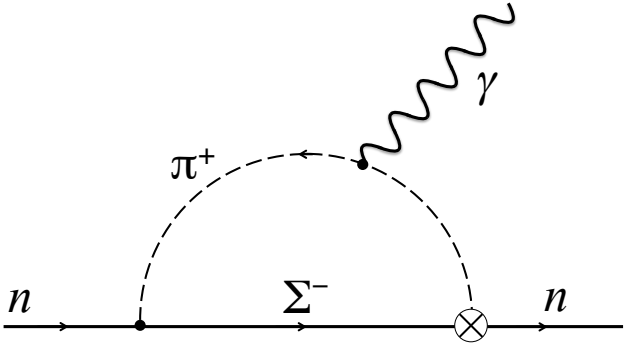}}
\vskip -0.75 truein
\caption{\label{fg:MDMEDMStrange} Representative chiral loop contribution to the neutron EDM arising from SM CKM CP-violation. The $\otimes$ indicates a CP-violating       $\Delta S=1$ vertex such as that shown in FIG.~\ref{fg:NeutronEDMVertex}, while the \textbullet\ corresponds to a CP-conserving $\Delta S=1$ interaction. Adapted from \textcite{Pospelov:2005pr}. }
\end{figure}

A significantly larger contribution to $d_n$ (and $d_p$) arises from \lq\lq long distance" meson-exchange contributions, for example that shown in FIG.~\ref{fg:MDMEDMStrange}, 
 where the CP-violating $\Delta S=1$ hadronic vertices are generated by the Penguin process of FIG.~\ref{fg:NeutronEDMVertex}, while the CP-conserving $\Delta S=1$ couplings arise from the tree-level strangeness-changing charged current interaction  (Eqn.~\ref{eq:ccint}). A full compilation of diagrams and corresponding results for SM neutron and proton EDMs based on heavy baryon chiral perturbation theory  is provided by~\textcite{Seng:2014lea}:
\be
|d_{n,p}|\approx (1-6)\times 10^{-32}\ \text{\rm \ecm}
\ee
where the range reflects the present uncertainty in various low energy constants that enter the heavy baryon effective Lagrangian and an estimate of the  higher order terms neglected in the heavy baryon expansion.

The CKM contribution enters the Schiff moment through the P-odd/T-odd $NN$ interaction mediated by kaon exchange~\cite{Donoghue:1987dd}.  \textcite{flambaum86} present an estimate of $S$ using the one-body effective P-odd/T-odd potential for a valence nucleon 
\be
{\hat W} = \frac{G_F}{\sqrt{2}}\ \frac{\eta_a}{2 m_N}\ {\vec\sigma}_a\cdot{\vec\nabla}\ \rho_A({\vec r})\ \ \ ,
\label{eq:dndp}\ee
where $\rho_A({\vec r})$ is the nuclear density and, for valence nucleon $a=n$ or $p$, the P-odd/T-odd coupling strength is
\bea
\eta_n &=&\left(N/A \right)\, \eta_{nn}+\ \left(Z/A\right)\, \eta_{np},\nonumber\\
\eta_p &=&\left(N/A \right)\, \eta_{pn}+\ \left(Z/A\right)\, \eta_{pp}.
\label{eq:etanetap}\eea
In the SM the $\eta_{a}$'s are proportional to 
$G_F {\bar\delta}$. 

For  $^{199}$Hg EDM, which has an unpaired neutron, the resulting SM estimate for the  Schiff moment and atomic EDM  are 
\ba
S(^{199}\mathrm{Hg})&\approx& -1.4\times 10^{-8}\ \eta_{np}\ e\ \mathrm{fm}^3\nonumber\\
d_A(^{199}\mathrm{Hg})&=& 3.9\times 10^{-25} \eta_{np}\ e\ \mathrm{cm},
\ea
where we have used $d_A(^{199}\mathrm{Hg})/S = -2.8\times 10^{-17}$cm/fm$^{3}$, which is given in Table~\ref{tb:SchiffCoef}. 
\textcite{Donoghue:1987dd} corrected an earlier computation of $\eta_{np}$ by properly taking into account the constraints from chiral symmetry resulting in  $|\eta_{np}| \lesssim 10^{-9}$ and
\be
|d_A(^{199}\mathrm{Hg})^\mathrm{CKM}|
 \lesssim 4\times 10^{-34}\ \ e\ \mathrm{cm}.
\ee

The EDMs of unpaired nucleons also contribute to the Schiff moment and atomic EDM. For $^{199}$Hg the unpaired neutron is dominant~\cite{dmitriev03}, and this contribrution can be estimated using the SM estimate for $d_n$ as
\be
d_A(^{199}\mathrm{Hg})^\mathrm{CKM(\it n)} \approx 4\times 10^{-4} d_n, 
\ee
resulting in
\be
|d_A(^{199}\mathrm{Hg})^\mathrm{CKM(\it n)}|
 \lesssim  2.4\times 10^{-35}\ \ecm.
 \ee

CP violation in the strong-interaction arises from the term in the QCD
Lagrangian formed by gluon field $G_{\mu\nu}$ combined with its dual  ${\tilde G}_{\mu\nu}
=\epsilon_{\mu\nu\alpha\beta}G^{\alpha\beta}/2$:
\begin{equation}
\mathcal{L}_{\bar\theta} =-\frac{\alpha_S}{16\pi^2} {\bar\theta} \,
\mathrm{Tr}\left(G^{\mu\nu}{\tilde G}_{\mu\nu}\right) \ ,
\label{thetaterm}
 \end{equation} 
where $\alpha_S$ is the strong coupling constant.\footnote{Following \cite{Grzadkowski:2010es}, $\epsilon_{0123}=1$ . This sign convention is opposite that used by~\textcite{Pospelov:2005pr} and elsewhere.  Consequently, $\mathcal{L}_{\bar\theta}$ carries an overall $-1$ compared to what frequently appears in the literature.} 
This will contribute to the neutron and proton EDM directly as well as induce a nuclear Schiff moment through the T-odd/P-odd (isospin-zero) pion-nucleon coupling~\cite{Pospelov:1999ha,Crewther:1979pi,rf:Crewther1979-erratum,Shindler:2015aqa}. For the neutron, the results fall in the range
\be
d_n^{\bar\theta}\approx -(0.9-1.2) \times 10^{-16}\bar\theta\ \text{\ecm}.
\ee
Recently~\textcite{Abramczyk:2017oxr} have observed the need to apply a correction to lattice QCD computations of the $d_n^{\bar\theta}$.

Thus experimental constraints on EDMs in hadronic systems can be used to set an upper bound on  ${\bar\theta}$. Assuming this interaction is the sole source of CP-violation, and neglecting uncertainties associated with the hadronic and nuclear physics, limits from  $d_n$ or from $d_A(^{199}\mathrm{Hg})$ imply  ${\bar\theta}\lessapprox 10^{-10}$. As we discuss in Sec.~\ref{sec:GlobalAnalysis}, allowing for multiple sources of CP violation can weaken this upper bound considerably, but the resulting constraint is nonetheless severe:   $\bar\theta\lessapprox 10^{-6}$. Either way, the tiny value allowed for a non-vanishing ${\bar\theta}$ parameter gives rise to the``strong CP problem." This may be addressed by the axion solution, which postulates an axion field $a(x)$  that couples to gluons with the  Lagrangian~\cite{Peccei:1977hh,Peccei:1977ur}
\begin{equation}
\mathcal{L}_{a} =\frac{1}{2}\partial^\mu a\partial_\mu a -V(a)- \frac{a(x)}{f_a}\frac{\alpha_S}{8\pi} G^{\mu\nu}{\tilde G}_{\mu\nu} \ .
\label{axionLagranian}
 \end{equation} 
The first term is the kinetic energy, $V(a)$ is the axion potential, the third term is the axion-gluon coupling, and $f_a$ the axion decay constant, which is analogous to the pion decay constant. The ground state is the minimum of the axion potential, which shifts the value of $\bar\theta\rightarrow \bar\theta+\frac{\langle a \rangle}{f_a}$ and could lead to cancellations that suppress $\bar\theta$.


Neutrino masses established by neutrino oscillations give rise to a $3\times3$ neutrino-mixing matrix with a single CP-violating phase analogous to the CKM phase. If neutrinos are Majorana particles,  two-loop contributions to $d_e$ are possible~\cite{rf:NgNg1996}. However, this turns out to make a small contribution unless the neutrino masses are very specifically tuned~\cite{Archambault:2004td}

\subsection{Beyond-Standard-Model Physics}

Observational and theoretical motivations for  Beyond-Standard-Model physics include the need to explain dark matter, non-vanishing neutrino masses, the observed matter-antimatter asymmetry, and considerations of naturalness, which require a mechanism to solve the \lq\lq hierarchy problem" associated with loop corrections to weak-scale physics. 
In general, BSM scenarios that address these issues
provide new mechanisms of CP-violation that also generate EDMs. 
Here, we discuss the EDM implications of a few representative BSM scenarios of current interest: supersymmetry (SUSY), left-right symmetric models, and extended Higgs sectors.

\begin{figure}[tb]
\vskip -0.5 truein
\centerline{\includegraphics[width= 4. truein]{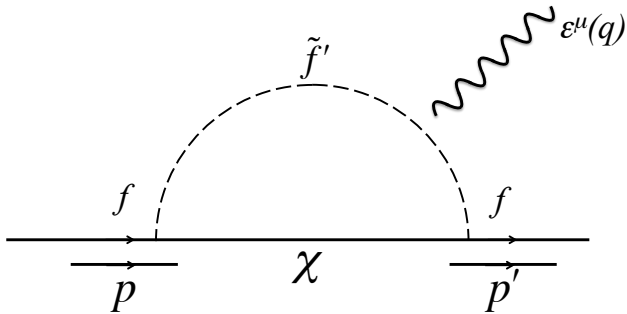}}
\vskip -0.75 truein
\caption{\label{fg:GenericSUSYEDM} One-loop fermion EDM generated by coupling to SUSY particles, $f^\prime$ and $\chi$. Adapted from~\textcite{Ellis:2008zy}).
}
\end{figure}
\begin{figure}[tb]
\vskip -0.5 truein
\centerline{\includegraphics[width= 4. truein]{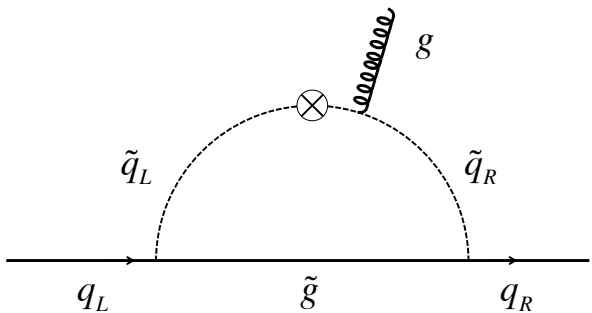}}
\vskip -0.75 truein
\caption{\label{fg:chromoEDM} An example of BSM couplings of the CEDM to the gluon field $g$. The crossed-circle indicates interactions that mix the left- and right-handed squarks. Figure from~\textcite{TardiffThesis}.
}
\end{figure}

SUSY introduces symmetry between fermions and bosons, postulating an extra Higgs doublet and a set of new particles  -  ``superpartners'' of the SM particles 
 called squarks, sleptons and gauginos. With this spectrum of new particles come new couplings and, most importantly, new CP-violating phases.
Though there is currently no direct experimental evidence for SUSY or SUSY particles, the theory is well motivated  by providing a mechanism for solving the hierarchy problem, unifying the gauge couplings, and by providing the new particles as potential dark matter candidates. In the MSSM minimal supersymmetric
extension, there exist 40 additional CP-violating phases, 
a subset of which can induce EDMs at the one-loop level. Representative one-loop contributions to the elementary fermion EDMs and quark CEDMs are shown in Figs.~\ref{fg:GenericSUSYEDM} and \ref{fg:chromoEDM}, respectively. In each case, the external gauge boson can couple to any internal superpartner carrying the appropriate charge (electric charge for the fermion EDM or color for the chromoEDM).

 It is useful to adopt several simplifying assumptions: 
 \begin{enumerate}[i.]
\item there is a single mass scale $M_\mathrm{SUSY}$ common to all superpartners;
\item there is a common relative phase $\phi_\mu$ between the supersymmetric Higgs/Higgsino mass parameter $\mu$ and the three SUSY-breaking gaugino masses, $M_j$ ($j=1,2,3$); 
\item the SUSY-breaking trilinear interactions involving scalar fermions and the  Higgs have a common phase, $\phi_A$. 
\end{enumerate}
The resulting one-loop EDMs
 and CEDMs 
following from Eqns.~(\ref{eq:nda1prime}-\ref{eq:nda1}) with $\Lambda\to M_\mathrm{SUSY}$, $Y_f$  the dimensionless Yukawa coupling for the fermion of interest, 
%
are \cite{Pospelov:2005pr,RamseyMusolf:2006vr}
\begin{eqnarray}
\nonumber
\delta_e & =&  -\frac{q\kappa_e}{\sqrt{32\pi^2}}\left[\frac{g_1^2}{12} \sin\phi_A + \left(\frac{5g_2^2}{24} + \frac{g_1^2}{24} \right) \sin\phi_\mu \tan\beta\right]\\
\delta_q & =&  -\frac{q_f\kappa_f}{\sqrt{32\pi^2}}\left[ \frac{\sqrt{2}g_3^2}{9} \Big( \sin\phi_\mu R_q - \sin\phi_A \Big) +\cdots\right] \nonumber \\
\tilde \delta_q & =& -\frac{\kappa_f}{\sqrt{32\pi^2}}\left[ \frac{5g_3^2}{18\sqrt{2}} \Big( \sin\phi_\mu R_q - \sin\phi_A \Big)+\cdots\right]\ ,\nonumber \\ 
\label{eq:MinimalSUSYEDMs}
\end{eqnarray} 
where we have followed the opposite sign convention for the trilinear phase $\phi_A$ compared to~\textcite{RamseyMusolf:2006vr}.
In Eqn.~\ref{eq:MinimalSUSYEDMs}, $f$ refers to the fermion (electron, $u$ and $d$ quark), $q_f$ is the fermion charge ($1$, 2/3, and -1/3, respectively for $e$, $u$, and $d$), $m_f$ the fermion mass, and $\kappa_f=\frac{m_f}{16\pi^2\Lambda}$. Also  $g_{1,2,3}$ are the gauge couplings, $\tan\beta=v_u/v_d$ is the ratio of the vacuum expectation values of the two Higgs doublets, the \lq\lq $+\cdots$" indicate  contributions from loops involving electroweak gauginos, 
%
$R_d = \tan\beta$, and $R_{u} = \cot\beta$ for down quarks and up quarks, respectively. 

Turning to the 4-fermion and 3-gluon operators: SUSY scenarios with large $\tan\beta$ can generate EDMs from CP- violating 4-fermion operators~\cite{Lebedev:2002ne,Demir:2003js}.
The Weinberg three-gluon operator  receives contributions at the two-loop level from squark-gluino
loops and at the three-loop level from diagrams involving the Higgs bosons. 
Na\"ively  $\sin\phi_{\mu}$ and $\sin\phi_{A}$ are expected to be $\mathcal{O}(1)$, and  electroweak baryogenesis typically requires larger phases and at least a subset  of the superpartner masses to be well below the  TeV scale~\cite{Morrissey:2012db}. 

An analysis of EDM results from $d_e$, $d_n$ and $d_A$($^{199}$Hg) by~\textcite{Pospelov:2005pr} found that $|\phi_{\mu,\, A}|\lessapprox 10^{-2}$ for  $M_{\rm SUSY}=500 $ GeV, {\it i.e.} much less than na\"ive expectations, leading to the so-called SUSY-CP problem \cite{Dimopoulos:1995ju}. 
A more general analysis that does not rely on the assumption of phase universality yields somewhat relaxed constraints but does not eliminate the SUSY CP-problem\cite{Li:2010ax}.
\textcite{Ibrahim:1998je,Ibrahim:1998je-erratum}  
have pointed out that the individual phases need not be small themselves if there are sufficient cancellations. However the dependence of $d_e$ on $\phi_\mu$ in Eqn.~\ref{eq:MinimalSUSYEDMs} makes this \lq\lq cancellation scenario" somewhat less plausible. A discussion within the context of $R$-parity violation is presented by~\textcite{Yamanaka:2014nba}.

\begin{figure}
\includegraphics[width=3.4 truein]{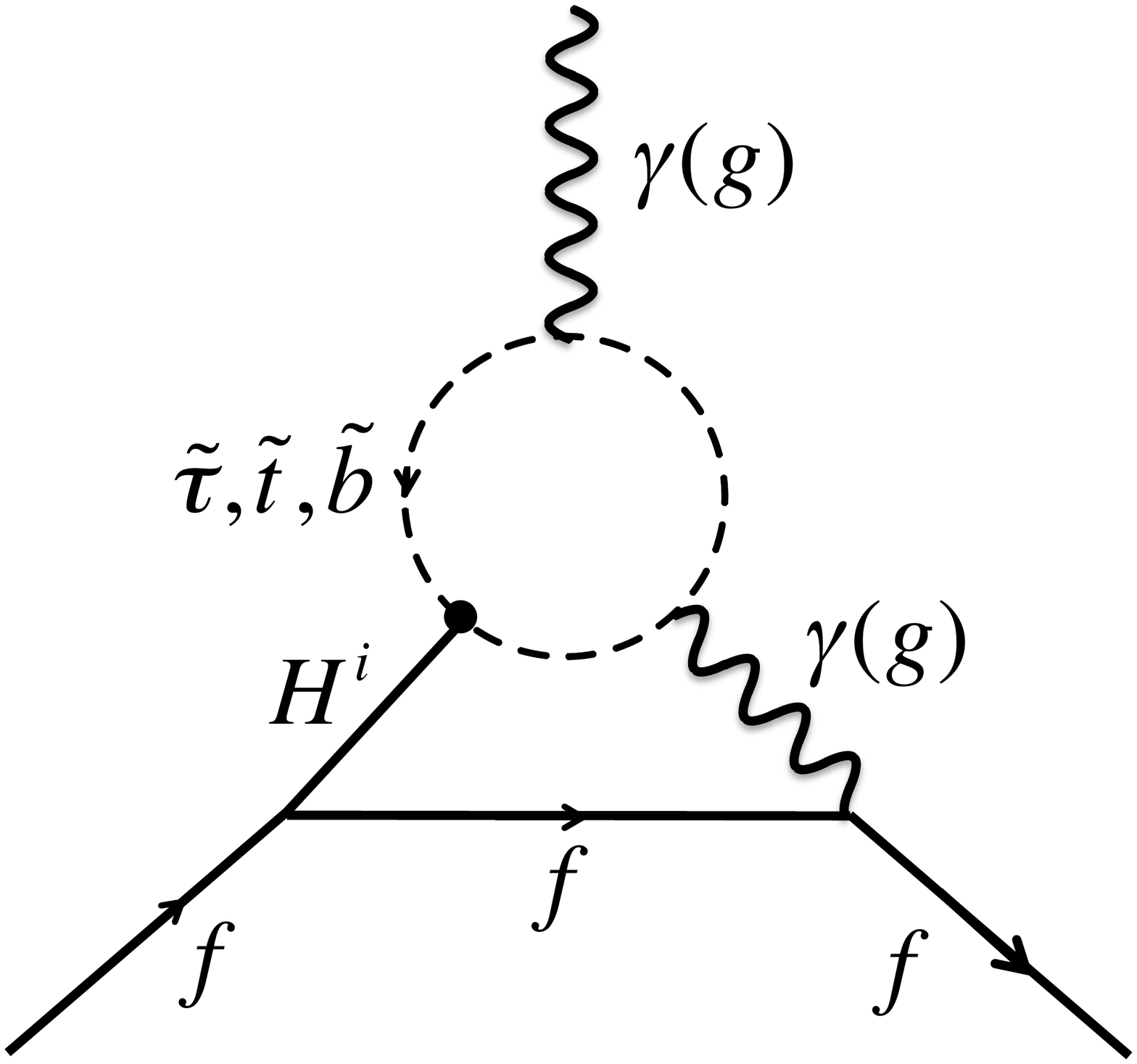} 
\vskip -.2 truein
\includegraphics[width=3.4 truein]{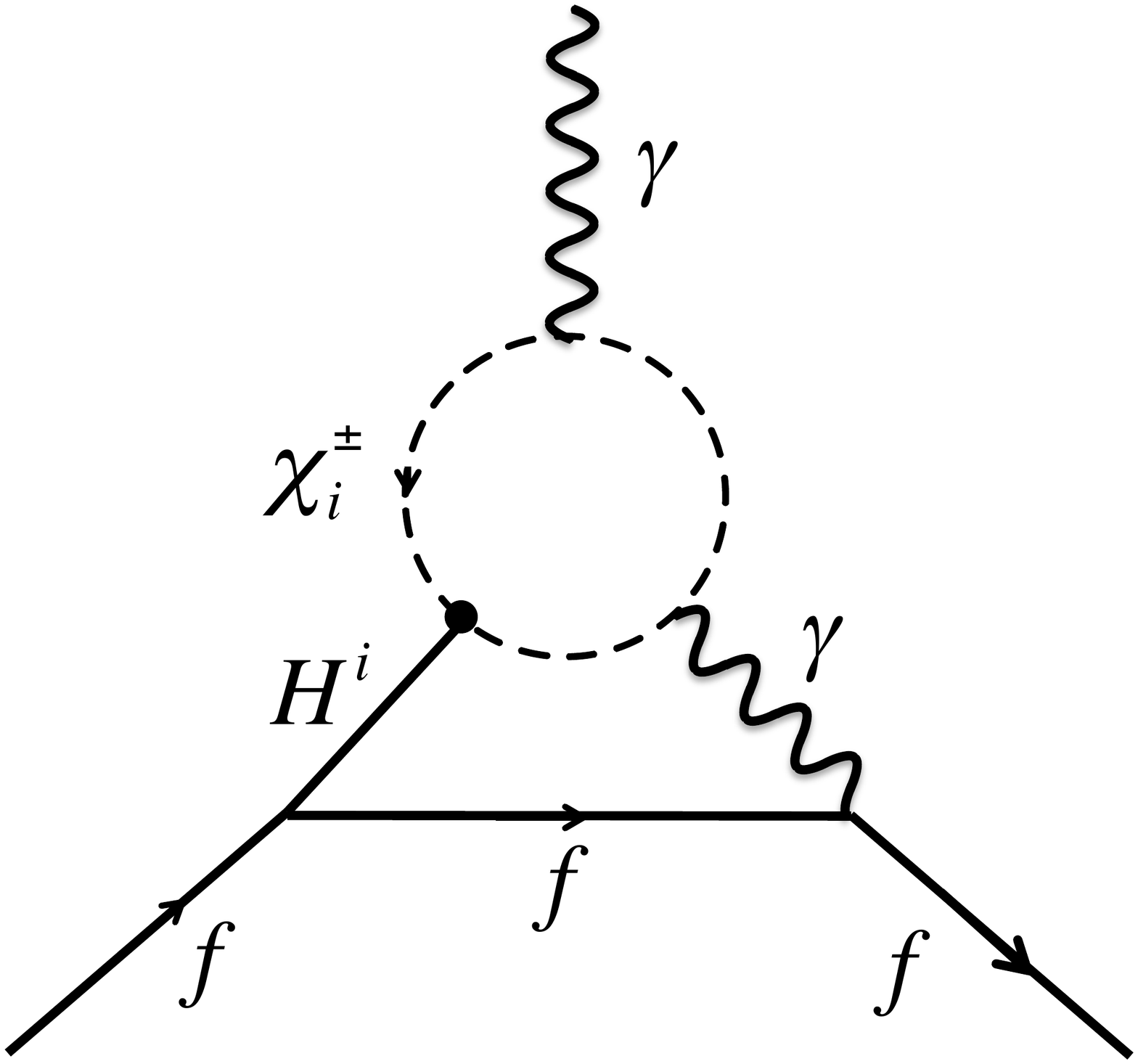}
\caption{Example two-loop Barr-Zee diagrams that give rise to a fermion EDM (coupling through $\gamma$) or CEDM (coupling through $g$).  Here $\tilde\tau$ is the $\tau$-slepton, $\tilde t$ and $\tilde b$ are squarks and $\chi$ is the chargino.
Adapted from~ \textcite{Barr:1990vd}.
\label{edm:fig:diagram2}}
\end{figure}

~\textcite{Giudice:2004tc,Giudice:2004tc-erratum} and \textcite{Kane:2009kv} have considered a scenario in which the squark and slepton masses are considerably heavier than 1 TeV, while the electroweak gauge bosons and Higgsinos remain relatively light, leading to less constrained phases. Present LHC constraints on squark masses are consistent with this possibility, though the LHC slepton mass bounds are much weaker. In this regime of heavy squarks and sleptons, electroweak baryogenesis proceeds via CP-violating bino and/or wino interactions in the early universe, while EDMs of first generation fermions arise at two-loop order through the chargino-neutralino \lq\lq Barr-Zee diagrams" shown in FIG.~\ref{edm:fig:diagram2}~\cite{Barr:1990vd,Barr:1990vd-erratum}. Applying this scenario to supersymmetric baryogenesis  and relaxing the phase universality assumption leads to the results given in FIG.~\ref{fg:MSSMdedn}, showing that improvements in the sensitivities to $d_e$ and $d_n$ by one and two orders of magnitude, respectively, would probe the entire CP-violating parameter space for MSSM baryogenesis~\cite{Morrissey:2012db,Cirigliano:2006dg,Li:2008ez,Cirigliano:2009yd}. Note that supersymmetric electroweak baryogenesis requires not only sufficient CP-violation, but also a strong first order electroweak phase transition. LHC measurements of Higgs boson properties now render this possibility unlikely in the MSSM~\cite{Curtin:2012aa,Katz:2015uja}; however, ~\textcite{Liebler:2015ddv} suggest that there are regions of parameter space that can satisfy both the observed Higgs mass and baryogenesis. On the other hand,  extensions of the MSSM with gauge singlet superfields presently allow for the needed first-order phase transition. In the context of these ``next-to-minimal'' scenarios,  CP-violating sources could give rise to the observed baryon asymmetry as indicated in FIG.~\ref{fg:MSSMdedn}.

%

The discovery of the  125 GeV Higgs boson has raised anew the possibility that it might be one of a number of scalars, and  a wide array of possibilities for the ``larger'' Higgs sector have been considered over the years. 
One scenario that has been studied extensively is the Two-Higgs-Doublet model, 
wherein the requirements of supersymmetry restrict the form of the scalar potential and the couplings of the two Higgs doublets to the SM fermions. In the more general context, the Two-Higgs-Doublet model allows for a variety of additional CP-violating phases that can give rise to EDMs. The phases may arise in the scalar potential and/or the amplitudes for scalar-fermion interactions. 
The implications of new CP-violating phases in the Two-Higgs-Doublet model have been analyzed by~\textcite{Inoue:2014nva}, who considered a potential that manifests a softly-broken $Z_2$ symmetry in order to avoid constraints from the absence of flavor-changing neutral currents. (A $Z_2$ symmetry is a discrete symmetry under phase reversals of the relevant fields.)
 In the absence of CP-violation the scalar spectrum contains two charged scalars, $H^\pm$, and three neutral scalars: the CP-even $H^0$ and $h^0$ and the CP-odd $A^0$. With the presence of CP-violating phases in the potential, the three neutral scalars mix to form the neutral mass eigenstates, $h_i$, one of which is identified with the 125 GeV  SM Higgs-like scalar. This CP-mixing translates into CP-violating phases in the couplings of the $h_i$ to SM fermions, thereby inducing EDMs. In a variant of the Two-Higgs-Doublet model considered by~\textcite{Inoue:2014nva}, the requirements of electroweak symmetry-breaking imply that there exists only one  CP-violating phase in the scalar sector $\alpha_b$, which is responsible for both CP mixing among the scalars and the generation of EDMs. The latter arise from the Barr-Zee diagrams shown in FIG.~\ref{edm:fig:diagram2}. 

\begin{figure*}[tb]
\vskip -1.25 truein
\includegraphics[width= 7 truein]{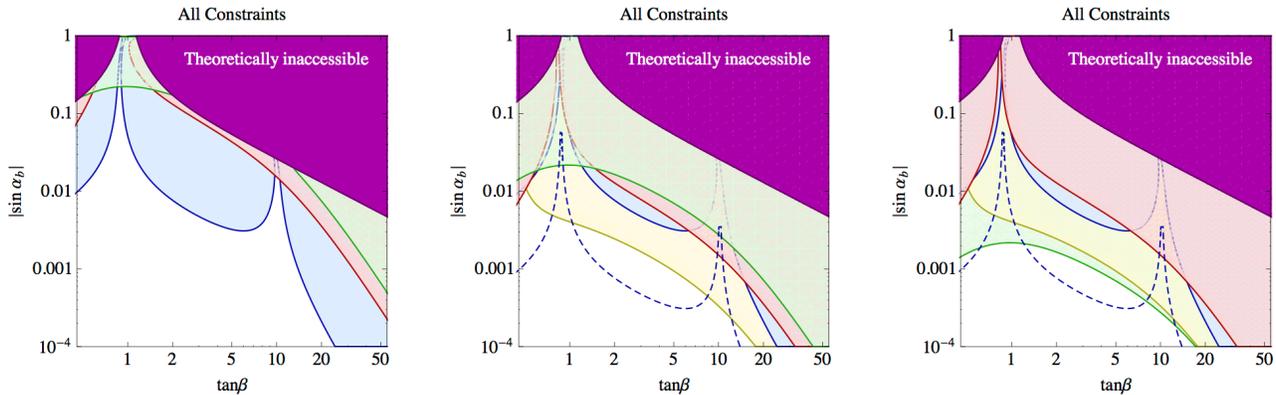}
\vskip -1.5 truein
\caption{\label{fg:combined_S}  (Color online) EDM results  for the Type II Two-Higgs-Doublet model. Horizontal axes show the ratio of up- and down-type Higgs vacuum expectation values. Vertical axes show the CP-violating phase that mixes of CP-even and CP-odd scalars. Purple regions are excluded by consistency with electroweak symmetry breaking. Left panel: current constraints from $d_e$ (blue), $d_n$ (green), and $d_A(^{199}\mathrm{Hg})$ (pink). Middle panel: impact of improving present constraints by one order of magnitude, where blue dashed line indicates the prospective reach of $d_e$. The yellow region indicates the reach of a future $d_A(^{225}\mathrm{Ra})$ with a sensitivity of $10^{-27}$ \ecm. Right panel gives same future constraints but with $d_n$ having two orders of magnitude better sensitivity than the present limit. 
Figure from~~\textcite{Inoue:2014nva}.
}
\end{figure*}

Constraints on $\alpha_b$ as a function of $\tan\beta$ set by present and prospective 
EDM results are  are shown in FIG.~\ref{fg:combined_S} for the \lq\lq type II" Two-Higgs-Doublet model (for an enumeration of several variants of the Two-Higgs-Doublet model, for exapmle see~\cite{Barger:1989fj}. The type II scenario has the same Yukawa structure as the MSSM.).
The $d_e$ limit from ThO is generally the most restrictive, except in the vicinity of $\tan\beta\sim 1$ and $\sim 10$. \textcite{Bian:2014zka} have pointed out that the vanishing sensitivity to  $d_e$  near $\tan\beta\sim 1$ arises from a cancellation between the effects of the induced CP-violating couplings of the Higgs-like scalar to the electron and the corresponding couplings to the $h F^{\mu\nu} {\tilde F}_{\mu\nu}$ operator associated with the upper loop of the Barr-Zee diagrams.
The neutron and $^{199}\mathrm{Hg}$ EDMs are not susceptible to the same cancellation mechanism as the electron and provide additional constraints near $\tan\beta\sim 1$.  The middle and far right panels show the sensitivity of prospective future EDM searches, including anticipated results of from $^{225}$Ra~\cite{PhysRevC.94.025501}. 
The reach of a ten times more sensitive $d_e$ search would extend somewhat beyond the constraints from  neutron and atomic searches, except in the cancellation regions. More optimistically, any non-zero result could indicate whether or not the observed EDM is consistent with CP violation in the Two-Higgs-Doublet model and help narrow the parameter space.\footnote{As indicated by \textcite{Inoue:2014nva}, there exist considerable hadronic and nuclear theory uncertainties associated with the $d_n$ and $d_A$ sensitivities.}

Left Right Symmetric Models postulate the existence of an SU(2)$_L\times$SU(2)$_R\times$U(1)$_{B-L}$ symmetry, in which parity-violation in the Standard Model arises from spontaneous breakdown of the SU(2)$_R$ symmetry at a scale above the electroweak scale  ($M_{W_R}>>M_{W_L}$)~\cite{Pati:1974yy,Pati:1974yy-erratum,Mohapatra:1974hk,Senjanovic:1975rk}. This gives rise to a  second CKM-like matrix for the right-handed charged-current couplings of $W_L$ with a new CP-violating phase. Spontaneous symmetry breaking induces mixing between left-handed $W_L$ and right-handed $W_R$ gauge bosons, and  the  mass eigenstates become a mixture of $W_L$ and $W_R$:
\begin{eqnarray}
\label{eq:LRSM1}
W_1^+ & = & \cos\xi W_L^+ + \sin\xi e^{-i\alpha} W_R^+\nonumber\\
W_2^+ & = & -\sin\xi e^{i\alpha} W_L^+ +\cos\xi W_R^+
\end{eqnarray}
where $\alpha$ is the CP-violating phase associated with the gauge boson mixing. This phase, along with the left- and right-handed CKM phases, can lead to one-loop quark EDMs arising from $W_{1,2}$ exchange. Retaining only the contribution from $\alpha$ and taking $|\sin\xi|\lessapprox 10^{-3}$ as implied by tests of CKM matrix unitarity, the resulting short-range contribution to the neutron EDM is
\begin{equation}
\label{eq:LRSM2}
| \bar d_n^{sr}|\lessapprox (3\times 10^{-14}\ e\ \mathrm{fm}) \left(1-\frac{M_1^2}{M_2^2}\right)\ \cos\theta_L\cos\theta_R\sin\alpha\ \ \ ,
\end{equation}
where $M_1$ and $M_2$ are the masses of $W_1$ ane $W_2$, respectively, and $\theta_{L,R}$ denote the left-handed and right-handed Cabibbo angles. The upper bound on the contribution to the neutron EDM  is an order of magnitude less than the current limits on $d_n$, though the analysis should be revisited to include  quark CEDM contributions. 

In addition, $W_L$-$W_R$ mixing gives rise to a unique, four-quark CP-violating operator that, in turn, generates the T-odd/P-odd $\pi-NN$ coupling $\gpbo$ discussed in Sec.~\ref{sec:LowEnergyParameters}.
\begin{equation}
\label{eq:LRSM3}
\gpbo\Big\vert^\mathrm{LRSM} \approx 10^{-4}\ \left(1-\frac{M_1^2}{M_2^2}\right) \sin\xi \cos\theta_L\cos\theta_R\sin\alpha\ \ \ .
\end{equation}
The resulting mercury atomic EDM  is
\begin{eqnarray}
\label{eq:LRSM4}
|d_A(^{199}\mathrm{Hg})| & \lessapprox & (1.1 \times 10^{-11}\ e\ \mathrm{fm})\\
\nonumber
&&\times  \left(1-\frac{M_1^2}{M_2^2}\right)\ \cos\theta_L\cos\theta_R\sin\alpha\ \ \ ,
\end{eqnarray}
where we have again used an approximate upper bound $|\sin\xi|\lessapprox 10^{-3}$ and take the $\gpio$ dependence of the Schiff moment as the midpoint of the range given in Table~\ref{tb:SchiffCoef}.
 Given the significantly larger coefficient in Eqn.~(\ref{eq:LRSM4}) compared to that in Eqn.~(\ref{eq:LRSM2}), together with the stronger mercury atom EDM bound, we observe that the atomic EDM results currently place the most severe constraints on the CP violation associated with $W_L$-$W_R$ mixing. There is, however, an important caveat: the contribution of $\gpbo$ to the $^{199}$Hg Schiff moment has significant nuclear theory uncertainties~\cite{rf:deJesus}, and it is possible that the sensitivity is considerably weaker than indicated in  Eqn.~\ref{eq:LRSM4}. On the other hand, the nuclear many-body computations for this contribution to the Schiff moments of other nuclei of experimental interest appear to be more reliable, providing motivation for active pursuit of improved experiments on $^{129}$Xe, $^{221/223}$Rn and $^{225}$Ra~\cite{dobaczewski05,ban10}.


New phases could in principle also affect CP-violation in flavor-violating process, such as meson mixing  or  rare $B$-meson decays, and give complementary information on the model
parameters~ \cite{Altmannshofer:2008hc} that could push the new
physics scale well beyond 10 TeV.
Even so, due to the generic flavor mixing, the  electron and neutron EDMs  are proportional to heavy-quark and lepton masses, and the experimental limits  probe scales of 1000 TeV in some cases.
An explicit example of such a case is given by the mini-split SUSY framework  for which current EDM bounds already probe masses up to 100~TeV~\cite{Hall:2011jd,Ibe:2011aa,Arvanitaki:2012ps,ArkaniHamed:2012gw,McKeen:2013dma,Altmannshofer:2013lfa}.

Recent work has also considered the constraints that EDMs may place on CP-violating couplings of other SM particles, such as the Higgs boson or top quark. \textcite{McKeen:2012av} computed constraints on the CPV Higgs-diphoton coupling, $h F_{\mu\nu} {\widetilde F}^{\mu\nu}$, and showed that the corresponding relative impact of this operator on the rate for the decay $h\to\gamma\gamma$ is at the $10^{-4}$ level,  well below the expected sensitivity at the LHC or future Higgs factories. This constraint may be weaker in specific models, such as those containing vector-like fermions (see also~\textcite{Chao:2014dpa} for the connection with baryogenesis). \textcite{Chien:2015xha} investigated the constraints on dimension-six operators that couple the Higgs boson to quarks and gluons, and found that the impact of hadronic and nuclear physics uncertainties is pronounced. \textcite{Cirigliano:2016njn,Cirigliano:2016nyn} and~\textcite{Fuyuto:2017xup} considered the constraints on the top quark EDM from $d_e$ and find that the bounds are three orders of magnitude stronger than obtained from other sources. 

\subsection{From theory to experiment}
\label{sec:EFTParameters}

Experiments probe P-odd/T-odd observables in systems that combine a number of scales as illustrated in FIG.~\ref{fg:EDMSubwayMap}. For the neutron and proton, the fundamental CP-violating interactions discussed above arise from two sources: a short range contribution (denoted by ${\bar d}_{n,p}^\mathrm{sr}$) and a long range contribution arising from the P-odd/T-odd pion-nucleon interactions. 
Storage ring experiments also have the potential to directly probe EDMs of light nuclei, namely, the deuteron ($^2$H$^+$) and helion ($^3$He$^{++}$) discussed in Sec.~\ref{sec:StorageRingEDMs}. The EDMs of these systems arise from the constituent nucleon EDMs as well as P-odd/T-odd nucleon-nucleon interactions arising from  pion-exchange and from four-nucleon contact interactions.  Paramagnetic atoms and molecules are most sensitive to the electron EDM and the nuclear-spin-independent electron-nucleus coupling. In diamagnetic atoms 
the dominant contributions are the nuclear-spin dependent electron-nucleus interaction and the Schiff moment, which also arises from long-range pion exchange and short range four-nucleon interactions. The following summarizes the contributions of these low-energy parameters to the experimentally accessible systems.

\vskip 0.1in

\noindent{\em Nucleons:}

\vskip 0.1in

\begin{figure}[tb]
\vskip -0.75 truein
\centerline{\includegraphics[width= 4.25 truein]{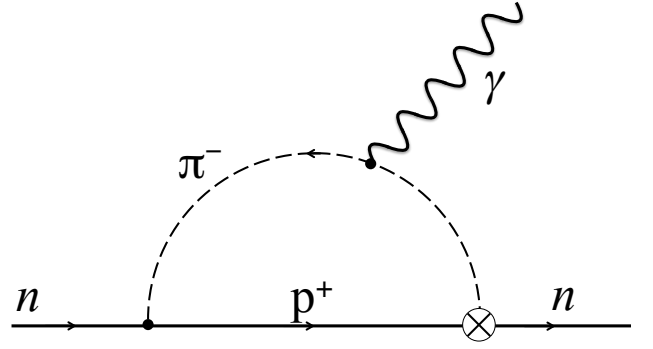}}
\vskip -0.5 truein
\caption{\label{fg:MDMEDM} Representative long-range, pion-exchange contributions to the neutron EDM.  The cross represents the CP-violating vertex, while the closed circle is the CP-conserving vertex. Adapted from \textcite{Pospelov:1999mv}.}
\end{figure}

Long-range strangeness-conserving pion-nucleon coupling contributions to the nucleon EDMs indicated in FIG.~\ref{fg:MDMEDM} have been computed using chiral perturbation theory.
The  magnitude of $\gpbt$ is expected to be suppressed by a factor of 100 or more relative to $\gpbz$ and $\gpbo$ based on chiral symmetry considerations  
and is typically neglected in the computation of the nucleon EDMs~
\cite{Chupp:2014gka}. 
The result to next-to-next-to-leading order is \cite{Seng:2014pba}:
\bea
\label{eq:dnfull}
d_n  =  {\bar d}_n^\mathrm{sr}-\frac{e g_A}{8\pi^2 F_\pi} &\biggl\{&\!\!\!\!\ \gpbz[\ln \frac{m_\pi^2}{m_N^2} -\frac{\pi m_\pi}{2 m_N}] \nonumber\\
&+&\gpbo\,\frac{(\kappa_1-\kappa_0)}{4}\frac{m_\pi^2}{m_N^2}\ln  \frac{m_\pi^2}{m_N^2} \biggr\} \nonumber\\
\label{eq:dpfull}
d_p  =  {\bar d}_p^\mathrm{sr}+\frac{e g_A}{8\pi^2 F_\pi} &\biggl\{&\!\! \gpbz[\ln \frac{m_\pi^2}{m_N^2}
-\frac{2\pi m_\pi}{m_N}]\nonumber\\
-\frac{\gpbo}{4}\bigl[  \frac{2\pi m_\pi}{m_N} \!\!\!&+& \!\! (\frac{5}{2}+\kappa_0+\kappa_1) \frac{m_\pi^2}{m_N^2}\ln  \frac{m_\pi^2}{m_N^2}\bigr]\biggr\} ,\nonumber\\
\eea
where $g_A$ is the nucleon isovector axial coupling,  $\kappa_0$ and $\kappa_1$ are the isoscalar and isovector nucleon anomalous magnetic moments, respectively, and the low-energy coefficients ${\bar d}_n^\mathrm{sr}$ and ${\bar d}_p^\mathrm{sr}$ account for remaining short range contributions.
Note that the ${\bar d}_{n,p}^\mathrm{sr}$ are linear combinations of the ${\bar d}_{0,1}$ given in Eqn.~(\ref{eq:pinn}). Computations of the ${\bar d}_{n,p}^\mathrm{sr}$ and $\gpbi$ in terms of the fundamental CP-violating interactions are reviewed by~\textcite{Engel:2013lsa}, \textcite{Shindler:2015aqa}, \textcite{Bhattacharya:2015esa}, \textcite{Seng:2016pfd},  and \textcite{Bouchard:2016heu}.

In particular, we point out that the QCD parameter $\theta$ contributes to $\gpbz$:  
\be
\gpbz\approx (0.015\pm 0.003)\bar\theta+\dots,
\label{eq:gpi0thetabar}
\ee
 where the ellipses indicate  BSM contributions~\cite{deVries:2015una}. 

\vskip 0.1in

\noindent{\em Light Nuclei:}

\vskip 0.1in

Experimental approaches to storage-ring measurements of the EDMs of the deuteron  and helion  are discussed in Sec.~\ref{sec:StorageRingEDMs}. For the deuteron, the EDM has contributions from the nucleon moments as well as the pion-exchange contribution, leading to
\bea
d_D&=&d_n+d_p\nonumber\\
&+&[(0.0028\pm 0.0003)g_\pi^0+(0.18\pm 0.02)\gpbo]\  e\ \text{fm}.\nonumber\\
\eea
For the helion ($^3$He$^{++}$), the proton spins are nearly completely paired~\cite{rf:FriarPayneGibson} and the neutron EDM dominates the one-body contribution:
\bea
d_h&=&0.9d_n-0.05d_p \nonumber\\
&+&[(0.10\pm0.03)\gpbz+(0.14\pm 0.03)\gpbo]\ e\ \text{fm}.\nonumber\\
\eea
Note that the contributions from the four-nucleon contact interactions have not been included here ~\cite{rf:deVriesDEDM} (see also \cite{Stetcu:2008vt,deVries:2011an,Song:2012yh,Wirzba:2016saz,Yamanaka:2016umw}. 

\vskip 0.2in

\noindent{\em Paramagnetic systems:}

\vskip 0.1in

In paramagnetic systems with one or more unpaired electrons, there is a net electric field $\vec{E}_\mathrm{eff}$ at the electron's average position that is generally much greater than a laboratory electric field (many V/\AA\ or GV/cm).
Consequently the EDMs  of paramagnetic atoms and P-odd/T-odd observables in polar molecules are dominated by the electron EDM and the nuclear-spin-independent electron-nucleon interaction, which couples to a scalar (S) component of the nucleus current.
Taking the nuclear matrix element of the interactions given in Eqn.~(\ref{eq:NSID}) and assuming non-relativistic nucleons lead to the atomic Hamiltonian
\be
{\hat H}_S = \frac{i G_F}{\sqrt{2}}\, \delta({\vec r})\, \left [(Z+N)C_S^{(0)}+ (Z-N)C_S^{(1)}\right]\gamma_0\gamma_5,
\ee
The resulting atomic EDM $d_A$ is given by
\be
d_A^\mathrm{para} = \rho_A^e d_e - \kappa_S^{(0)}\, C_S,
\ee
where
\be
C_S\equiv C_S^{(0)} +\left( \frac{Z-N}{Z+N}\right) C_S^{(1)},
\ee
and  $\rho_A^e$ and $\kappa_S^{(0)}$ are obtained from atomic and hadronic computations. 

For polar molecules, the effective Hamiltonian is
\be
{\hat H}_\mathrm{mol} = \left[W_d\, d_e +W_S\,  (Z+N)C_S\right] {\vec S}\cdot{\hat n}+\cdots\ \ \ ,
\ee
where ${\vec S}$ and ${\hat n}$ denote the unpaired electron spin and the unit vector along the intermolecular axis, respectively. The quantities $W_d\propto E_\mathrm{eff}$ and $W_S$ that give the sensitivities of the molecular energy to the electron EDM and electron-quark interaction are obtained from molecular structure calculations~\cite{Ginges:2003qt,Meyer:2008gc,rf:Skripnikov2013,kozlov98,Petrov2007,Fleig2013,Skripnikov2017}. The resulting ground state matrix element in the presence of an external electric field ${\vec E}_\mathrm{ext}$ is
\be
\bra{\mathrm{g.s.}} {\hat H}_\mathrm{mol} \ket{\mathrm{g.s.}} = \left[W_d\, d_e +W_S\, (Z+N)C_S\right] \, \eta(E_\mathrm{ext}),
\ee
with 
\be
\eta(E_\mathrm{ext}) = \bra{\mathrm{g.s.}}{\vec S}\cdot{\hat n}  \ket{\mathrm{g.s.}} _{E_\mathrm{ext}}\ \ \ .
\ee
This takes into account the orientation of the internuclear axis and the internal electric field with respect to the external field, {\it i.e.} the electric polarizability of the molecule. This leads to the observable, a  P-odd/T-odd  frequency shift measured in molecular experiments discussed in Sec.~\ref{sec:PolarMolecules}.


\vskip 0.1in

\noindent{\em Diamagnetic atoms and molecules:}

\vskip 0.1in

The EDMs of diamagnetic atoms of present experimental interest arise from the nuclear Schiff moment  and the nuclear-spin-dependent electron-nucleon interaction, which couples to the tensor (T) nuclear current. 
The Schiff moment, accounting for both contributions from the EDM's of unpaired nucleons  and the long-range pion-nucleon coupling, can be written
\begin{equation}
S=s_{N} d_{N} + \frac{m_N g_A}{F_\pi}\left[a_0 \gpbz + a_1 \gpbo	+	a_2\gpbt\right] \ \ ,
\label{eq:SchiffMoment}
\end{equation} 
where contributions from the unpaired nucleon EDM's are given by $s({d_{N}})=s_n d_n+s_p d_p$  \cite{Dzuba1985,dmitriev03,ban10,Yoshinaga_doi:10.1143/PTP.124.1115}. 
Values of $a_{0,1,2}$ from Eqn.~\ref{eq:SchiffMoment} for $^{199}$Hg, $^{129}$Xe, $^{225}$Ra, and TlF are presented in Table~\ref{tb:SchiffCoef}. These  depend on the details of the assumed nucleon-nucleon interaction. However note that there is no single consistent approach for all nuclei of interest. 
As discussed above, each isospin component may be particularly sensitive to a subset of the possible CP-violating interactions. For example,  the QCD parameter $\bar\theta$ contributes most strongly to $\gpbz$, while the effect of $W_L$-$W_R$ mixing in the Left-Right Symmetric Model shows up most strongly in $\gpbo$. 

The nucleon EDM  long-range and short-range contributions to the Schiff moment can be separated using Eqn.~\ref{eq:dnfull} to write
\bea
S=   s_N \bar d_N^{sr}+\left[\frac{m_N g_A}{F_\pi} a_0\right.&+&\left.s_N\alpha_{n\gpiz}\right] \gpiz \nonumber\\
&+&  \left[\frac{m_N g_A}{F_\pi} a_1+ s_N\alpha_{n\gpio}\right] \gpio , \nonumber\\
\eea
 where the coefficients $\alpha_{N\bar g_\pi^{(0,1)}}$, given in Table~\ref{tb:diamagnetics}, are the factors multiplying $\gpiz$ and $\gpio$ in Eqn.~\ref{eq:dnfull} , and  the smaller $\gpbt$ pion-nucleon contribution to $S$ has been dropped.

Contributions from the electron-nucleus interaction are revealed in the Hamiltonian resulting from Eqn.~(\ref{eq:NSD}):
\be
{\hat H}_T = \frac{2i G_F}{\sqrt{2}}\, \delta({\vec r})\, \left [C_T^{(0)}+ C_T^{(1)}\tau_3\right]\, {\vec\sigma}_N\cdot{\vec\gamma}\ \ \ ,
\ee
where the sum over all nucleons is again implicit;  $\tau_3$ is the nucleon isospin Pauli matrix, ${\vec\sigma}_N$ is the nucleon spin Pauli matrix, and ${\vec\gamma}$ acts on the electron wave function. Including the effect of ${\hat H}_T$, the individual nucleon EDMs $d_N$, and the nuclear Schiff moment $S$ (Eqn.~\ref{eq:SchiffDef}), one has
\be
\label{eq:diamag}
d_A(\mathrm{dia}) = 
\kappa_S S - \left[k_{C_T}^{(0)} C_T^{(0)} + k_{C_T}^{(1)} C_T^{(1)}\right]\ \ \ ,
\ee
where $\kappa_S$ and  $k_{C_T}^{(0,1)}$ give the sensitivities of the $d_A^\mathrm{dia}$ to the Schiff moment and the isoscalar and isovector electron-quark tensor interactions and are provided in Tables~\ref{tb:diamagnetics} and~\ref{tb:SchiffCoef}. 
As indicated in Eqn.~\ref{eq:CSi}, the isoscalar and isovector tensor couplings depend on the same Wilson coefficient $\mathrm{Im}\ C_{\ell e qu}^{(3)}$, so their values differ only due to the different nucleon tensor form factors. 
Until recently, there has been limited information on the nucleon tensor form factors $g_T^{(0,1)}$. Computations using lattice QCD have now obtained $g_T^{(1)}=0.49 (03)$ \cite{Bhattacharya:2016zcn} and $g_T^{(0)} = 0.27 (03)$~\cite{Bhattacharya:2015wna}\footnote{Note that the numerical values of the  tensor couplings given in these references are two times larger than quoted here, owing to differences in normalization conventions.}. For the diamagnetic atoms of experimental interest (Hg, Xe, Ra) the nuclear matrix elements are dominated by the contribution from a single, unpaired neutron while for TlF, to a good approximation the proton is unpaired. We therefore replace the last term in brackets in Eqn.~(\ref{eq:diamag}) with  $k_T^{n} C_T^{n}$ or $k_T^{p} C_T^{p}$, for an unpaired neutron or proton, respectively, where
\bea
C_T^n &=& - \left[ g_T^{(0)} - g_T^{(1)}\right]\, \left(\frac{v}{\Lambda}\right)^2\,\mathrm{Im}\, C_{\ell e q u}^{(3)} \approx - 0.76 C_T^{(0)}\nonumber \\ 
C_T^p &=& - \left[ g_T^{(0)} + g_T^{(1)}\right]\, \left(\frac{v}{\Lambda}\right)^2\, \mathrm{Im}\, C_{\ell e q u}^{(3)}\approx +2.45 C_T^{(0)}. \nonumber \\ 
\label{eq:CTs}
\eea
 Table~\ref{tb:diamagnetics} provides the coefficients for the dependence of $d_A^{dia}$ on $\gpiz,\ \gpio$, and $\bar d_n^{sr}$.

In summary, contributions to the EDMs in systems accessible to experiment can be expressed in terms of the following set of low-energy parameters:
\begin{enumerate}

\item The lepton EDMs; the electron EDM $d_e$ contributes in first order to the EDMs of paramagetic atoms and molecules. 

\item  Two isospin components of the nuclear-spin-independent $eN$ coupling $C_S^{0,1}$. Since most of the heavy-atom systems have a roughly equal ratio of neutrons to protons this can be reduced to a single average $\bar C_S=C_S^0-\frac{(N-Z)}{A}C_S^1\approx C_S^0$.

\item The nuclear-spin-dependent $eN$ coupling labeled by $C_T^{(0,1)}$, most important in diamagnetic atoms and molecules.

\item The short-range contribution to the nucleon EDMs $\bar d_{n,p}^{sr}$.

\item The pion-nucleon couplings labeled $\gpbi$ that contribute to the nucleon and nuclear EDMs and to the Schiff moments of nuclei. Given that the sensitivity of $\gpbt$ to the CP-violating interactions  is highly suppressed, we will omit it in the following.

\end {enumerate}

We therefore separate paramagnetic atoms and molecules from diamagnetic systems and also separate nucleon and  fundamental-fermion EDMs, 
as follows:

\medskip
\noindent
\centerline
{\bf Paramagnetic atoms}
\begin{equation}
d_A(\mathrm{para})=\eta_{d_e}d_e + k_{C_S} \bar C_S
\label{eq:ParamagneticEDMs}
\end{equation}

\noindent
\centerline
{\bf Polar molecules}
\begin{equation}
\Delta\omega^{\not P\not T}= \frac{-d_e E_{eff}}{\hbar} + k^\omega_{C_S} \bar C_S
\label{PolarMoleculeDeltaOmega}
\end{equation}

\noindent
\centerline
{\bf Diamagnetic atoms} 
\begin{equation}
d_A(\mathrm{dia})=\kappa_S S(\bar g_\pi^{0,1},d_N) + k_{C_T^{(0)}} C_T^{(0)} + ...
\label{eq:DiamagneticAtoms}
\end{equation}

\noindent
\centerline
{\bf Nucleons}
\begin{equation}
d_{n,p}=d_{n,p}^{lr}(\bar g_\pi^{0,1} )+{\bar d}_{n,p}^{sr}
\end{equation}

\noindent
\centerline
{\bf Charged leptons}
\begin{equation}
d_e, d_{\mu}, (d_\tau)
\end{equation}

\noindent 
The coefficients $\eta$, $k$ and $\kappa$ are presented in Tables~\ref{tb:paramagnetics},~\ref{tb:diamagnetics} and~\ref{tb:SchiffCoef}.

Note that the other contributions enter the atomic and molecular systems at higher order, but are less important.  However due to the exquisite sensitivity of the $^{199}$Hg EDM measurement, the higher order contribution of the electron EDM $d_e$ does have an impact. 
Additionally, experiments in paramagnetic solid-state systems with quasi-free electrons are directly sensitive to $d_e$.




\begin{table*}[h]
\centering
\begin{tabular}{|c|c|c|c|c|}
\hline\hline
System & $\alpha_{d_e}=\eta_e$  & $\alpha_{C_S}=W_S$ & $\alpha_{C_S}/\alpha_{d_e}$ & ref.\\
\hline
Cs  & 123  & $7.1\times 10^{-19}$ \ecm &  $5.8\times 10^{-21}$    (\ecm) & $a$, $b$,$c$ \\
&$(100$-$138)$ &  $(7.0$-$7.2)$  &  $(0.6$-$0.7)\times 10^{-20}$  &   \\
\hline
Tl  & -573 &  $-7\times 10^{-18}$ \ecm &  $1.2\times 10^{-20}$   (\ecm) &  $a$, $b$ \\
& $-(562$-$716)$ &  $-(5$-$9)$      &  $(1.1$-$1.2)\times 10^{-20}$   &\\
 \hline
YbF  & $-3.5\times 10^{25}$ $\frac{\rm rad/s}{\ecm}$ & $-2.9\times 10^5$ rad/s &  $8.6\times 10^{-21}$   (\ecm) & $d$ \\
 &$-(2.9$-$3.8)$& $-(4.6$-$6.8)$    &  $(8.0$-$9.0)\times 10^{-21}$   &  \\
\hline
ThO &$-1.6\times 10^{26}$ $\frac{\rm rad/s}{\ecm}$ &$-2.1\times 10^6$ rad/s &  $1.3\times 10^{-20}$  (\ecm)  &$e$, $f$, $g$  \\
 & $-(1.3$-$1.6)$ & $-(1.4$-$2.1)$  &  $(1.2$-$1.3)\times 10^{-20}$  &   \\
\hline
HfF$^+$ & $3.5\times 10^{25}$ $\frac{\rm rad/s}{\ecm}$ &$3.2\times 10^5$ rad/s &  $8.9\times 10^{-21}$  (\ecm)   &$h$, $i$, $j$  \\
 & $-(3.4$-$3.6)$ &  $(3.0$-$3.3)$ &$(8.3$-$9.7) $  &  \\
\hline\hline
\end{tabular}
\vskip -0.1 truein
\caption{Sensitivity  to $d_e$ ($\alpha_{d_e}$) and $C_S$ ($\alpha_{C_S}$) and the ratio $\alpha_{C_S}/\alpha_{d_e}$  for observables in paramagnetic systems  based on atomic theory calculations.  Ranges (bottom entry) for coefficients $\alpha_{ij}$ representing the contribution of each of the T-odd/P-odd parameters to the observed EDM of each system. For atomic systems, the atom EDM is measured, whereas for molecular systems the P-odd/Todd frequency is measured, from which $d_e$ and $C_S$ are determined from the tabulated $\alpha's$. (Note that for YbF and ThO, $\alpha_{d_e}=eE_{eff}/\hbar=\pi W_d$, with $W_d$ given in \cite{Engel:2013lsa}; for HfF$^+$, $\alpha_{d_e}=eE_{eff}/\hbar$~\cite{Cairncross:2017fip} and $\alpha_{C_S}=W_S=W_{T,P}\frac{Z+N}{Z}$ with $W_{T,P}$ given by~\cite{Skripnikov2017}.) References: $a$~\cite{Ginges:2003qt}; $b$~\cite{Engel:2013lsa}; $c$~\cite{Nataraj2008}; $d$~\cite{rf:Dzuba2011,rf:Dzuba2011-erratum}, $e$~\cite{Meyer:2008gc}; $f$~\cite{rf:Dzuba2011,rf:Dzuba2011-erratum}; $g$~\cite{rf:Skripnikov2013}; $h$~\cite{Petrov2007}; $i$~\cite{Fleig2013}; $j$~\cite{Skripnikov2017}. \label{tb:paramagnetics}}
\end{table*}

 \begin{table*}[h]
\centering \renewcommand{\arraystretch}{1.5}
\begin{tabular}{|c|c|c|c|c|c|c|c|}
\hline\hline
System & $\partial d^{exp}/ \partial d_e$  & $\partial d^{exp}/\partial C_S$ ({\it e}~cm)  & $\partial d^{exp}/\partial C_T^{(0)}$ ({\it e}~cm)  & $\partial d^{exp}/\partial \gpiz$ ({\it e}~cm) &   $\partial d^{exp}/\partial \gpio$ (\ecm) &
 $\partial d^{exp}/\partial \bar d_n^{sr}$\\
\hline
neutron    &  & & &  $1.5\times 10^{-14}$  & $1.4\times 10^{-16}$ & $1$  \\
\hline
$^{129}$Xe  & -0.0008  & $-4.4\times 10^{-23}$  & $-6.1\times 10^{-21}$   & $-0.4\times 10^{-19}$  & $-2.2\times 10^{-19}$ & $1.7\times 10^{-5}$  \\
 &    &    $-4.4$-$(-5.6)$                           & $-6.1$-$(-9.1)$   &    $-23.4$-$(1.8)$  & $-19$-$ (-1.1)$  & $1.7$-$2.4$  \\
\hline 
$^{199}$Hg & -0.014  & $-5.9\times 10^{-22}$  & $3.0\times 10^{-20}$   & $-11.8\times 10^{-18}$  & 0 & $-5.3\times 10^{-4}$ \\
 &    $-0.014$-$0.012$ &  & $3.0$-$9.0$   &   $-38$-$(-9.9)$  & $(-4.9$-$1.6)\times 10^{-17}$ & $-7.7$-$(-5.2)$  \\
\hline
$^{225}$Ra  &   & $ $  & $5.3\times 10^{-20}$   & $1.7\times 10^{-15}$  & $-6.9\times 10^{-15}$ &   \\
 &    &                               &  5.3-10.0 &  $6.9$-$0.9$     &   $-27.5$-$(-3.8)$ & $(-1.6$-$0)\times 10^{-3}$ \\
\hline
TlF  & 81  &   $2.9\times 10^{-18}$  & $2.7\times 10^{-16}$  & $1.9\times 10^{-14}$  & $-1.6\times 10^{-13}$ & $0.46$ \\
 &    &                                 & & $0.5$-$2$ &  & $-0.5$-$0.5$ \\
 \hline\hline
\end{tabular}
\vskip -0.1 truein
\caption{Coefficients for P-odd/T-odd parameter contributions to EDMs  for diamagnetic systems and the neutron. The second line for each entry is the reasonable range for each coefficient.   The $\partial d^{exp}/ \partial d_e$  and  $\partial d^{exp}/ \partial C_S$ are from~\cite{Ginges:2003qt} and are based on~\cite{rf:Martensson1985} and~\cite{rf:Martensson1987} for $^{129}$Xe, and $^{199}$Hg. Also see~\cite{Fleig:2018bsf} for $^{199}$Hg. The $\partial d^{exp}/ \partial d_e$ and  $\partial d^{exp}/ \partial C_S$   for TlF are compiled in~\cite{rf:Cho1991}. The $\partial d^{exp}/ \partial C_T^{(0)}$ are adjusted for the unpaired neutron in $^{129}$Xe, $^{199}$Hg and $^{225}$Ra using $k_T$ from ~\cite{Ginges:2003qt} and is consistent with~\cite{Sahoo:2016zvr}.  For $^{225}$Ra $\partial d^{exp}/ \partial C_T^{(0)}$ is from~\cite{Dzuba2009,Singh:2015aba}. The $\gpbz$, $\gpbo$ and $\bar d_n^{sr}$ coefficients for atoms and molecules are based on data provided in Table~\ref{tb:SchiffCoef}; the range for $^{225}$Ra corresponds to $0\le s_n\le 2$ fm$^2$. 
For TlF, the unpaired neutron is replaced by an unpaired proton  and the ``best value'' assumes $\bar d_p^{sr}=-\bar d_n^{sr}$, {\it i.e.} mostly isovector in analogy to the anomalous magnetic moment, while the range is defined by $|\bar d_p^{sr}|\le |\bar d_n^{sr}|$ . \label{tb:diamagnetics}}
\end{table*}

\begin{table*}[h]
\begin{center}
\begin{tabular}{|c|c|c|c|c|c|}
\hline\hline
System & $\kappa_S=\frac{d}{S}$ (cm/fm$^3$)  & $a_0=\frac{S}{13.5\gpbz}$ ($e$-fm$^3$) & $a_1=\frac{S}{13.5\gpbo}$ ($e$-fm$^3$) & $a_2=\frac{S}{13.5\bar g_\pi^(2)}$ ($e$-fm$^3$) & $s_N$ (fm$^2$) \\
\hline
$^{129}$Xe &	   $0.27\times 10^{-17}$  (0.27-0.38) & $-0.008 (-0.005$-$(-0.05))$ & $-0.006 (-0.003$-$(-0.05))$ & $-0.009 (-0.005$-$(-0.1))$ & 0.63 \\
\hline
$^{199}$Hg &   $-2.8\times  10^{-17}$( $-4.0$-$(-2.8)$) & 0.01 $(0.005$-$0.05)$ & $\pm$0.02 $(-0.03$-$0.09)$ & $0.02 (0.01$-$0.06)$ & $1.895\pm 0.035$\\
\hline
$^{225}$Ra &   $-8.5\times 10^{-17}$ ($-8.5$-$(-6.8)$)  & $-1.5\ (-6$-$(-1))$ & $+6.0$  $(4$-$24)$ & $-4.0\ (-15$-$(-3))$ & \\
\hline
TlF &  $-7.4\times 10^{-14}$  & -0.0124 & 0.1612 & -0.0248 & $0.62$\\
\hline\hline
\end{tabular}
\end{center}
\vskip -0.1 truein
\caption{\label{tb:SchiffCoef} Ranges and ``best values'' used in~\textcite{Chupp:2014gka} for atomic EDM sensitivity to the Schiff-moment and dependence of the Schiff moments on $\gpbz$ and $\gpbo$;  $\kappa_S$ and $s_N$.  References:  TlF: \cite{rf:TlFTheory}; Hg:\cite{rf:DzubaPRAv60p02111a2002,rf:FlbmKhrNuclPhysAp449a1986,Singh:2014jca}; Xe: \cite{rf:DzubaPLBv154p93a1985,rf:DzubaPRAv60p02111a2002,Teruya:2017don}; Ra: \cite{rf:SpevakPRCv56p1357a1997,rf:DzubaPRAv60p02111a2002,Singh:2015aba}. Values for $a_0$, $a_1$ and $a_2$ are compiled in \cite{Engel:2013lsa}. The value of $s_n$  is from~\cite{Dzuba1985} for $^{129}$Xe and from~\cite{dmitriev03} for $^{199}$Hg; there is no available calculation of $s_n$ for $^{225}$Ra. The value for $s_p$ for TlF is derived from~\cite{rf:Cho1991}.
}
\end{table*}

\section{Experimental techniques}
\label{sec:Techniques}

The crux of any EDM measurement is  to measure the effect of the coupling to an electric field in the background of much larger magnetic effects using the unique P-odd/T-odd signature.
Most EDM experiments using beams or cells are magnetic resonance approaches that measure the energy or, more commonly, frequency given in Eqn.~\ref{eq:EDMFreqEquation1} of transitions between magnetic sublevels in the presence of a well-controlled magnetic field $\vec B$, and electric field $\vec E$ aligned either parallel or anti-parallel to $\vec B$. Storage-ring experiments with charged particles measure the result of the additional torque on the spin due to $\vec d\times\vec E$, where $\vec E$ may arise in part from the motional field $\vec v\times\vec B$. In solid-state electron-EDM experiments, the observable is proportional to $\vec B\cdot\vec E$, where only one field is applied and the other measured - for example a strong electric field $\vec E_{\rm applied}$ would polarize electron spins in the material giving rise to an observable magnetic field $\vec B_{\rm observed}$.
 
Because every system of interest has a magnetic moment, the magnetic environment is crucial and the magnetic field must be characterized in space and time. Magnetic shielding, magnetic sensors external to the EDM volume, and comagnetometers that monitor the magnetic field within the EDM volume during the EDM measurement are all essential elements of past and future experiments. Comagnetometer species are chosen because they are less sensitive to P-odd/T-odd effects than the key species. For example a $^{199}$Hg comagnetometer was used for the neutron-EDM experiment~\cite{Baker:2006ts}, and Na was used as a comagnetometer for Cs~\cite{rf:Weisskopf1968} and Tl~\cite{Regan:2002ta}. The measurement of the $^{129}$Xe EDM utilized $^3$He as the comagnetometer species~\cite{rf:Rosenberry2001}. In the case of polar molecules discussed in Sec~\ref{sec:PolarMolecules}, the comagnetometer can be effected with one molecular species using combinations of transitions~\cite{Hudson:2002az,Baron:2013eja,Cairncross:2017fip}. 
 
 
EDM measurements in many systems require determining frequency differences with precision of nHz ($10^{-9}$ Hz) or less with measurement times much less than $10^9$ seconds.
Another interesting feature for stored atoms and neutrons  is the need to correct for the Earth's rotation as well as the accumulated quantum phase or Berry's phase that arises due to the combination of motional magnetic field $(\vec v\times\vec E)/c^2$ with magnetic-field gradients.
 
\subsection{Magnetic shielding}
\label{sec:MagneticShielding}
A critical component of EDM experiments is the magnetic shielding, which mitigates electromagnetic distortions in time and space. Neutron and proton EDM experiments in particular require large volumes with stringent magnetic shielding requirements. For example, improving the current limit on the neutron EDM in the next generation experiments 
by two orders of magnitude requires magnetic-field gradients less than nT$/$m (see sec.~\ref{sec:systematics}). The temporal stability of the magnetic gradient must be better than 100~fTm$^{-1}$s$^{-1}$ over the  ~100-300~s neutron-storage time. This requires  strong damping  of external perturbations at extremely low frequencies, {\it i.e.} between 1-100~mHz. It is also crucial to reverse the magnetic field orientation in the lab, which requires magnetic shields that can tolerate large changes of the field {\em inside} the shield.
Recent advances in active and passive magnetic shielding techniques for  next-generation room-temperature EDM  experiments are reported by~\textcite{rf:JAPv117p233903y2015,Altarev:2014wka,rf:SunDynamicShielding}. Cryogenic  magnetically shielded environments have  been demonstrated, {\it e.g.} by \textcite{Cabrera:1989qr,ref:romalis_ferrite}, and these concepts have been extended to larger volumes for the SNS cryogenic nEDM experiment~\cite{Slutsky:2017mbn}. 
 

Passive magnetic shielding is based on surrounding the volume of interest with a high-magnetic-permeability  material generally called mu-metal or permalloy. Permalloy is applied in thicknesses of order 1-4 mm rolled into welded cylinders or cones~\cite{rf:BudkerMSRCones} or assembled as sheets.
A passive  shield is best characterized by  the reduction in the amplitude of magnetic field variations {\it i.e.} the frequency-dependent damping factor or shielding factor $SF(f)$. 
A second crucial characteristic is  the residual field and gradient inside the shield, which is affected by a procedure called degaussing or equilibration.

It is useful to provide analytic approximations for static fields as guidelines for cylindrical shields composed of permalloy cylinders with permeability $\mu$, thickness $t$ and radius $R$~\cite{SumnerMagneticShielding1987}, with the caveat that actual results for damping factors at low frequencies, in particular for multi-layer shields, may differ significantly.  The transverse damping factor for a single shield is 
\begin{equation}
SF^T\approx 1+\frac{\mu t}{2R}.
\end{equation}
The transverse shielding factor for multiple layers is a product of damping factors;  for $n$ layers the transverse shielding factor is approximately
\be
SF^T\approx SF^T_n\prod_{i=1}^{n-1} SF_i^T \left(1-(\frac{R_i}{R_{i+1}})^2\right).
\ee 
The air-gaps between layers can be optimized for a given material, thickness, and number of layers.
Axial shielding is generally less effective and depends on the ratio of the length of the cylinder to the radius $a=L/R$ and the empirically determined distribution of magnetic flux over the cylinder end caps and walls, which is characterized, respectively, by quantities  $\alpha$ and $\beta$  ($\alpha\le 1$ and $1\le\beta\le 2$): 
\be
SF^A \approx 1+\left(\frac{2\kappa(a)}{1+a+\alpha a^2/3}\right) SF^T,
\ee
where
\begin{eqnarray}
\kappa(a)&=&\left(1+\frac{1}{4a^3}\right)\beta-\frac{1}{a}\nonumber\\
&+&2\alpha\left\{\ln(a+\sqrt{1+a^2})-2\big(\sqrt{1+\frac{1}{a^2}}-\frac{1}{a}\big)\right\}.\nonumber\\ 
\end{eqnarray}
While these expressions are illustrative, in general, effective design of shields  is aided by simulations of Maxwell's equations using finite-element approaches.

Additional considerations for passive shields include penetrations (holes), the temperature dependence of the magnetic properties of the shielding material,  and the applied internal field, which couples to the shield and may cause a temperature dependence of fields and gradients for the experiment~\cite{ANDALIB2017139}.  Holes up to 130 mm  do not notably change the damping factor. 
 Also,  any conductor close to the experiment produces Johnson current noise, which in turn causes magnetic-field noise~\cite{LeeSKRomalis2008}. This includes the permalloy and is also a consideration for the non-magnetic ({\it e.g.} aluminum)  RF-shielding layer.
Temperature differences also cause slowly changing magnetic fields in many conductors, which put additional constraints on the design of the experimental apparatus located inside shields.

A comparison of  leading magnetic-shield installations, is presented in Table~\ref{fig:sf} and Fig.~\ref{fg:ShieldComparison}. The Boston Medical Center shield (Boston), used for biomagnetism research, is composed of three permalloy layers and three aluminum layers resulting in a large damping factor at the relatively high freqeuency of 1 Hz~\cite{ref:boston}. The Berlin-Magnetically-Shielded Room (BMSR-II) is a large-scale/walk-in user facility with small residual fields~\cite{ref:bmsr2}.  The TU-M\"unchen shield (TUM-Shield), consists of six layers plus a 1 cm thick aluminum layer for RF shielding.The outer layers form a rectangular box and the inner most layer is a permalloy cylinder~\cite{Altarev:2014wka}. 

\begin{table}
\begin{ruledtabular}
\begin{tabular}{|l|r|c|r|}
\textrm{Shield} &      \textrm{$f$~$[ Hz ]$} &  \textrm{B$_{ext.}$~$[ \mu T_{rms} ]$} &  $\textrm{SF}(f)$ \\
\colrule
\colrule
BMSR-II      &                                0.01     &   1                                   &           75000          \\
BMSR-II       &                                1     &       1                                   &           2,000,000          \\
\colrule
Boston    &                                0.01     &       1                                &           1630          \\
Boston       &                                1     &         1                                  &           200,000          \\
\colrule
TUM MSR $+$ Insert      &   0.01        &             4.5                 &     $\sim$2,000,000          \\
TUM MSR $+$ Insert       &   1.25        &             22                   &     $>$ 16,700,000          \\
\end{tabular}
\end{ruledtabular}
\caption{\label{fig:sf} Measured  damping factor $SF(f)$ of three shield-installations  for different external excitation strengths B$_{ext.}$(either peak-to-peak or root mean square) and frequency $f$. The BOSTON shield ~\cite{ref:boston}, BMSR-II is the Berlin Magnetically-Shielded Room at PTB~\cite{ref:bmsr2} were developed primarily for biomagnetism and magneto-medicine. The TUM shield at the Technical University of M\"unchen, was developed for neutron EDM measurements~\cite{Altarev:2014wka,rf:JAPv117p233903y2015}. 
}
\end{table}

\begin{figure*}[tb]
\center
\includegraphics[width= 7 truein]{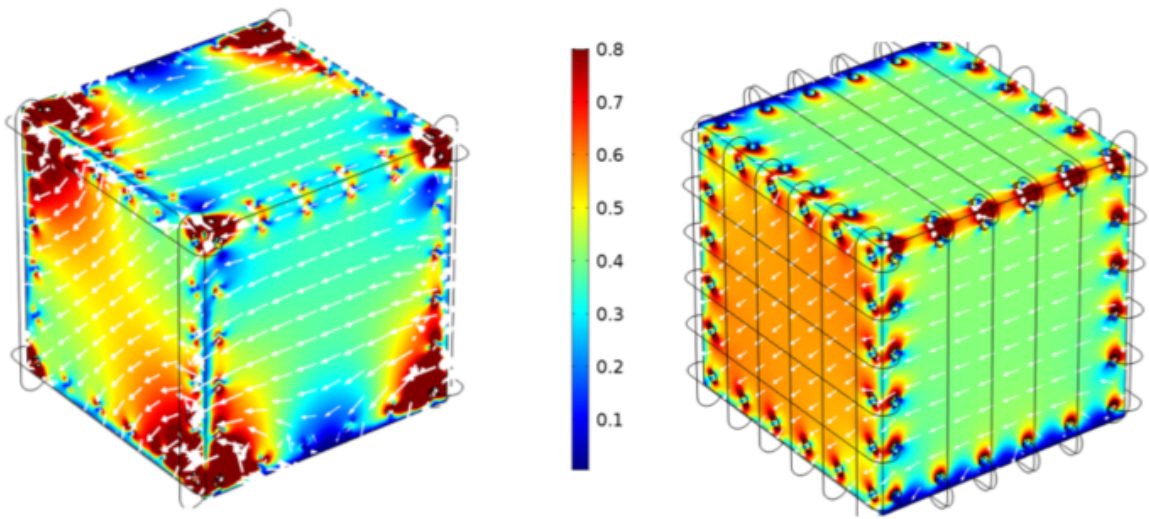}
\vskip -2 truein
\caption{\label{fg:LShapedCoils}  (Color online)
Magnetization of a permalloy box with equilibration coils. Left: Original L-shaped configuration with coils at the corners. Right: Distributed L-shaped configuration with coils arranged over the surfaces. The coils are solid lines. The color scale shows the magnetization in the permalloy in arbitrary units. Figure provided by Z. Sun. 
}
\end{figure*}

Magnetic equilibration is a procedure based on commonly known degaussing techniques developed to achieve extremely small residual fields and gradients~\cite{rf:FThielRSIv78p035106y2007,rf:JAPv117p233903y2015,rf:SunDynamicShielding,10.2478/mms-2013-0021}. 
Magnetic equilibration based on the developments of~\textcite{10.2478/mms-2013-0021} used coils wound around the edges of each shell of a cuboid shield to generate strong magnetic flux with a sinusoidal AC current of several Hz frequency and linearly or exponentially decreasing amplitude. 
 The result of the procedure is a superposition of damped fields from outside the shield, residual magnetization of the shield material, distortions caused by holes and other imperfections, and applied internal fields. A residual field gradient of order 1~nT$/$m over about 1~m$^3$ was reached after applying the equilibration procedure in all three directions for typically 100~s in each direction.
%
%
%
Recently~\textcite{rf:JAPv117p233903y2015} implemented the L-shaped coil arrangement shown in FIG.~\ref{fg:LShapedCoils}, which enabled magnetic saturation of all shells at once in a much shorter total time  ($<$~50~s) with similarly small residual fields inside the shield. The speed and reproducibility are a benefit to EDM measurements, because it is necessary  to reverse the direction of the magnetic holding field and equilibrate the shields regularly  during an experiment to control systematic effects.
Numerical simulations have been compared to measurements showing that equilibration can be successfully fully modeled~\cite{rf:SunDynamicShielding}.
%
%
%
%
%
%
%
%
%
 
%

\begin{figure}[tb]
\vskip -0.5 truein
\includegraphics[width=3.5 truein]{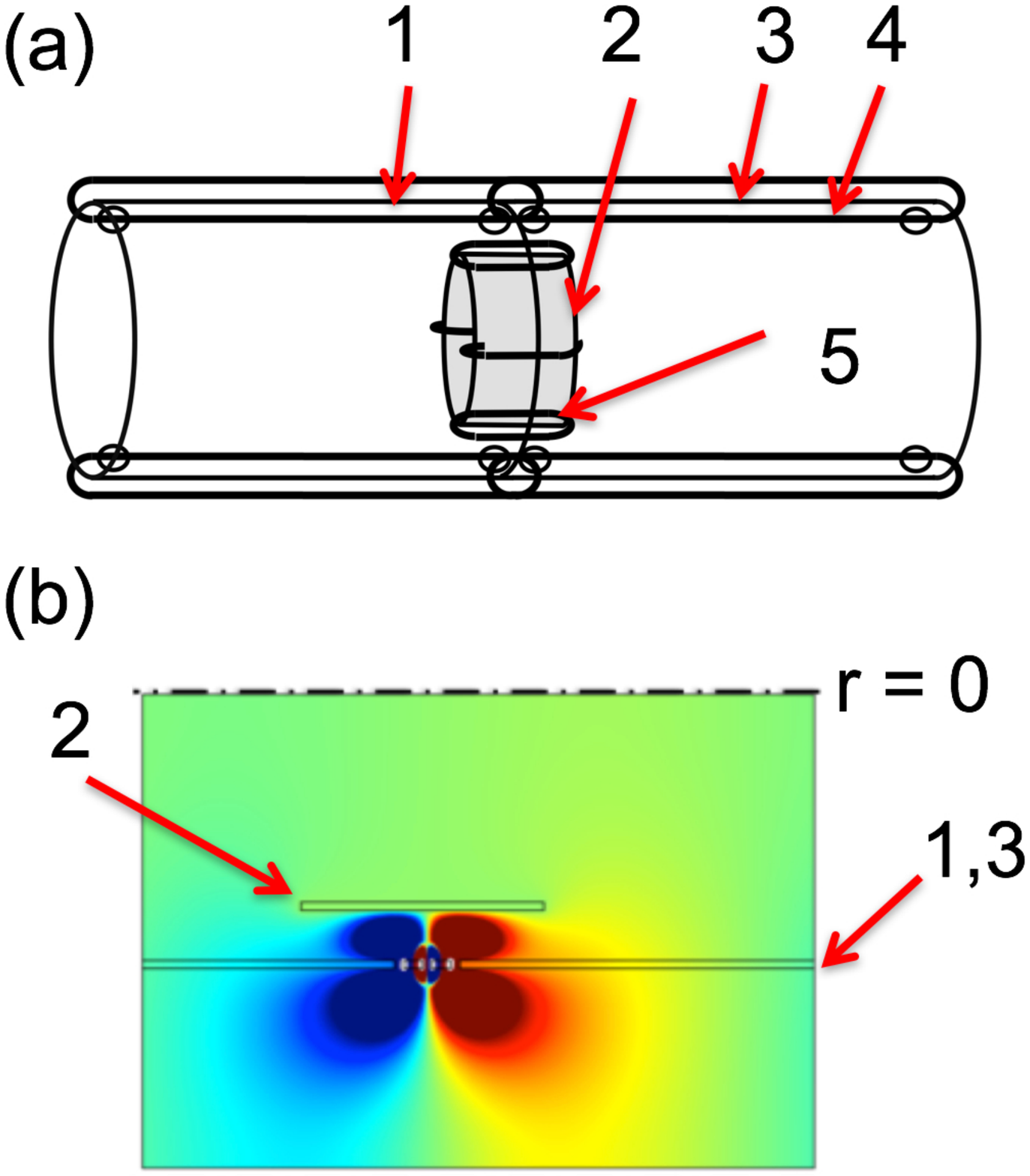}
\vskip -0.5 truein
\caption{\label{fig:pedmshield}  (Color online)
Shielding designed for a proton EDM experiment: (a) Permalloy and compensation coil configurations. The outer layers (1,3)  are connected in segments. A correction ring (2) compensates for the magnetic distortions caused by the magnetic equilibration coils (4). The compensation ring equilibration coils are labeled (5).  (b) Simulation of magnetic distortions caused by  outer-cylinder equilibration coils mitigated by the compensation ring. 
}
\end{figure}

A proposed experiment to measure the EDM of the proton, described in Sec.~\ref{sec:StorageRingEDMs}, requires magnetic shielding of an 800~m circumference electrostatic storage ring with  magnetic less than a few nT at any point~\cite{Anastassopoulos:2015ura}. 
An effective solution is a toroidal shield made up of individual three meter long cylinders illustrated in FIG.~\ref{fig:pedmshield}. 
%
%
Magnetic equilibration of a shield of this length would require  degaussing individual cylindrical sections, but
residual magnetization may build up near the equilibration coils due to non-uniformity of the magnetic flux at the ends of the cylinders. This can be compensated by the short ring of permalloy  placed inside the shield,.
%
%
%
%
 
%
%
%
%
%
 
%
%
%
%
%
%
%


\subsection*{Cryogenic shields}
 
Cryogenic shielding, based on the Meissner effect with Type-1 superconductors  {\it e.g.} Pb, has been envisioned for low-temperature EDM experiments. In contrast to permalloy-based passive shields, cryogenic shields stabilize both external perturbations and instabilities in the applied magnetic field. Also, the residual magnetic field inside a cryogenic shield is frozen during the transition to superconductivity. \textcite{Slutsky:2017mbn} developed a prototype cylindrical cryogenic shield 4 m long with 1.2 m diameter, which  provided a gradient less than 1 nT/m  over over a 0.1 m$^3$ volume.
An additional consideration for  cryogenic shielding is that the magnetic field cannot be reversed without warming up the shield.
 
\begin{figure}[tb]
\includegraphics[width=3.75 truein]{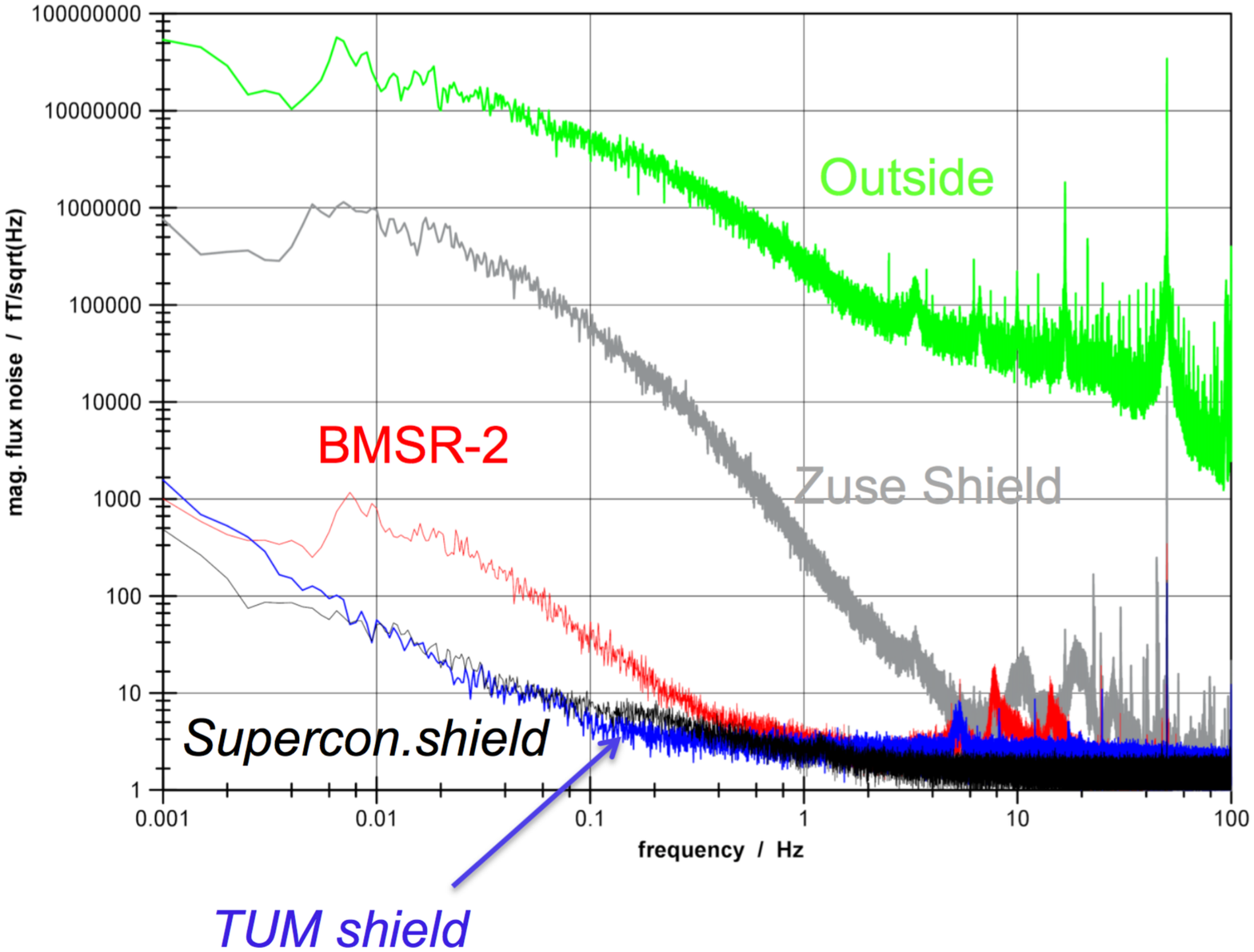}
\vskip -0.25 truein
\caption{\label{fg:ShieldComparison}  (Color online)
Comparison of noise in different shields measured with SQUIDs. Supercon. refers to a superconducting cylinder cooled below the transition temperature inside the low magnetic field environment BMSR-II. The Zuse shield is also at PTB-Berlin~\cite{10.2478/mms-2013-0021}. 
At very low frequencies, the intrinsic noise of the (different) SQUID systems combined and the integration time to record data for this plot dominate the performance; at high frequencies, the noise level is dominated by the experimental setup to perform the measurement. Figure provided with permission by M. Burghoff.
}
\end{figure}

In operation, a superconducting shield is surrounded by a room-temperature shield so that it can be cooled below the critical temperature in a small external field.  Another approach is a  combination of room-temperature and superconducting materials, for example a cylinder wrapped with METGLAS\footnote{METGLAS is commercial product: Metglas¨, Inc.
440 Allied Drive, 
Conway, SC 29526-8202,
www.metglas.com.}~\cite{PG11,Slutsky:2017mbn}. Magnetic-field noise less than 100 fT/$\sqrt{\rm Hz}$ at $f=0.01$ Hz has been achieved in hybrid configurations measured by~\textcite{rf:burghoff_kreuth_presentation}. Figure~\ref{fg:ShieldComparison} provides a comparison of damping factors for several shield configurations including the cryogenic shield placed inside the Berlin MSR-II.


\subsection{Magnetometers}
\label{sec:Magnetometry}
%
%
%
%
Any uncompensated change in the magnetic field between two subsequent or spatially separated  measurements with opposite electric field will  appear as a false EDM:
\begin{equation}
d_\mathrm{false} = \frac{\mu \Delta B }{ E}.
\label{eq:sigma_b}
\end{equation}
For a random-noise spectrum of magnetic field variations, this effect will be reduced in proportion to $\sqrt{M}$ for $M$ subsequent electric-field reversals. However, any correlation of the electric and magnetic field, for example due to leakage currents across the storage cell, will not average towards zero. %
In either case, monitoring the magnetic field is essential to a successful measurement. 
%
As an example, for a measurement precision of $\sigma_{d} = 10^{-28}$\ecm, an electric field $E = 10^4$~V/cm, and $M$ = 10$^4$, the required effective magnetic field measurement precision is $\delta B\approx 2$ fT.
%
%
The magnetic field can  be measured (i) with magnetometers surrounding the experiment to estimate the full flux entering and leaving the experiment at the time of the measurement, or (ii) directly at the position of the measurement and at the same time, with a comagnetometer. External magnetometer measurements can be used to estimate the magnetic flux through the EDM experiment volume,  {\it i.e.} 4$\pi$ magnetometry~\cite{Nouri:2014epa,Nouri201492,Nouri:2015xva,LinsThesis2016} without the complications of injecting the comagnetometer species into the EDM measurement chambers. However external magnetometry requires that the magnetic environment inside  the EDM chamber be well characterized and that any changes, for example magnetized spots generated by sparks  can be monitored.

%
%

Comagnetometers were first deployed in the electron-EDM Cs beam measurement by~\textcite{rf:Weisskopf1968} and the Tl beam measurement by~\textcite{Regan:2002ta}. For atomic-EDM measurements, the concept relies on comparable magnetic moments but very different EDMs due to the $Z$ dependence  contributions due to the electron EDM, electron-nucleus couplings and  the Schiff moment.  Generally, these scale approximately as $Z^2$ for diamagnetic atoms and $Z^3$ for paramagnetic atoms. In principle, however, an experiment really measures the {\em difference} of the species' EDMs. In contrast to external magnetometers, comagnetometers should have coherence times comparable to the storage/interrogation times of milliseconds for atomic beams to  several hundred seconds for UCN to thousands of seconds for the $^{129}$Xe EDM experiments.

%
%
%
%
%
%

\begin{figure}[tb]
\includegraphics[width=3.5 truein]{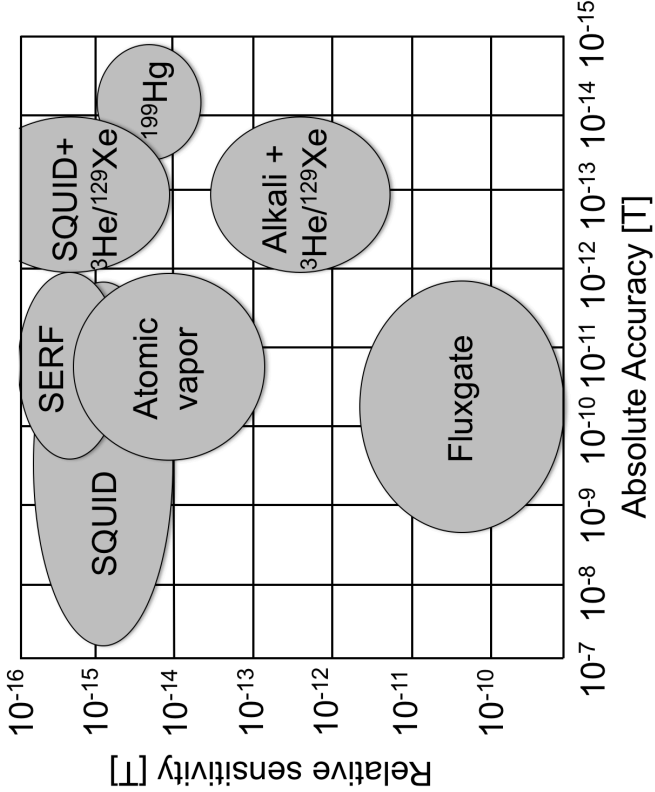}
\caption{\label{fig:magnbandwidths} 
Sensitivity vs. accuracy of sensors used for low-field magnetometry in EDM measurements in typical state-of-the-art configurations: flux-gate (FG), $^3$He, alkali-metal magnetometers with Cs and K, spin-exchange-relaxation-free (SERF) mangetometers and combinations of $^3$He and Cs or SQUID readout. The term ``sensitivity'' is used to describe the response of the sensor to changes in the magnetic field, whereas ``accuracy'' is used here to describe the  absolute accuracy for measuring a fields, which is a measure of the stability, which is crucial for EDM experiments.  
}
\end{figure}

%
A comparison of the sensitivity and accuracy of commonly used magnetometers is presented in FIG.~\ref{fig:magnbandwidths}. Sensitivity characterizes the smallest change in the magnetic field that can be detected and generally improves with the measurement time, at least for short times. Sensitivity is clearly important for frequency stability and for monitoring systematic effects such as leakage currents. Accuracy, which is essentially  calibration stability, is required for two (or more) separated magnetometers used, for example, to determine static and time-changing magnetic-field gradients.

\subsection*{Rb and Cs magnetometers}
Alkali-metal  magnetometers (usually Rb or Cs) have been developed and implemented since the inception of optical pumping, and their sensitivity and stability has been improved and optimized for a variety of experiments. Recently, Cs has been the main focus of magnetometers for EDM measurements because sufficient vapor density is attainable at low temperatures and due to the availability of optical components including diode lasers and optical fibers~\cite{AfachBisonWeisPSIMagnetometry}. Typically a Cs optical magnetometer uses glass cells with spin polarized or aligned vapor. The magnetic field is determined from the frequency of a resonance or free-precession signal read out via transmission of resonant polarized light or optical rotation of off-resonant light.

Several magnetometer schemes are feasible, and the most common are called $M_x$ magnetometers~\cite{rf:BloomOriginalMangetometer,PhysRevLett.6.280} and NMOR sensors~\cite{ref:pustelny}.  $M_x$ sensors monitor the Larmor frequency of atomic spins in a static field, {\it e.g.} along $\hat z$, using a perpendicular  oscillating field along $\hat  x$ tuned to the magnetic-resonance frequency of the atomic species. 
Though $M_x$ magnetometers generally have simpler design and better stability over longer times, the  RF magnetic fields may lead to cross talk among multiple sensors placed in close proximity to each other~\cite{rf:AleksandrovCsMx}. Metal cans are used as Faraday shields to mitigate these cross talk effects, but introduce magnetic-field  Johnson noise. 
%
NMOR  refers to Nonlinear Magneto-Optical Rotation of linear polarization, which can be used for magnetometry when the light is modulated in frequency or intensity at the Larmor frequency.
NMOR is a fully optical technique, and sensors can be built without any metallic components. Additionally, the atoms can be prepared with an alignment, a distribution of magnetic sub-levels with a magnetic quadrupole moment but no magnetic dipole moment, and  magnetic cross talk to other sensors is significantly reduced.
%

For Cs magnetometers, the frequencies for a typical magnetic field of 1 $\mu$T are 3.5 and 7~kHz for $M_x$ and NMOR modes respectively.  
Typical response times are on the order of 10-100~ms. Operation modes include continuous pumping at the resonance frequency, self-oscillation, and free precession decay, depending on the type of information needed~\cite{magnetometry_modes}.
%
%
%
In particular, free precession decay is systematically cleaner due to smaller interactions with the pump laser, whereas forced oscillation or self-oscillation may be affected by light shifts, a modification of the atomic Hamiltonian in the presence of the near-resonant light, which takes the form of an effective magnetic field~\cite{ref:cohen-tann}. 
For example, a drift of the laser power, frequency, or polarization would change the light shift leading to instability of the magnetometer~\cite{Grujfa2015}.
A quantitative study of light shifts for Cs magnetometers has been undertaken by~\textcite{brian_lightshifts}.

The walls of the evacuated alkali metal vapor cells are generally coated with paraffin or other materials to improve wall-relaxation times~\cite{SingCsCoating1972}. Transverse spin-coherence times $T_2^*$ of 1-2~s, were observed for paraffin~\cite{ref:alexandrov,ref:alexandrov-erratum}, and   as long as 60~s in alkene-coated cells~\cite{ref:balabas}.\footnote{In this review, we characterize longitudinal relaxation due to wall interactions, magnetic field gradients, collisions ({\it e.g.} dipole-dipole), and weakly bound molecules as $T_1$. Observed decay of spin coherence or transverse relaxation, which may be due to longitudinal effects as well as magnetic field gradients and collisions, as $T_2^*$.}   Spin exchange between atoms is a fundamental limitation that has been suppressed by increasing the alkali-metal density or attaining very high polarization in spin-exchange relaxation-free or SERF magnetometers~\cite{RomalisSERFNature}.  Even the longest observed $T_2^*$'s are much less than the duration of a typical EDM measurement, and many independent measurements are in effect  added incoherently to obtain the magnetic-field ($B$) sensitivity for a measurement time $\tau$:
\be
\sigma_B\approx  \frac{1}{2\pi\gamma}\sqrt{\frac{1}{N_A T_2^*\tau}},
\label{eq:CsSensitivity}
\ee
where the gyromagnetic ratio is $\gamma=3.5$ kHz/$\mu$T for Cs and $N_A$ is the effective  number of spins observed in the time $T_2^*$.  For observation times $\tau$ up to about 10 s, the typically observed sensitivity of alkali-metal magnetometers is~\cite{ref:csintrinsicsensitivity}
\be
\sigma_B\sim~1-10~\frac{\rm fT}{\sqrt{\rm Hz}}\times\frac{1}{\sqrt{\tau}}.
\ee 
The dependence on $\tau$ of Eqn.~\ref{eq:CsSensitivity} does not hold for times greater than about 10-20 seconds due to many sources of instability, {\it i.e.} drifts in Cs density,  laser intensity, light polarization, fibers and electronics as well as, for example temperature dependent interactions of the Cs spins with the vapor cell walls.

Magnetometers that measure a single frequency proportional to the magnitude of a magnetic field provide information that is intrinsically a scalar.
Several techniques to extract vector information have been developed by  \textcite{ref:pustelny} and \textcite{Afach:15}, among others. In addition, approaches that use light shifts along different directions to modulate the vector information are very promising~\cite{ref:lightshift_cs}.
Laser-driven Cs atomic magnetometers can also be  operated in an array, driven by the same laser which can mitigate common-mode noise and drifts, though the possibility of cross talk between magnetometers requires care in the deployment.

\subsection*{Nuclear-spin magnetometers: $^3$He and $^{199}$Hg}

Optically pumped external and internal magnetometers with either $^3$He or $^{199}$Hg have also been proposed for neutron EDM measurements by~\textcite{rf:Ramsey1984} and studied by \textcite{GREEN1998381} and~\textcite{BORISOV2000483}. Several  planned future room-temperature neutron EDM experiments also plan to use  $^{199}$Hg as a comagnetometer, and the SNS nEDM experiment will use $^3$He as a comagnetometer as well as the detector, as described in Sec.~\ref{sec:NeutronEDM}.

 For nuclear-spin magnetometers, the spins are prepared using optical pumping techniques and then set to precess freely in the magnetic field. For $^{199}$Hg, the free precession is monitored by the transmission of linearly-polarized or circularly-polarized light~\cite{GREEN1998381}. For $^3$He, the precessing magnetization at $\approx$300 $\mu$T has been monitored by inductive pickup~\cite{rf:Rosenberry2001}. For the lower fields used for neutron EDM measurements non-inductive sensors  such as SQUID magnetometers~\cite{Allmendinger:2013eya,Kuchler:2016eik} or Cs magnetometers~\cite{ref:heil3he} have been used to monitor $^3$He precession. 
%
%
%

%
The practical sensitivity limit of $^{199}$Hg magnetometers is a few fT for a 100~s integration time using both linearly and circularly polarized light. For $^{199}$Hg, the Larmor frequency is 7.79~Hz$/$$\mu$T, 
 which limits the bandwidth for monitoring magnetic field variations to about 1 Hz. 
%
%
In the ILL-Sussex-Rutherford neutron EDM  $^{199}$Hg,   the $^{199}$Hg  coherence time $T_2^*$ was observed to drop  to about 60~s when the high voltage was applied~\cite{GREEN1998381,nEDMHarrisPhysRevLett.82.904}. 
Cleaning the walls with an oxygen discharge at 1~Torr 
 with regular reversal of E-field  restored the $^{199}$Hg $T_2^*$ to 400 s~\cite{GREEN1998381}. 
Other ways to improve this behavior are being investigated, including adding helium as a buffer gas to reduce the Hg mean free path which reduces the rate of depolarizing wall collisions. Unfortunately, introducing helium gas also reduces the high voltage breakdown strength of the storage cell. 
%
%
%
%
The use of a $^{199}$Hg comagnetometer further reduces the choice of wall coatings for neutron storage, as any contact with metallic surfaces will generally cause loss of $^{199}$Hg polarization. 
Light shifts during readout can also affect $^{199}$Hg magnetometers. For readout using a resonance lamp, the width of the emission spectrum averages over the light shift. For laser readout,  light-shift effects for Hg can be mitigated in several ways including reducing the readout laser power, free precession or relaxation in the dark, for which the laser interrogates the $^{199}$Hg phase for short times ({\it e.g.} 15 s) at the beginning and end of the free-precession period, or detuning the laser to the  zero light shift point~\cite{Griffith:2009zz}.
\label{sec:LightShift}

%
%
%
%
%
For $^3$He transverse-spin lifetimes $T_2^*\sim 10^3$~s are easily possible and have been demonstrated at a few mbar pressure even for cells with volume comparable to the dimensions of a neutron EDM chamber~\cite{ref:heil3HeT2}. Higher $^3$He pressure is generally necessary for  detection of the polarization by a SQUID sensor or Cs magnetometer. 
%
For neutron EDM sensitivity of 10$^{-28}$\ecm,  $^3$He pressure of few mbar is a possible compromise. 
In the planned SNS neutron EDM experiment, the density of highly polarized $^3\mathrm{He}$ is much less, as discussed in Sec. IV.A.
%
%
%

%

Two-photon magnetometry was originally suggested for $^{129}$Xe as an alternative to $^{199}$Hg due to the smaller neutron-absorption cross section~\cite{SkylerThesis}. Since the 147 nm single photon transition in Xe is too far in the UV, two 256 nm photons could be used for the free-precession measurement~\cite{SkylerThesis}. For $^{129}$Xe magnetometry, the spin polarized gas sample would be prepared by SEOP~\cite{rf:RosenRSI}. 
Ultrafast two-photon spectroscopy~\cite{FreqComb} has the advantages of high peak intensity (two-photon absorption is proportional to the intensity squared) and coherence, which allows two photons of different frequencies to combine thus probing all atoms within the Doppler profile. Homogeneous broadening due to buffer gas collisions also helps cover the Doppler profile. 
Both the magnetometry signal and the efficiency for harmonic generation of the UV light depend on the laser's peak intensity, resulting in a considerable enhancement for an ultra-fast pulsed laser compared to a continuous-wave laser of the same average power. Picosecond pulses are a good compromise between high peak intensity and lower damage to crystals and optics since the dominant damage mechanisms scale with average intensity~\cite{SkylerThesis}. 
One particularly important feature of two-photon-magnetometry is the spatial resolution possible due to the quadratic dependence of the scattering rate on laser intensity for two-photon transitions.  
Since the laser beam can be focused to a small waist and high intensity along the propagation direction, the scattering rate is highly position dependent and can be used to map out a magnetic field with resolution on the order of 1 mm.
 
\subsection{Magnetic field-coil design and current sources}
 
\begin{figure}[tb]
\includegraphics[width=3.6 truein]{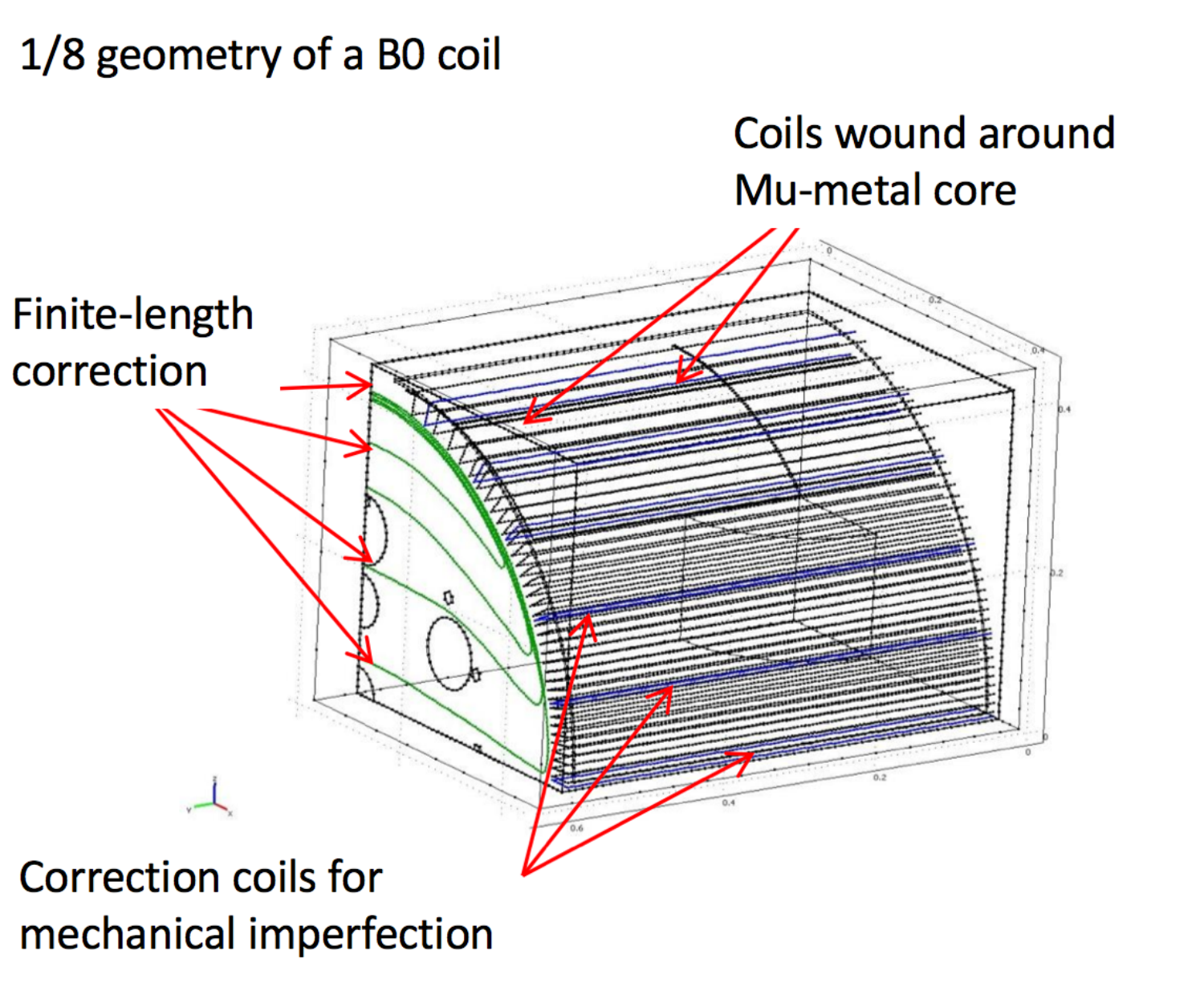} 
\caption{\label{fig:B0Coil}   (Color online) $B_0$ coil from the panEDM apparatus. The coil consists of a closed box of mu-metal with a mu-metal cylinder 
inside the shield. The main coil is wound around the cylinder (see text), which induces azimuthal magnetization in the cylindrical shell; 
correction coils at the ends of the cylinder compensate for the finite length of the windings and for return current paths. Additional correction coils indicated account for imperfections in the geometry.
}
\end{figure}

The stability and uniformity of the applied magnetic field is essential, and 
 EDM experimenters have brought new innovations to the design of magnetic field coils and current sources. 
 The basic cylindrical coil is often called the cosine-theta coil, which is based on the principle that the surface current density for a uniformly magnetized cylinder with magnetization $\vec M$ is $\vec J=\vec M\times \hat n$, {\it e.g.} for $\vec M=M\hat y$,  $|J_z|\propto|\hat y\times \hat n|=\sin(\frac{\pi}{2}-\theta)=\cos(\theta)$ ($\theta$ is the angle measured from the $x$-axis). For a finite length cylinder, the end-cap wires can be positioned to reduce end effects as determined by calculations. Often permalloy is used as a flux return, for example as shown in Fig~\ref{fig:B0Coil}, which has the feature that the field can be aligned to $\approx$~0.1 mrad for $\mu$T fields~\cite{PatentIAltarevetal}.

Another very interesting approach to coil design for a source free volume ($\vec\nabla \times \vec H=0$) is provided by solving Laplace's equation for a magnetic scalar potential $\phi_B$ that satisfies the requirement, for example, of a uniform field $\vec B=B_0\hat z$ within a cylindrical volume. The currents on the surface of the volume correspond to equipotential lines, {\it i.e.} the current-carrying wires should run along equally spaced equipotential contours. For the cylinder with uniform field, this of course also corresponds to the cosine-theta configuration; however this approach is particularly useful for different shapes of coil form, {\it e.g.} a rectangular box and other field profiles~\cite{2017NIMPA.854..127M,rf:ChrisCrawfordPC}.

 Ultra-stable current sources designed for very specific currents and loads  have been developed using standard techniques of PID (proportional-integral-differential) or PI feedback. Current sensing with low-temperature-coefficient resistors compared to ultra-stable voltage references, selecting  discrete components to optimize offset drifts, tuning the gain-bandwidth product for different stages of the circuit and temperature stabilization of crucial components can provide routine performance of $10^{-7}$ and better. Feedback from magnetometers has been used to effectively stabilize magnetic fields to one part in 10$^{11}$ over 1000 second time scales~\cite{rf:Rosenberry2001}.

 
\subsection{Ultra-cold neutron sources}
\label{sec:UCNSources}
Ultra-cold neutrons (UCN), introduced by \textcite{ref:originalucn,ref:originalucn-ru} have velocities less than about 7 m/s, corresponding to kinetic energies $E_\mathrm{kin}< \sim$ 260~neV, temperatures of mK and wavelengths of tens of nm~\cite{book:golub_ucn}. As a result of the long wavelength, the interaction of UCN with material surfaces is characterized by a potential energy called the Fermi energy $V_F$ that is positive for most materials. Neutrons with kinetic energy less than $V_F$  are repelled from the chamber walls for any angle of incidence and thus can be stored in a bottle or cell. In Table~\ref{tb:UCNProperties} we list UCN properties. The gravitational and magnetic potential energies  for maximum UCN kinetic energy correspond to 2.5 meters height and magnetic field 4.3 Tesla, respectively. Thus, for example, neutrons can be stored in a gravitational bottle a few meters deep and can be polarized by reflecting one spin state from magnetic field barrier of $\approx$ 5 T.
%
%
%
%

\begin{table}
\begin{centering}
\begin{tabular}{|l|c|c|c|}
\hline\hline
UCN Velocity & $v_\mathrm{UCN}$ &$<$ 7 &m/s \\
\hline
UCN Kinetic Energy & KE & $<$ 260 & neV  \\
\hline
Gravitational energy& $m_n g$ & 102 & neV$/$m \\
\hline
Magnetic moment & $\mu_n$ & $-60.4$& neV/T \\
\hline
Gyromagnetic ratio & $\gamma_n=\frac{2\mu_n}{\hbar}$ & $2\pi\times$ 29.16 & MHz/T  \\
\hline
\hline
 \end{tabular}
\end{centering}
\caption{\label{tb:UCNProperties} Properties of UCN and neutrons relevant to the neutron EDM experiments. The negative magnetic moment indicates that the strong magnetic field seeking neutrons are ``spin down" with respect to the magnetic field.}
\end{table}

UCN are in principle present in any moderated neutron source. In thermal equilibrium, the neutron-density spectrum is
\begin{equation}
\frac{d\rho(v_n)}{ dv_n}= 2\Phi_0\frac{v_n^2}{v_T^4}\exp(-v_n^2/v_T^2)
\label{eq:thermalflux}
\end{equation}
where $v_T = \sqrt{2 k_B T / m_n}$.
Integrating the density up to the velocity corresponding to the Fermi energy for $v_F<<v_T$  results in a UCN density of
\begin{equation}
\rho_\mathrm{UCN}=\int\limits_0^{v_{F}}d\rho(v_n) \approx \frac{2}{3}\frac{\Phi_0}{v_T}\left(\frac{v_F}{v_T}\right)^{3},
\label{eq:inteflux}
\end{equation}
for $v_F=7$ m/s.
For a thermal neutron source $T=300$K, the core flux is $\Phi_0 \sim 10^{15}$~cm$^{-2}$~s$^{-1}$, and $\rho_\mathrm{UCN}\approx 100$ cm$^{-3}$.  
For cold moderators, typically at 20K, $\Phi_0$ decreases by about an order of magnitude and the UCN in the moderator is increased to 1000-2000~cm$^{-3}$. 
%
%

\begin{figure}[tb]
\vskip -.35 truein
\centerline{\includegraphics[width=5truein]{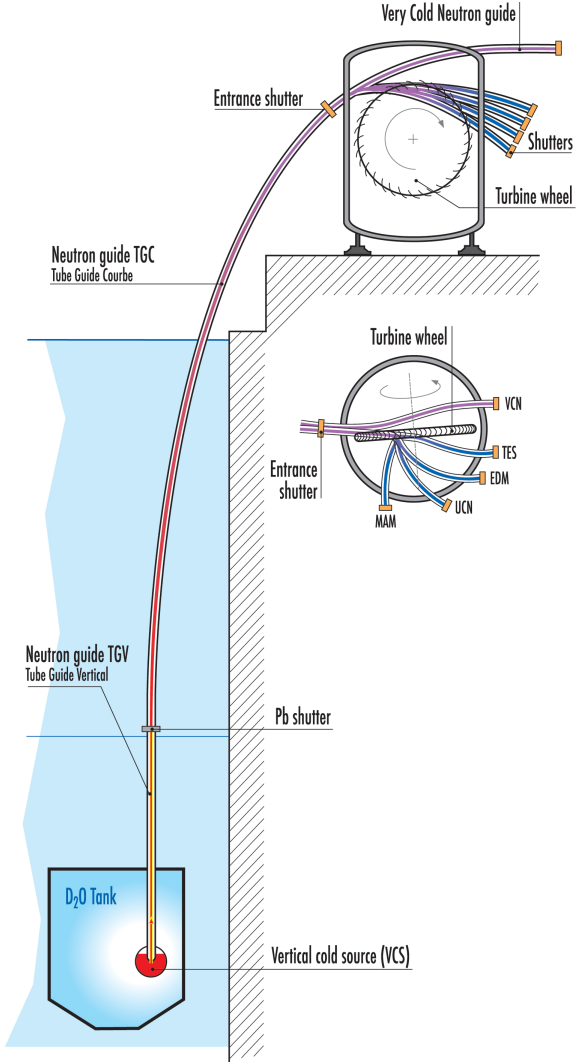}}
\vskip -.35 truein
\caption{\label{fg:ILLUCNTurbine}  (Color online) UCN source at PF2 at the Instiut Laue Langevin in Grenoble, France (ILL). Neutrons from the low-energy tail of the cold-neutron spectrum are guided upward and lose energy in the gravitational potential. The turbine further shifts the  spectrum to longer wavelengths to produce UCN that are provided to a number of experiments including the EDM. Figure provided by ILL~\cite{rf:ILLYellowBook}.
}
\end{figure}

The world's best thermal UCN source is   illustrated in FIG.~\ref{fg:ILLUCNTurbine} \cite{steyerl1986}. 
Neutrons with $v_n < 20$ m/s are extracted from a $T=$20K liquid D$_2$ moderator through a vertically mounted neutron guide, slowing as they rise 17 m in the gravitational potential. The neutron guide is a tube that transports neutrons using total reflection from the surface, with the criterion being the normal component of the velocity satisfies $v_\bot \leq \sqrt{2V_F/m_n}$. At the end of the guide, the neutrons reflect from the receding copper blades of a``turbine." The turbine blades act as moving mirrors that shift the neutrons to lower velocity. After several recoils, neutrons exit the turbine. Densities up to 10 UCN~cm$^{-3}$ are available for experiments.

\begin{figure}[tb]
\begin{tabular}{l}
\hskip-0.5 truein
\includegraphics[width=4 truein]{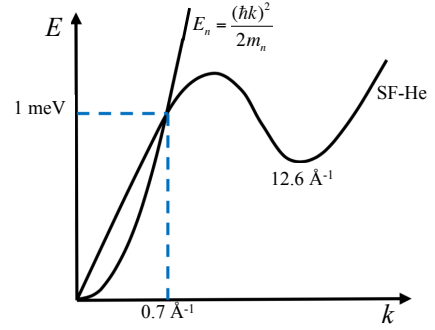}\\
\end{tabular}
\vskip -0.55 truein
\caption{\label{fig:4hedispersion}  (Color online) Single phonon dispersion curve for SF-He with a minimum at ~$k\approx$ 2 nm$^{-1}$ and the free neutron $E_n=(\hbar k)^2/2m_n$. The two curves intersect at $E_n=0$ and $E_n$=1 meV corresponding to $k$=0.7~\AA$^{-1}$ or $\lambda_0=8.9$ \AA.
}
\end{figure}

Most modern UCN sources  are based on
superthermal conversion introduced by~\textcite{ref:golub_pendlebury_superthermal}. This achieves higher phase-space density than thermal sources  using a medium that is not in thermal equilibrium with the neutrons.
For example, FIG.~\ref{fig:4hedispersion} shows the dispersion curves for a free neutron and for thermal excitations in superfluid-helium (SF-He), a typical choice for a superthermal source material. The curves cross at two points: $E_0=0$ and $E_1=E_0 + \Delta$, where $\Delta\approx 1$ meV corresponding to neutron wavelength $\lambda_0= 8.9$ \AA~\cite{book:golub_ucn}. This is effectively a two-state system, and neutrons at $E_1$ can resonantly transfer their energy to the SF-He resulting in final UCN energy $E_{UCN}\approx E_0$ with a small spread  due to the width of the excitations. The process $E_\mathrm{UCN} +\Delta  \rightarrow E_\mathrm{UCN}$, called ``down-scattering," is effectively independent of SF-He temperature $T$. The reverse process - a UCN absorbing energy $\Delta$ from the SF-He thermal bath (``up-scattering")  is exponentially suppressed for $\Delta \geq k_BT$ according to  
the principle of detailed balance, which gives the ratio of the up-scattering and down-scattering cross sections~\cite{book:golub_ucn}:
\be
\frac{\sigma(E_\mathrm{UCN} \rightarrow E_\mathrm{UCN} +\Delta)} {\sigma(E_\mathrm{UCN} +\Delta  \rightarrow E_\mathrm{UCN})}= 
\frac{E_\mathrm{UCN}+\Delta}{E_\mathrm{UCN}}e^{-\Delta/k_\mathrm{B}T}.
\label{eq:detaieledbalance}
\ee
%
The accumulation of UCN and increase of the phase space density of the neutrons does not violate Liouville's theorem, because the UCN and excitations in SF-He are both part of the same thermal system.
Producing SF-He requires temperatures $T<2.17$~K. However due to up-scattering $T\leq$~0.6~K is optimal for UCN production.  
%
%
%

The equilbrium UCN density for a given UCN lifetime $\tau_\mathrm{tot}$ inside an SF-He source is $\rho_\mathrm{UCN}= \tau_\mathrm{tot} R_I$, where the UCN production rate per unit volume for SF-He density $\rho_\mathrm{SF}$ and incident cold-neutron differential flux  $\frac{d\Phi_0}{d\lambda}$  is 
\begin{equation}
R_I=\rho_\mathrm{SF} 
\int\ \frac{d\Phi_0}{d\lambda} \sigma \left ( \lambda \rightarrow \lambda_\mathrm{UCN} \right ) \ d\lambda.
\end{equation}
For incident neutron wavelength near $\lambda_0$=8.9~\AA, and assuming a chamber with $V_F$= 252 neV ({\it i.e.} Be), the theoretical UCN production rate based on the combined calculations of \textcite{ref:golub_pendlebury_superthermal} and \textcite{YOSHIKI2003} is $R_I = \left(4.55 \pm 0.25 \right) \!\times\! 10^{-8} d\Phi_0 /d\lambda |_{\lambda_0}$ cm$^{-3}$ s$^{-1}$, for $d\Phi_0 /d\lambda$  in units of  neutrons cm$^{-2}$ s$^{-1}$ \AA$^{-1}$. 
\textcite{baker_phys_lett_a_308_67_2003} measured the production rate for a narrow-band neutron beam near 9~\AA, which, when combined with the measured incident flux, is interpreted as $R_I = \left ( 3.48 \pm 0.53 \right ) \!\times\!10^{-8}d\Phi_0 /d\lambda |_{\lambda_0}$ cm$^{-3}$ s$^{-1}$.

%
 %

Minimizing losses during UCN production is critical for achieving high UCN densities.
The  loss rate is ultimately limited by the neutron lifetime $\tau_n$ with additional contributions for a total loss rate:
\begin{equation}
\frac{1}{\tau_\mathrm{tot}} = \frac{1}{\tau_\mathrm{up}} + \frac{1}{\tau_\mathrm{walls}} + \frac{1}{\tau_\mathrm{slits}} + \frac{1}{\tau_\mathrm{abs}} + \frac{1}{\tau_{n}}
\label{eq:helium_loss_rate}
\end{equation}

\begin{enumerate}
\item The thermal up-scattering rate $1/\tau_\mathrm{up}$ is small at the typical operating temperature of about 0.8~K (see Eqn.~\ref{eq:detaieledbalance}), and  below 0.6~K performance does not further improve~\cite{piegsa2014}.\\
\item The wall collision losses $1/\tau_\mathrm{walls}  = \mu \nu$ are defined by the energy dependent parameters, loss probability per wall collision $\mu \approx 10^{-4}$  and the wall collision frequency in the trap $\nu \sim 10-50$ s$^{-1}$,  given in Table~\ref{tb:UCNMaterials}.\\
\item $1/\tau_\mathrm{slits}$ is related to the mechanical precision of the trap, which defines the leakage of UCN out of the trap, which has been reduced with low-temperature Fomblin oil (a fluorinated, hydrogen free fluid with low-neutron-absorption)~\cite{Serebrov2008}. \\
\item $\tau_\mathrm{abs}$ is the UCN absorption, due mostly to $^3$He in the trap. Less than part-per-billion $^{3}$He contamination in the superfluid is essential due to the strong absorption of neutrons on $^3$He.
\\
\end{enumerate}
In a demonstration by \textcite{piegsa2014},  total storage times of greater than 150 s and UCN densities of $\mathrm{220\ cm^{-3}}$ have been achieved by using low temperature Fomblin oil to close gaps and slits in the trap volume. The Fermi-potential $V_F\approx$ 100 neV for Fomblin oil results in a lower UCN energy spectrum. The loss of higher energy UCN is balanced by potential advantages, because losses, depolarization, and some velocity-dependent systematic effects may be smaller.

Many factors limit UCN-source performance including low Fermi potential of wall materials,  other contributions to storage lifetime, heat input due to the geometry of the incident neutron beam extraction guides, source volume dimensions, beam size, and beam divergence. 
UCN densities can be increased to 10$^3$~cm$^{-3}$ with an optimized 8.9~\AA~neutron beam, larger storage volumes, and surface-coating  improvements.

\begin{figure}[tb]
\vskip 0.25 truein
\includegraphics[width=3.5 truein]{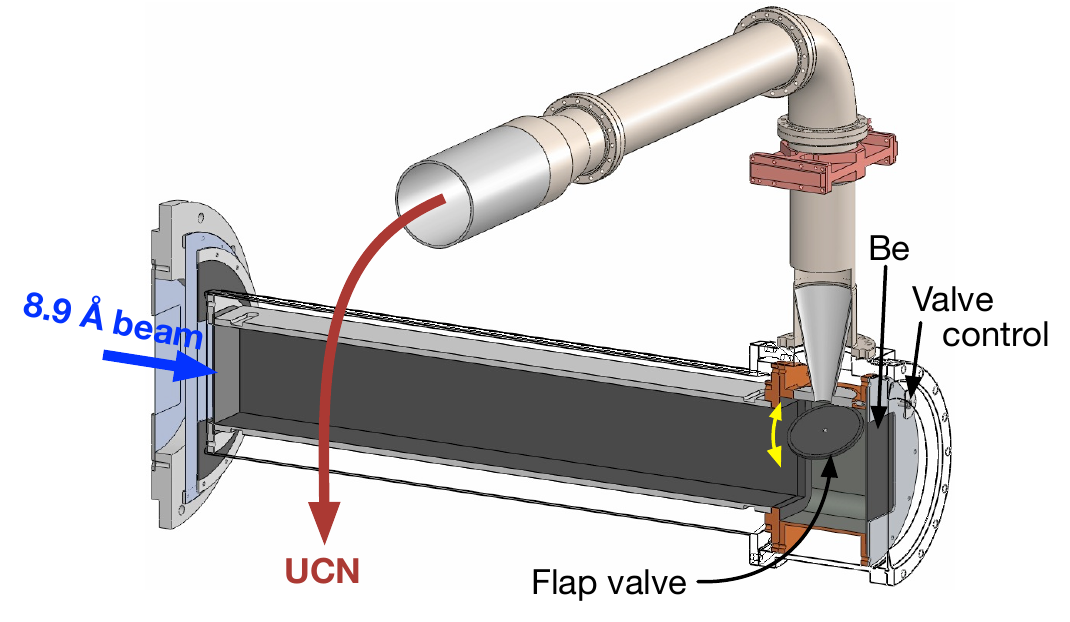}
\caption{\label{fg:SUN1ASchematic}  (Color online) Schematic diagram of the ILL  SUN2 source described in the text. Figure adapted from~\textcite{Sun2PRC} by  K. Leung and O. Zimmer.}
\end{figure}

Extraction of the UCN from the source to the experiment~\cite{piegsa2014,Masuda:2002dy}
requires either a window or a vertical exit from the superfluid He chamber such as indicated in FIG.~\ref{fg:SUN1ASchematic}.
For a superfluid He source, the UCN are accumulated in the source volume during production and filling of the experimental volume, which  reduces the density in proportion to the ratio of the source volume relative to the total volume of the source, guides, and experiment. The experiment should therefore be placed close to the source, and the ratio of source volume to experiment volume should be as large as possible.

A second-generation SF-He  source at ILL (SUN2) is shown in FIG.~\ref{fg:SUN1ASchematic}~\cite{Sun2PRC}. The SF-He is held in  a container with inner surfaces that have a  large  Fermi-potential {\it e.g.} for beryllium $V_F= 252$ neV, or with the help of magnetically enhanced confinement \cite{Zimmer:2013xja}. The cryogenics package (not shown) dissipates 60 mW at 0.6K. 
Neutrons with 8.9~\AA~wavelength  enter the SF-He container through one of the UCN reflecting walls.
Neutron beams available for such sources are typically cold beams that are guided to experiments 10's of meters away from the neutron source.
Neutron fluxes at ILL for the H172A beamline are  $\Phi_0=2.62 \times 10^7$~neutrons cm$^{-2}$\,s$^{-1}${\AA}$^{-1}$~\cite{piegsa2014}.
 %

UCN production is a coherent phenomenon in superfluid helium. Thus the polarization of an incident neutron beam is preserved, and it is possible to produce polarized UCN from a polarized cold-neutron beam or with a magnetic reflector surrounding the UCN production volume~\cite{Zimmer:2013xja}.
The alternative - polarizing UCN after extraction from the source  rejects at least half the neutrons.
%
 
The possibility of directly performing a UCN storage experiment inside the source has led to the SNS nEDM experimental concept described in Sec.~\ref{sec:NeutronEDM}. This would also enable the use of higher electrostatic voltages due to the  dielectric properties of the SF-He. Breakdown strengths exceeding 100~kV$/$cm have recently been demonstrated in a 1 cm electrode gap~\cite{ref:los_alamos_hvtests}.
%
%

\begin{table*}
\begin{centering}
\begin{tabular}{|c|c||c|c||c|c|}
\hline\hline
    Material    &V (neV)	&Loss per bounce  & Ref		& Depolarization & Ref  \\
\hline
DPe (300K)        		  &214 		&$1.3\times 10^{-4}$ & $a$  &$4\times 10^{-6}$ & $b$ \\
\hline
DLC on Al substrate (70K)  &270 		&$1.7\times 10^{-4}$ &	$c$	&$0.7\times 10^{-6}$ & $c$\\
\hline
DLC on Al substrate (300K) &270 		&$3.5\times 10^{-4}$ &	$c$	&$3\times 10^{-6}$& $c$\\
\hline
DLC on PET substrate (70K)&242 		&$1.6\times 10^{-4}$ &	$c$	&$15\pm \times 10^{-6}$& $c$\\
\hline
DLC on PET substrate (300K)&242 		&$5.8\times 10^{-4}$ &	$c$	&$(14\pm 1)\times 10^{-6}$& $c$\\
\hline
Fomblin 300K   			      &106.5    	&$2.2\times 10^{-5}$ & $d$ &$1\times 10^{-5}$& $e$\\
\hline
Be (10 K) 				      &252 		&$3\times 10^{-5}$    & $d$ 	    &$1.1\times 10^{-5}$& $e$\\
\hline
Be (300K) 			      &252 		&$(4-10)\times 10^{-5}$& $d$  &$1.1\times 10^{-5}$& $e$\\
\hline
NiP 			      &213 		& $1.3\times 10^{-4}$ & $f$  & $< 7\times 10^{-6}$ & $g$\\
\hline
$^{58}$Ni 			      &335		&& $h$ & strong  &\\
\hline
Fe/steel/stainless			      & 180-190	&	&  $h$ & strong  &\\
\hline\hline
\end{tabular}
\caption{Properties of materials for UCN production, storage and transport showing loss per bounce and depolarization per bounce. DPe is deuterated polyethylene; PET is Polyethylene Terephthalate; DLC is diamond-like carbon. $^{58}$Ni and steel alloys are magnetic, resulting in strong depolarization of UCN. References $a$: \cite{BrennerAPL2015}; $b$:\cite{Ito2015}; $c$ \cite{Atchison2007}; $d$: \cite{Serebrov2005}; $e$: \cite{Serebrov2003}; $f$:\cite{Pat17}; $g$: \cite{Tan16}; $h$ \cite{book:golub_ucn}.\label{tb:UCNMaterials}}
\end{centering}
\end{table*}

\begin{table*}
\begin{centering}
\begin{tabular}{|l|l|l|rl|l|}
\hline\hline
Source & Type  & Converter & &\hskip -0.17 truein UCN/cm$^{3}$  & Ref\\
\hline
\hline
ILL PF2 & reactor cold source   &   (turbine) & 2 & polarized UCN based on detected UCN  & $a$\\
\hline
LANL & spallation & sD$_2$  & 40  &  polarized UCN observed in a test chamber & $b$\\
\hline
PSI & spallation & sD$_2$ & 22 & unpolarized UCN in a ``standard'' storage bottle & $c$\\
\hline
TRIGA Mainz & pulsed reactor & sD$_2$  & 10 & unpolarized & $d$\\
\hline
ILL SUN-II & reactor cold neutron beam & SF-He & 10 & polarized UCN source production, dilution, and polarization & $e$\\
\hline
J-PARC & spallation-VCN & rotating mirror & 1.4 & unpolarized, at source & $f$\\
\hline\hline
\end{tabular}
\caption{Currently operating UCN sources with UCN densities relevant to EDM experiments based on data reported in the cited references. For unpolarized UCN, the reported densities should be decreased by  a factor of at least two  
 to make a consistent comparison. 
References: $a$~\cite{Baker:2006ts}; $b$~\cite{Ito:2017ywc}; $c$~\cite{Becker201520} ; $d$~\cite{LauerTh2013}; $e$~\cite{Sun2PRC}; $f$~\cite{doi:10.1093/ptep/ptv177}.
\label{tb:UCNSourceComparisonTable}}
\end{centering}
\end{table*}

 Alternative superthermal converters include solid deuterium~\cite{Golub1983}, deuterated methane or oxygen, as well as other (para)magnetic/spin polarized substances.~\cite{superthermal_materials}.  The choice of materials for superthermal  UCN production is based on increasing the phase space density and modes of UCN extraction and are discussed in~\textcite{ref:ZimmerCooling,nesvishevski,nesvishevski-ru}.
 
Solid deuterium (SD$_2$) molecules consisting of two spin-one deuterons combine to total nuclear spin $I=0$ or 2 for ortho-deuterium ($o$D$_2$) and $I=1$ for para-deuterium ($p$D$_2$). The  rotational ground-state for  spin-one bosons is $o$D$_2$.(For H$_2$, the ground state is parahydrogen with $I=0$.)
Rotational excitations in SD$_2$ for rotational quantum number $J$ are given by:
\begin{equation}
E_J = hcB J(J+1)=3.75\ {\rm meV}\times J(J+1),
\label{eq:sd2_rot_levels}
\end{equation}
where $B$ is the rotational constant~\cite{sd2-rotational-states}. For low-lying excitations, the rotational energy is 
comparable to cold neutron energies. 
At low temperature, the six $I=0,2$ $o$D$_2$  states and the three $I=1$ $p$D$_2$ states are approximately equally populated, so the relative population of ortho and para states is 2-to-1. For UCN production, an $o$D$_2$ concentration $\gtrsim$~95$\%$ is necessary to suppress upscattering of UCN, absorbing energy from $o$D$_2$ to $p$D$_2$ transitions. This requires low temperatures as well as a magnetic converter to enhance  the spin-flip rate for deuteron spins in a separate apparatus outside the source. One possible converter material is hydrous iron (III) oxide, cooled to below the triple point of deuterium~\cite{ref:sd2_ortho_para_converter}. 
The converted $o$D$_2$ is then transferred to the UCN source in a gaseous state without losing the ortho-state configuration due to very long thermal relaxation times in the absence of a converter.

The conversion cross section  per unit volume from cold neutrons to UCN for a deuterium density $\rho_{\rm H_2}$ is given by~\cite{AtchisonPRL2005,Atchison2005,ATCHISON2009252}
\be
\sigma^{CN \rightarrow UCN}_{{\rm sD}_2,8K} \rho_{\rm H_2} \sim \left( 1.11\pm 0.23\right) \times 10^{-8}~{\rm cm}^{-1},
\ee
The momentum acceptance for cold neutrons is much greater for sD$_2$ than for SF-He, leading to a higher production rate in the source.  The UCN lifetime in the sD$_2$ source is given by
\begin{equation}
\frac{1}{\tau_\mathrm{UCN,sD_2}}= \frac{1}{\tau_\mathrm{up}} + \frac{1}{\tau_{o/p}} + \frac{1}{\tau_{\rm D-abs}} + \frac{1}{\tau_{\rm H-abs}} + \frac{1}{\tau_{\rm cryst}},
\label{eq:sd2ucndensity}
\end{equation}
where $1 / \tau_{up}$  accounts for the losses due to thermal up-scattering (elastic scattering of UCN with mK temperature on nuclei at few K temperatures in the source)~\cite{Liu2000R3581},  $1 / \tau_{o/p}$ accounts for  losses due to scattering on $p$D$_2$, $1/\tau_{\rm D-abs}$, $1/\tau_{\rm H-abs}$, and $1/\tau_{\rm cryst}$ accounts for absorption losses on deuterium, hydrogen and scattering in the crystal, respectively~\cite{Yu1986137}.
Most loss channels can be controlled to better than  $(150\ \text{ms})^{-1}$, resulting in an overall UCN lifetime of $\tau_\mathrm{UCN,sD_2} \approx 75\ \mathrm{ms}$ in the perfect
 $o$D$_2$ crystal  at 4~K.

UCN densities of 10$^4$~cm$^{-3}$ are in principle feasible inside the source. In contrast to the SF-He source, which requires accumulation in the source for several hundred seconds followed by dilution into the guides and experimental volume,  sD$_2$ provides an effectively continuous source of UCN, provided  efficient  extraction to the experiment  within a time $\tau_{\rm UCN, sD_2}$. The optimum source thickness based on diffusion of UCN in this time is $\sim$~1~cm, not taking into account pre-moderation or thermal engineering issues that may dictate increased  thickness.  Also, UCN extracted from the source should be prevented from  diffusing back into the sD$_2$, for example by closing off the source with a valve.
%

%
%

%
An sD$_2$ source can provide a steady flux of UCN from the source and can, in principle, be interfaced directly to the neutron-guide vacuum. However, any UCN that diffuse back to the D$_2$ will be lost due to the short UCN storage time.
On the other hand, the UCN density provided by a continuous sD$_2$ source does not depend on the size of the experiment (to first order), and large experiment volumes can be filled with UCN.
%
%
For EDM experiments, the duty cycle of the source can be less than 10\%, as the filling times of UCN storage chambers are typically 10-100~s and the cycle time $\approx$~1000~s,
%
%
which is important for accelerator-based spallation neutron sources that share the primary accelerator beam.%

At the Los Alamos National Lab, spallation  neutrons produced by a pulsed 800-MeV proton beam striking a tungsten target are thermalized by beryllium and graphite moderators at ambient temperature and further cooled by a cold moderator that consists of cooled polyethylene beads~\cite{Sau13}. The cold neutrons are converted to UCN by down-scattering in an sD$_2$ crystal at 5 K.  UCN are vertically extracted  in a 1 meter long $^{58}$Ni guide that compensates for the 100-neV boost that UCN receive when leaving the sD$_2$.  A valve between the source and vertical guide is closed except when the proton beam pulse is incident in order to keep the UCN in the guide from returning to the sD$_2$ and being absorbed.
When in production, the peak proton current from the accelerator was typically 12 mA, delivered in bursts of 10 pulses each 625~$
\mu$s long at 20 Hz, with a 5 s delay between bursts; the average current delivered to the spallation target is 9 $\mu$A. 
After recent upgrades, the UCN density measured at the end of a 6 meter long stainless steel guide was $\left(184\pm32\right)$  cm$^{-3}$, and a measurement made with {\em polarized} UCN in a bottle, similar to what would be used for an EDM measurement, yielded  $\left(39\pm 7\right)$ UCN cm$^{-3}$~\cite{Ito:2017ywc}.
SF-He sources are also planned at PNPI in St. Petersburg, which calculations suggest may produce 10$^4$ UCN/cm$^3$~~\cite{Serebrov:2017gea}.

At PSI~\cite{Anghel2009272}, the cyclotron provides  a 2.2 mA,  590 MeV proton beam incident on a heavy metal spallation target for 4~s with a duty cycle of 1$\%$. Measurements with a ``standard'' storage bottle showed 22 UCN/cm$^3$ of unpolarized UCN~\cite{Becker201520}.
%
%
 Another pulsed sD$_2$ source is installed at the pulsed TRIGA reactor in Mainz, where a 200~MW-20~ms neutron pulse produces UCN, which leave the source quickly and can - with proper timing - be captured in an experiment using mechanical shutters. A density of order 10~UCN-cm$^{-3}$ in a few-10's of liter volume was achieved~\cite{LauerTh2013}.


Though the typical lifetimes in sD$_2$ and the energy dependent production rate have been well determined experimentally, the  performance of solid deuterium based sources is typically lower than expected for operating sources (see Table~\ref{tb:UCNSourceComparisonTable}). Several issues related to the $\mathrm{sD_2}$ crystals have been considered in an effort to explain this discrepancy. For example, in practical sources, shrinkage of the crystal during the cool down may cause cracks and may also reduce thermal contact to the cooling apparatus.
Cracks in the crystal also change the neutron mean-free path in the source and thus may affect the extraction efficiency.
Thermal stress and geometrical alignment of the source may also require preparation of the crystal from the gas phase instead of freezing from the liquid state.
 Another important consideration is guiding the neutrons from the sD$_2$ source to an experiment.
Until recently, UCN transmission in long guides had been a limitation, but ~\textcite{ref:zechlau}   measured transmission greater than $>$~50$\%$ for   80 mm ID,  22 m long guides coated with NiMo (in the ratio 85:15) and including three 90$^\circ$ bends.

In Table~\ref{tb:UCNSourceComparisonTable}, we list the currently operating UCN sources and an estimate of the performance relevant to the neutron EDM experiments based on measurements reported in the papers cited. 
In some cases, the number of polarized neutrons was reported, as noted; for unpolarized UCN, the reported densities should be decreased by at least a factor of two to account for the loss of one spin state.

A number of additional sources based on both SF-He and sD$_2$ are under construction or planned. The  operating SF-He sources are in principle working at the level of theoretical estimates. For these sources, scaling to larger neutron flux would result in larger heating from the primary beam and require increased cryogenic capacity, but  is not expected to significantly change the performance. At ILL, the SuperSUN source has significantly larger converter volume, cryogenic capacity, and will incorporate the magnetic reflector~\cite{Zimmer:2013xja} with the expectation of up to 1000 polarized UCN/cm$^3$. 
 At KEK, a SF-He based source using proton-spallation production of cold neutrons provide 0.7 unpolarized UCN/cm$^3$ with 78 Watts of proton power~\cite{Masuda:2002dy}. This source has been moved to TRIUMF, where orders of magnitude higher proton-beam power will be available. Upgrading the cryogenics to handle higher power is predicted to provide 100's of UCN/cm$^3$ for experiments.  
 The concept of the nEDM experiment under development for the Spallation Neutron Source at Oak Ridge National Lab is to produce the UCN within the EDM experiment as demonstrated by~\textcite{OShaughnessy:2009amn} for a neutron-lifetime measurements as discussed in detail in Sec.~\ref{sc:nEDMprospects}.
 
Plans to adapt the converter design of the sD$_2$ source at the Mainz TRIGA reactor to the strong core-flux from the FRM-II reactor at the Technical University of M\"unchen FRM-II, are expected to produce~1000 UCN/cm$^3$.  The TRIGA reactor operates at a lower average power; therefore scaling to higher power and radiation dose needs to be explored. An sD$_2$ source is under development at the  PULSTAR reactor at North Carolina State University that is predicted to provide 40 UCN/cm$^3$ for 1 MW~\cite{KOROBKINA2014169}.   

%
 The next major neutron source will be the European Spallation Source (ESS). The ESS will provide an opportunity for new UCN source possibilities with even greater densities for future experiments, and a number of UCN source concepts have been developed~\cite{PENDLEBURY201478,Lychagin:2015spa,Klinkby:2014ria,Nesvizhevsky:2014dya,ZIMMER201485}. 
   The goal of these proposals is 10$^{3}$-10$^{4}$ UCN/cm$^{3}$.
  

\section{Experiments}

In this section we review the current status and prospects of EDM experiments. Our goal is to describe in some detail  technical aspects of the experiments, systematic errors, near-term improvements, and new concepts under development. Guided in part by the historical procession of experiments discussed in the introduction, we begin with the neutron and move onto Cs and other paramagnetic atom and molecule experiments. Diamagnetic atoms and molecules including octupole-enhanced nuclei follow. Finally we describe the development of storage ring experiments to measure EDMs of light nuclei.
The results so far are summarized in Table~\ref{tb:EDMResults}.

%
%

%
%
%
%
%
%



\subsection{The Neutron }
\label{sec:NeutronEDM}

Here we discuss the neutron EDM experiment from the original beam measurements to the UCN experiments. The dominant systematic errors are discussed as well as brief descriptions of the several efforts underway to improve the sensitivity by one to two orders of magnitude.

\begin{figure}[tb]
\includegraphics[width=3.6 truein]{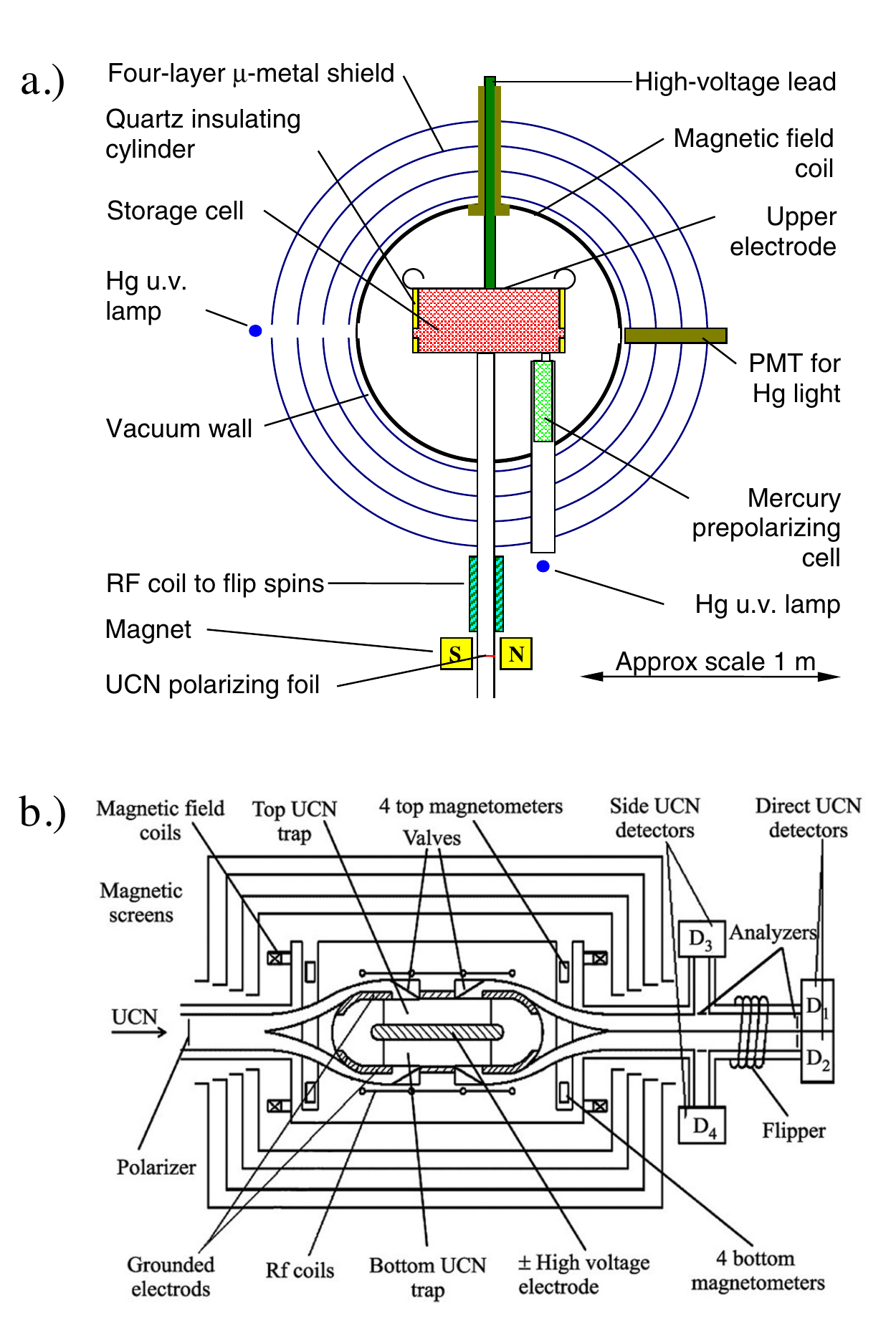}
\caption{\label{fig:nedm_history}  (Color online) Neutron EDM apparatus from (a) the ILL-Sussex-Rutherford experiment with one chamber for UCN and a comagnetometer~\cite{Baker:2006ts}; (b) the Gatchina apparatus, as set up at ILL, with two neutron storage chambers so that parallel and anti-parallel $E$ and $B$ field orientations are measured simultaneously~\cite{Serebrov:2015idv}. Both experiments are run with the storage chamber in vacuum at room temperature. Figures used with permission.
}
\end{figure}

The first experiments to search for the neutron EDM used a cold neutron beam~\cite{rf:Smith1957}, but 
by 1980, the beam measurements became systematics limited due to complications arising from beam divergence and motional ($\vec v\times \vec E$) effects~\cite{rf:Dress1977}. These systematic issues led to experiments using stored UCN, trading higher neutron density  for  longer observation times and reduced velocity~\cite{Altarev1980}. 
%
%
%
Two recent stored-UCN experiment 
configurations are shown in FIG.~\ref{fig:nedm_history}.  
For the ILL-Sussex-Rutherford experiment (FIG.~\ref{fig:nedm_history}a)~\cite{Baker:2006ts} a single chamber is used along with a comagnetometer, while the Gatchina experiment (FIG.~\ref{fig:nedm_history}b)~\cite{Altarev1992,Altarev:1996xs} employs a pair of chambers with opposite electric fields and a common magnetic field. In the latter scheme, common-mode time dependent magnetic field variations are rejected to the degree to which the common magnetic field is uniform.
%
%
%
The ILL experiment is analyzed as a spin-clock comparison using a variation of Ramsey's separated oscillatory field technique to measure  two spin-polarized species (neutrons and a $^{199}$Hg comagnetometer) in the same volume at the same time. 
%
%
%
%
%
%
In the Gatchina approach, both chambers have very similar systematics and the velocities of the UCN are small enough that many systematic effects are negligible at the 10$^{-26}$\ecm~ level.
%

\begin{figure}[tb]
\includegraphics[width=3.5 truein]{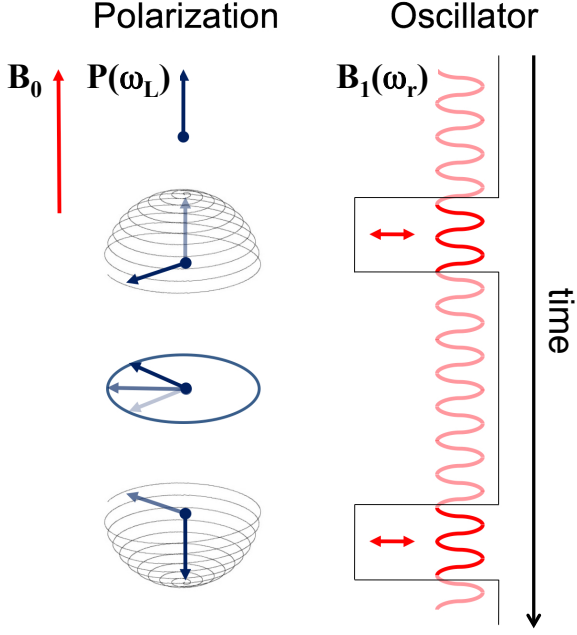}
\vskip -1 truein
\caption{\label{fig:ramseytechnique}  (Color online) Ramsey's technique of separated oscillatory fields. The experiment starts out with polarized particles in a stable and uniform magnetic field B$_0$, with a stable external oscillator at frequency $\omega_R$ near the Larmor frequency $\omega_L$ of the particles in the B$_0$ field. First a $\pi/2$ pulse of oscillating magnetic field ($B_1$) rotates the polarization into the plane normal to B$_0$, creating a superposition of spin-up and spin-down states. The spins and external clock evolve independently until a second $\pi / 2$ pulse is applied. The second pulse measures the phase difference between the oscillator and precessing spins that accumulates during the free precession interval. Time evolves from top to bottom in the figure.
}
\end{figure}

%
In Ramsey's technique~\cite{RevModPhys.62.541}, an interferometer in time is realized by comparing the phase from a spin-clock with frequency $\omega_L$, the Larmor precession frequency, with the phase of a reference clock  with frequency $\omega_R$ after a fixed measurement time $\tau$ as illustrated in FIG.~\ref{fig:ramseytechnique}.
The optimal observation time $\tau$ is based on the UCN storage time and  the polarization lifetimes ($T_2^*$) of the UCN and the comagnetometer.
%
%
The  phases of the spins and the clock evolve at the different frequencies, and the phase difference after a time $\tau$ is $\Delta=(\omega_L-\omega_R)\tau$. This is read out using the polarization $P_z$, {\it i.e.} the projection of the spin along the $B_0$ field  after the second pulse is applied. In terms of the number of neutrons detected with spin  parallel ($N_\uparrow$) and antiparallel ($N_\downarrow$) to $B_0$ at the end of the free-precession cycle, the polarization is: 
\begin{equation}
P_{z} = \frac{N_\uparrow - N_\downarrow}{N_\uparrow + N_\downarrow}.
\label{eq:polarization}
\end{equation}
For $\Delta=0, \pi/2, \pi$, $P_z/P_0=-1, 0, 1$, respectively, where $P_0$ is the maximum magnitude of the polarization.
To maximize sensitivity to a change of frequency, $ \Delta=\pm \pi/2$ is chosen to provide the maximum slope of the fringes:
\begin{equation}
\left. \frac{\partial P}{\partial \omega} \right|_{\Delta = \pi/2} = \tau P_0. 
\label{eq:ramseyphase}
\end{equation}
The EDM frequency shift leads to a polarization shift

\be
\delta P_z = \frac{2d_nE}{\hbar} P_0 \tau.
\ee
The statistical precision of the Ramsey measurement is:
\begin{equation}
\sigma_{d_n}= \frac{\hbar\sigma_{P_z}}{2EP_0\tau},
\label{eq:sigmad1}
\end{equation}
%
%
where the uncertainty of each  $P_z$ measurement, for a small frequency shift, is 
\be
\sigma_{P_z}\approx 1/\sqrt{N_\uparrow+N_\downarrow}.
\ee
%

%

%
%

For the ILL-Sussex-Rutherford experiment, 
$^{199}$Hg was chosen as the comagnetometer for several reasons: the upper limit on the $^{199}$Hg  EDM is much smaller than the neutron-EDM sensitivity (see Sec.~\ref{sec:XeHg}); the coherence time of $^{199}$Hg can be 100 s or more; the coherence time combined with the large signal-to-noise ratio of the optically pumped atoms provides high sensitivity to magnetic field variations~\cite{GREEN1998381} . %
%
The $^{199}$Hg vapor was polarized with 254~nm light from a discharge lamp in a pre-polarizing chamber adjacent to the neutron EDM apparatus. 
After the UCN were loaded, the polarized Hg vapor was added to the neutron storage volume at a pressure of  about $10^{-4}$~mbar, the vapor pressure of Hg at room temperature. At this low pressure, the macroscopic magnetic field due to the $^{199}$Hg did not appreciably shift the spin precession of the neutrons. Spin-precession was initiated with a  resonant $\pi$/2 pulse that rotated the Hg spins into the plane transverse to $\vec B_0$. The $^{199}$Hg precession frequency was determined by the time-dependence of the absorption of a weak probe beam of circularly polarized 254 nm light directed through the storage chamber continuously during the UCN measurement. 
The polarization and readout light intensities were balanced to optimize the performance of the magnetometer. With a mean velocity of $v_\mathrm{rms}\approx 193$ m/s, the $^{199}$Hg atoms sampled the entire storage volume in a time short compared to a  Larmor cycle of 128  ms.

A reanalysis of the~\textcite{Baker:2006ts} experiment  by~\textcite{Afach:2015sja} led to the final  result
 \be
d_n=(-0.21\pm 1.82)\!\times\! 10^{-26}\ \text {\ecm}.
\ee
For the Gatchina experiment, the combination of results from PNPI~\cite{Altarev:1996xs} and ILL~\cite{Serebrov:2015idv} are presented as a sensitivity of $3\times 10^{-26}$ \ecm and an upper limit at 90\% confidence level of $5.5\times 10^{-26}$ \ecm~\cite{Serebrov:2015idv}.
Major systematic effects are discussed in the next section.



\subsection*{Neutron-EDM systematics}
\label{sec:neutron_systematics}
\label{sec:systematics}

Improved sensitivity of EDM experiments led to the discovery of new systematic effects that have been incorporated by~\textcite{Afach:2015sja} into the analysis of the~\textcite{Baker:2006ts} result. 
%
%
%
%
These include both direct false effects, such as stray magnetic fields correlated with the electric field polarity due to leakage currents generated by the high voltage apparatus near the EDM measurement cells, and indirect false effects, such as the geometric phase effect discussed below. 
%
To put this in perspective, with an applied electric field of 10 kV/cm, an EDM of 10$^{-26}$ \ecm~corresponds to a frequency shift of $\Delta\omega\approx 2\pi\!\times\! 50$ nHz, equivalent to a magnetic field change of 2 fT. 

A leakage current due to the high voltage applied across the storage cell would be correlated with the electric field and would produce a false EDM signal. For example,  the storage chamber used by~\textcite{Baker:2006ts} has radius $\approx 0.2$~m, so a  leakage current of 1  nA making a full turn around the cell would produce a 3 fT field. 
In principle, this would be compensated or monitored by the comagnetometer or external magnetometers. However, as noted below, the UCN and $^{199}$Hg positions are separated by about 3 mm and the non-uniform magnetic field due to the leakage current  would not be perfectly compensated.  For a storage cell of height of 20 cm the cancellation to first order was estimated to be 0.3/20 resulting in a false EDM of  $d_\mathrm{false}^\mathrm{leak}\approx~2\times 10^{-28}$ \ecm~ for $E$= 10 kV/cm.


%

\begin{figure}[tb]
\includegraphics[width=2.9 truein]{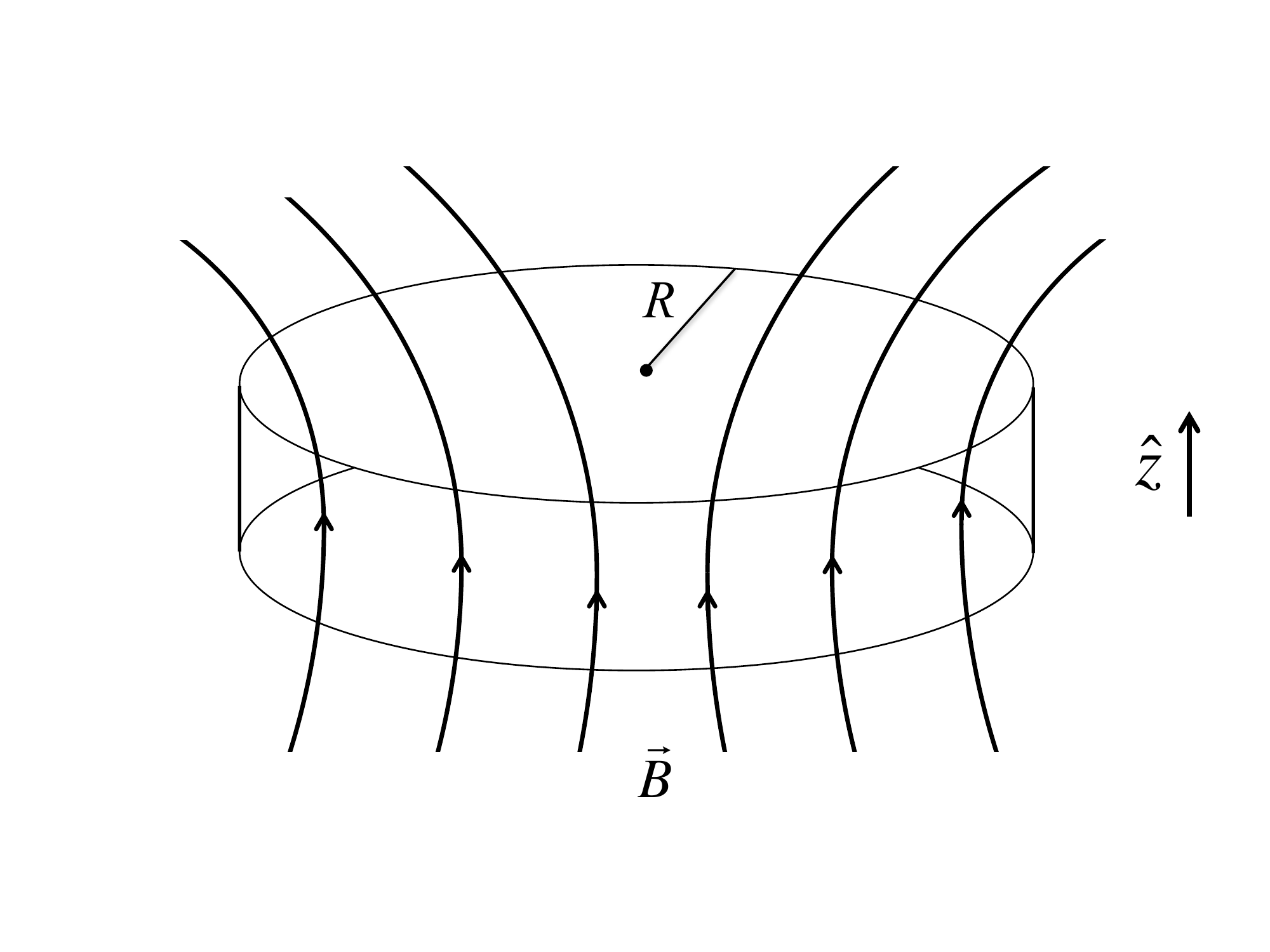}
\\
\vskip -0.25 truein
\includegraphics[width=3.5 truein]{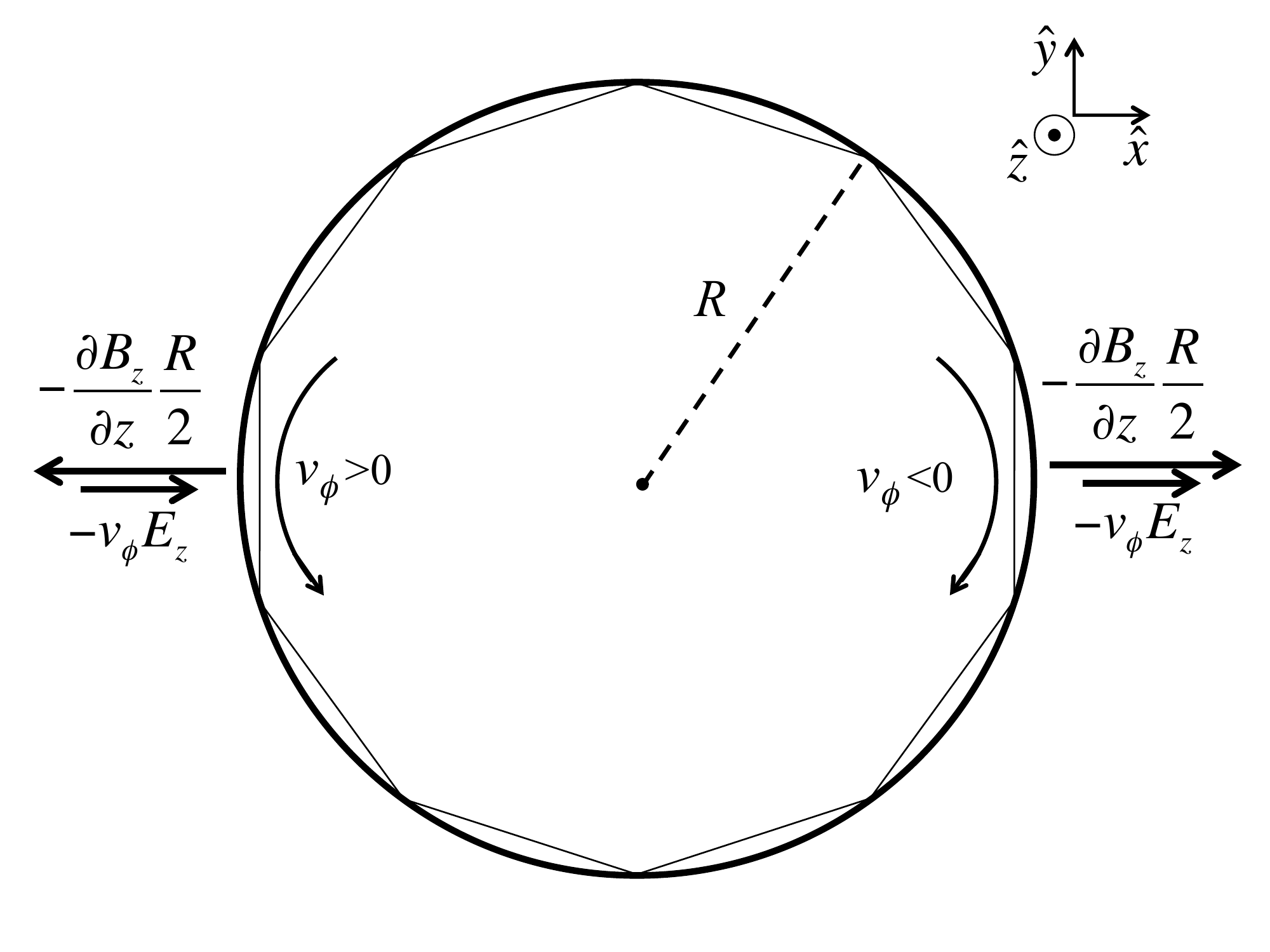}
\vskip -0.2 truein
\caption{\label{fg:nGeomPhase} Contributions to the geometric phase effect. Top: a non-uniform magnetic field that leads to radial magnetic field components.
Bottom: illustration of particle trajectories and contributions to $B_\rho$ from the gradient and motional magnetic field. The left side shows a positive azimuthal velocity, and the right side shows a negative azimuthal velocity. The average frequency shift for the two counter-propagating trajectories is given by Eqn.~\ref{eq:falseD_GP}. Adapted from~\textcite{rf:PendleburyGP}.}
\end{figure}

The geometric phase or $\vec E\times\vec v$ effect~\cite{rf:ComminsAJPBerryPhase,rf:PendleburyGP,rf:GolubLamoreauxGP} is the extra phase accumulated by a quantum system due to the rotation of the quantization axis, in this case due to the combination of motional  fields and gradients of the magnetic field. To illustrate the origin of this effect, following~\textcite{rf:PendleburyGP},  consider the situation shown in  FIG.~\ref{fg:nGeomPhase} for a particle moving in a nearly circular trajectory near the wall of the radius $R$ chamber with an axial magnetic field $\vec B_0$ nominally directed out of the page ($\vec B_0\approx B_z\hat z$) and a magnetic field gradient, which produces radial components. With $\vec E\times\vec v=-E_z v_\phi \hat \rho$, and assuming cylindrical symmetry
\begin{equation}
 B_{\rho} = - \frac{\partial B_{z}}{\partial z} \frac{R}{2}-\frac{v_\phi E_z}{c^2},
\label{eq:perpfield}
\end{equation}
where the azimuthal velocity component $v_\phi$ is positive for counter-clockwise and negative for clockwise rotation. Note that $B_\rho$ changes magnitude when $v_\phi$ changes sign.
In the frame of the particle, the transverse magnetic field of magnitude $B_\rho$ {\em rotating} at a frequency $\omega_\mathrm{Rot}=v_\phi/R$ causes a shift of the Larmor frequency analogous to the Bloch-Siegert shift of NMR~\cite{BS40} as generalized by Ramsey~\cite{rf:RBSShift}: 
\begin{equation}
\Delta \omega = \omega_L - \omega_0 \\
= \sqrt{(\omega_0-\omega_\mathrm{Rot})^2 + \omega_\rho^2} - (\omega_0-\omega_\mathrm{Rot}).
\label{eq:geomphase0}
\end{equation}
Substituting $\omega_0=\gamma B_z$ and $\omega_\rho = \gamma B_\rho$, where $\gamma$ is the particle's gyromagnetic ratio, and taking  $v_\phi$ positive, the first order shift
($\omega_\rho^2\ll(\omega_0-\omega_\mathrm{Rot})^2$ and $\omega_0>\omega_\mathrm{Rot}$) is%
%
%
\begin{equation}
\Delta\omega_{v_\phi}=\frac{(\gamma B_\rho)^2}{2(\gamma B_z-v_\phi/R)}.
\label{eq:RBSShift}
\end{equation}
Because $B_\rho$ changes magnitude when $v_\phi$ changes sign, this shift does not average to zero for the two signs of $v_\phi$. Combining Eqns.~\ref{eq:perpfield} and~\ref{eq:RBSShift} the average shift for a positive and negative $v_\phi$ and a trajectory of radius $R$ is
\begin{equation}
\Delta\omega_{avg}=
\frac{1}{2}\frac{\gamma B_z[(\gamma\frac{\partial B_{z}}{\partial z}\frac{R}{2})^2+(\frac{\gamma v_\phi E_z}{c^2})^2]
+\gamma^2\frac{\partial B_{z}}{\partial z} v_\phi^2 \frac{E_z}{c^2}}
{(\gamma B_z)^2-(v_\phi/R)^2}.
\end{equation}
When $E_z$ is reversed, the last term in the numerator changes sign,  producing a false-EDM signal
\begin{equation}
d_\mathrm{false}^\mathrm{GP}\approx    \frac{\hbar\gamma^2\frac{\partial B_{z}}{\partial z} v_\phi^2R^2/c^2}{4(v_L^2-v_\phi^2)},
\label{eq:falseD_GP}
\end{equation}
where the ``Larmor'' velocity is $v_L=\gamma R B_z$, the size of the trajectory is $R$, and the effective velocity $v_\phi$. Note that the denominator in Eqn.~\ref{eq:falseD_GP}  goes to zero as $v_\phi\rightarrow v_L$, which has led to characterization of an {\em adiabatic} regime  with $|v_L|>>|v_\phi|$  and a non-adiabatic regime, e.g. when $|v_L|\approx |v_\phi|$. In the adiabatic regime, the spins track the magnetic field in their frame.  

As a numerical example, we take $v_\phi= 7$ m/s and $v_L\approx 200$ m/s for UCN and $v_\phi= 200$ m/s and $v_L\approx 50$ m/s for room-temperature $^{199}$Hg.  For a gradient  $\frac{\partial B_{z}}{\partial z}=0.3$ nT/m, and $R=$ 0.2~m, this results in $d_\mathrm{false}^\mathrm{GP} \approx 1\times 10^{-28}$\ecm~for UCN and $d_\mathrm{false}^\mathrm{GP} \approx  1\times 10^{-26}$\ecm~for $^{199}$Hg.
Consequently the geometric phase effect is actually much more significant for the $^{199}$Hg comagnetometer~\cite{Baker:2006ts}. More refined modeling~\cite{pignol,steyerl3} as well as numerical studies~\cite{rf:Bales2016} result in similar estimates of this effect for more realistic trajectories and field maps.

Approaches 
to reduce the geometric phase effect for future experiments include reducing the magnetic field gradients, making the chamber smaller, and manipulating the effective velocity and radius of the trajectories, for example with a buffer gas that would change the mean-free path. The residual field gradient in the  TUM\"unchen magnetic shield is $<100$ pT/m, sufficiently small to suppress the geometric phase  effect to about $1\times 10^{-28}$\ecm. However the ultimate limitation will likely be due to distortions from magnetized components of the EDM apparatus, for example  the valves for the UCN and $^{199}$Hg,  magnetic contamination of the surface, or magnetization of electrodes resulting from HV discharges.
A magnetic dipole inside the chamber producing a  10~pT field at 2~cm is estimated to produce a false EDM  of $10^{-28}$\ecm. A thorough treatment of magnetic dipole sources and their effect on the geometric phase can be found in discussions by \textcite{pignol,harris,rf:GolubLamoreauxGP,steyerl2,steyerl3}. 

\begin{figure}[tb]
\centerline{\includegraphics[width=4.25 truein]{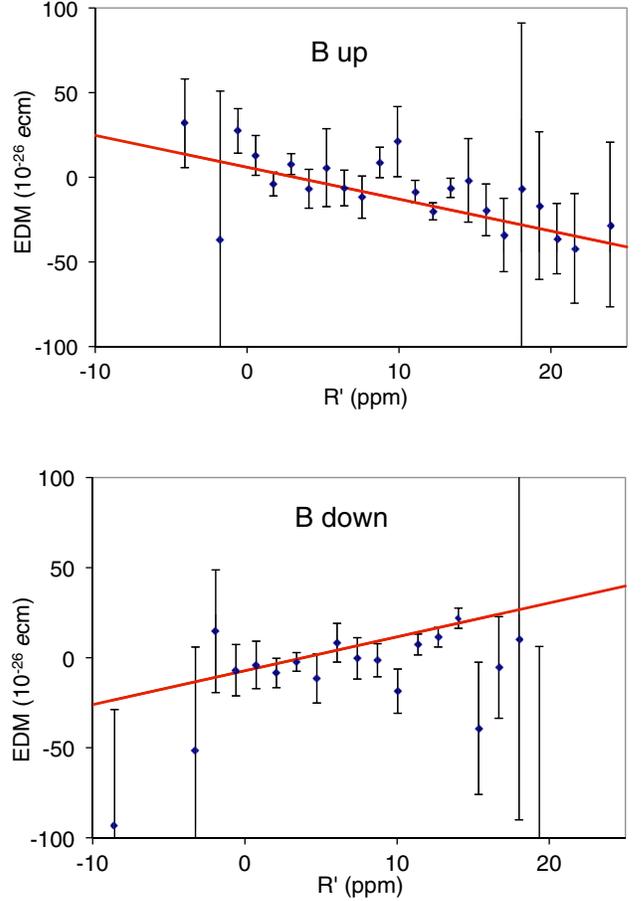}}
\vskip -0.4 truein
\caption{\label{fig:baker_edm} (Color online)  Dependence of the measured $E_z$-odd  EDM signal  on the magnetic field gradient, represented by $R\,^\prime$, for two $B_z$ orientations. The $y$-axis labeled ``EDM" refers to  $d^{meas}$ in Eqn.~\ref{eq:dnmeas}. The data points with error bars are from a typical  data run. The solid red lines are linear fits to Eqn.~\ref{eq:dmeas} for  the data set  with $B$-up and $B$-down. Figure from ~\textcite{Baker:2006ts}. 
}
\end{figure}

\textcite{Baker:2006ts}  developed a scheme to mitigate the geometric-phase effect with an applied gradient $\partial B_z / \partial z|_\mathrm{appl}$, taking advantage of the fact that the average position of the UCN's  is lower than the room-temperature $^{199}$Hg by about 2.8~mm.
A pair of frequency measurements with $\vec E$ and $\vec B$ oriented parallel and antiparallel can be combined into an $E_z$-even geometric phase and the $E_z$-odd EDM signal. The geometric phase signal is linear in the gradient, while the EDM signal  is independent of the first-order magnetic field gradient. 
Taking the frequencies to be positive quantities, the EDM signal is the $E_z$-odd combination of neutron and comagnetometer frequencies
\bea
d^{meas}&=&\frac{\hbar}{4|E_z|} [(\omega_n^{\uparrow\uparrow} - \omega_n^{\downarrow\uparrow})- (\omega_{\rm Hg}^{\uparrow\uparrow} - \omega_{\rm Hg}^{\downarrow\uparrow})\bigl |\frac{\gamma_n}{\gamma_{\rm Hg}}\bigr |]\nonumber\\
&\approx& d_n+d_{\rm Hg}^{GP}+\dots,
\label{eq:dnmeas}
\eea
where  ($\uparrow\uparrow$) and ($\uparrow\downarrow$)  refer to $\vec E$ and $\vec B$ parallel and antiparallel, respectively, and the $+\cdots$ indicates additional $E_z$-odd false-EDM effects. 
Assuming a linear gradient and taking $B_z^0$ as the magnetic field at $\langle z_{\rm Hg}\rangle=0$
\bea
\frac{1}{2} (\omega_{\rm Hg}^{\uparrow\uparrow} + \omega_{\rm Hg}^{\downarrow\uparrow})&=&\bigl | \gamma_{\rm Hg}\bigr | B_z^0\nonumber \\
\frac{1}{2}(\omega_n^{\uparrow\uparrow} + \omega_n^{\downarrow\uparrow})\ &=&\bigl | \gamma_n\bigr | \bigl [ B_z^0+ \frac{\partial B_z}{\partial z} (\langle z_n\rangle - \langle z_{\rm Hg}\rangle)\bigr ]
\eea
These can be combined into the ratio
\bea
 R^\prime&=& \frac{\omega_n^{\uparrow\uparrow}+ \omega_n^{\downarrow\uparrow}}{\omega_{\rm Hg}^{\uparrow\uparrow} + \omega_{\rm Hg}^{\downarrow\uparrow}} \big |\frac{\gamma_{\rm Hg}}{\gamma_n}\bigr |-1\nonumber\\
&\approx&  \frac{1}{B_z^0} \frac{\partial B_z}{\partial z} (\langle z_n\rangle - \langle z_{\rm Hg}\rangle).
\label{eq:Rprime}
\eea
Here $B_z^0$ can be  positive or negative, and $\langle z_n\rangle<\langle z_{\rm Hg}\rangle$. Noting that $d_\mathrm{false}^\mathrm{GP}$ is propotional to $\frac{\partial B_z}{\partial z}$ while $d_n$ is independent of the gradient,  Eqns.~\ref{eq:falseD_GP},~\ref{eq:dnmeas}, and~\ref{eq:Rprime} are combined into:
\be
d^{meas}=d_n^\pm \pm kR^\prime,
\label{eq:dmeas}
\ee
where  $+/-$ refer to $B_z$ up/down, $kR^\prime$ is $d_\mathrm{false}^\mathrm{GP}$,  and the slope $k$ is  assumed to have the same magnitude for $B_z$ up and down. 
In FIG.~\ref{fig:baker_edm}, $d_n^{meas}$ is plotted vs $R^\prime$ for $B_z$ positive and negative. These data were fit to determine $d_n^+$, $d_n^-$, and $k$. The average $\frac{1}{2}(d_n^++d_n^-)$, corrected for additional systematic effects, is the final $d_n$ result.%

{

Additional  systematic effects are the light shift of $^{199}$Hg, discussed in Sec.~\ref{sec:LightShift} and   
 the changing energy distribution of the trapped UCN sample during the Ramsey measurement coupled with the energy dependence of UCN losses on the walls. 
A thorough investigation of systematic effects for the ILL-Sussex-Rutherford experiment has  been reported by~\textcite{Afach:2015sja}.

%
The control of systematics for future experiments will rely on  intentionally varying and/or amplifying the effects, for example varying gradients, changing temperatures to affect the effective velocities, or changing buffer gas pressures.
 Ideas include a double chamber EDM experiment with a surrounding magnetometer array requiring that the magnetic gradient measured by the difference of neutron or comagnetometer frequencies in the two chambers  be consistent with the gradient measured with a surrounding (4-pi) magnetometer. Any inconsistency would be due to internal magnetization, {\it e.g.} magnetization caused by a spark of the high voltage. We also anticipate that most future EDM experiments will introduce blind analysis techniques.
%


\subsection*{Neutron-EDM prospects}
\label{sc:nEDMprospects}

Several neutron EDM experiments are under development with the prospects of improving the sensitivity by one and eventually two orders of magnitude. In addition to the novel idea of probing P-odd/T-odd neutron scattering in a crystal, several experiments use UCN from a variety of different sources listed in Sec.~\ref{sec:UCNSources}. In the case of the SNS EDM experiment, the experiment is the source. 

\noindent{\bf The PSI experiment}

For the neutron EDM program at the Paul-Scherrer Institute (PSI) near Zurich a large collaboration will use the UCN source described in Sec.~\ref{sec:UCNSources}. In a first generation effort, the original ILL-Sussex-Rutherford apparatus~\cite{Baker:2006ts} was moved and rebuilt with significant improvements to the neutron storage bottle lifetime, neutron polarization detection,  magnetic shielding, Hg comagnetometer and  the addition of Cs magnetometers. With these improvements, a result with sensitivity at the $1\times 10^{-26}$\ecm~ level is expected. Plans are underway for an improved experiment with a double-EDM chamber that is expected to extend the sensitivity to as low as $10^{-27}$\ecm.

\noindent{\bf The PNPI-ILL-PNPI experiment}

The Gatchina EDM experiment, shown in FIG.~\ref{fig:nedm_history}, has been running at ILL, where it is expected to improve on the result of~\textcite{Serebrov:2015idv}. Over the longer term, the UCN source at the PNPI research reactor WWRM is expected to provide up to 10$^4$ UCN/cm$^3$~\cite{Serebrov:2017gea} and  could extend the sensitivity to $2\times 10^{-28}$ \ecm.

\noindent{\bf The FRM-II/PanEDM experiments}
The EDM experiment, originally developed at the FRM-II reactor \cite{Altarev:2012uy}, has been moved to ILL to couple to the SuperSun UCN source and renamed PanEDM.
%
It is a modular concept with UCN confined in two adjacent room-temperature chambers with opposite electric field.  Ramsey's technique will be applied to both cells simultaneously, and the EDM phase shift will have the opposite sign in the two chambers.
%
%
Magnetometry will be effected by  two $^{199}$Hg magnetometers above and below the EDM chambers. In addition, an array of Cs atomic-vapor magnetometers will be placed near the EDM chambers.  Plans call for two phases: Phase 1 will use the SuperSun source without the magnetic reflector with the sensitivity goal of 10$^{-27}$ \ecm; Phase 2 will employ the magnetic reflector, and a factor of 3-4 improvement is expected. With no comagnetometer, the requirements on magnetic shielding and external magnetometry are more stringent, but it is expected that the electric field can be increased by 50\% or more compared to what could be applied with a Hg comagnetometer.

\noindent{\bf The LANL Room Temperature EDM experiment}

The recent four-fold increase in the LANL UCN source performance~\cite{Ito:2017ywc} has provided motivation to develop a  nEDM experiment. The concept is a double cell with a Hg comagnetometer and external magnetometry. There is also the possibility to study  $^3$He as a comagnetometer monitored with either a SQUID or by detection of UCN capture on $^3$He by detecting recoil protons with scintillators.  The plan is to deploy state of the art  magnetic shielding and to optimize storage volumes and UCN guide volumes to the LANL UCN source. The sensitivity goal is 1-3$\times 10^{-27}$ \ecm.

\noindent {\bf The TRIUMF/KEK experiment}

The TRIUMF/KEK experiment is based on a SF-He source coupled to the high-intensity proton beam from the TRIUMF cyclotron~\cite{Masuda:2002dy}. Due to the dilution of UCN when such a source is coupled to the experiment volume, the EDM chamber will be smaller than in other experiments.
A further option discussed for this room-temperature neutron EDM experiment is the use of a low-pressure $^{129}$Xe-based comagnetometer~\cite{rf:TRIUMFEDM}  with  two-photon readout~\cite{ref:xenon_twophoton}, introduced in Sec.~\ref{sec:Magnetometry}, as well as  a $^{199}$Hg comagnetometer designed to address the geometric-phase and leakage-current systematic effects.
%
%

\noindent {\bf The SNS experiment}

A major effort underway in the US is the cryogenic SNS-nEDM experiment. The concept introduces  a number of novel features based on the proposal by~\textcite{rf:GolubLamoreaux}. The plan is to produce UCN in SF-He from a cold neutron beam within the EDM-storage volumes. Cryogenic superconducting shielding and the high dielectric strength of SF-He are expected to provide reduced systematics and higher electric fields than room-temperature vacuum experiments.  
A small amount of highly polarized $^3$He will be introduced into the EDM chambers to serve both as a comagnetometer and a spin analyzer. UCN are captured by $^3$He in a spin-singlet state due to an unbound resonance in $n$+$^3$He with emission of a proton  and triton with a total energy of  764 keV ($n+^3{\rm He}\rightarrow\ ^1{\rm H}+^3{\rm H}$).  Thus, effectively, only neutrons with spin opposite the $^3$He spin are captured,  and the rate of emission of the  proton and triton   will  measure  the projection of the relative spin orientation of the neutron and $^3$He. The proton and triton will be detected by scintillation in the SF-He due to the formation and decay of He$^*$ molecules and emission of an 84~nm photons. 
The UV photons are converted to longer wavelengths for detection by photomultipliers after absorption on the surface coating of  deuterated tetraphenyl-butadiene (dTPB)  polymer matrix, chosen due to the low loss of UCN on the deuterium). 
As the UCN and $^3$He spins precess the absorption is modulated at the difference of precession rates. The change of modulation frequency of the scintillation light with electric field would signal a difference of the two species' EDMs and, assuming the $^3$He atomic EDM is much smaller  due to Schiff screening~\cite{Dzuba:2007zz}, the neutron EDM. 

The precession frequency of $^3$He is 32.43~MHz/T, compared to 29.16~MHz/T for the neutron, so any systematic effect (false EDM) due to a change of magnetic field correlated with the electric field is suppressed by a factor of about 10. However it is possible to shift the precession frequencies of both species with an oscillating magnetic field via the Ramsey-Bloch-Siegert effect (Eqn.~\ref{eq:geomphase0}), with $\omega_{Rot}$ set between the two free-precession frequencies to ``dress'' the spins~\cite{rf:CohenTannoudjiDressing}. This results in potentially orders of magnitude more sensitive comagnetometry.
Alternatively, the $^3$He spin-precession can be independently monitored with SQUID sensors to signal any changes in the magnetic field. 
The precessing $^3$He magnetization can be monitored with SQUID magnetometers~\cite{IEEE6905771KimClayton}. The geometric phase  will affect $^3$He, and the mean free path of the $^3$He atoms depends on temperature, which is nominally 0.4-0.5 K, but can be adjusted to scan the geometric-phase effect.
After each measurement, the  depolarized $^3$He must be removed from the superfluid helium using a heat-flush/phonon wind  technique~\cite{rf:HeatFlushRefs}.
%

%
%
%
%
%

\noindent {\bf Possible EDM experiments at the European Spallation Source}

A fundamental-neutron-physics program has been proposed for the European Spallation Source~\cite{Pignol20143}. In addition to possible UCN EDM experiments, a reprised beam experiment has been proposed~\cite{Piegsa2013}.
  In this concept, the pulsed structure of the ESS would provide velocity discrimination and could be used to monitor $\vec v\times \vec E$ effects
with the much greater statistical power provided by the cold-neutron beam.

\noindent{\bf Crystal EDM}

P-odd/T-odd rotation of the neutron spin in Laue diffraction of polarized neutrons incident on a  crystal is sensitive to the neutron EDM interaction with the strong interplanar electric field. The first experiment on CdS was undertaken by~\textcite{ShullCrystalEDM}, who found $d_n=(2.4\pm3.9)\times 10^{-22}$\ecm.  \textcite{Fedorov:2010sj} carried out an experiment at the ILL cold neutron beam facility PF1B measuring the spin-rotation of  monochromatic polarized neutrons incident on a quartz crystal. The effective electric field was estimated to be of order $10^8$ V/cm. The final neutron spin directions were analyzed for different incident neutron spin to separate the EDM effect from the Mott-Schwinger interaction with atomic electrons in the crystal. The  result from about one week of data was $d_n=(2.5 \pm6.5\ {\rm (stat)}  \pm5.5\ {\rm (sys)})\times 10^{-24}$\ecm. Prospects for an improved setup suggest that the sensitivity can be improved to $2\times 10^{-26}$\ecm~ for 100 days of data taking~\cite{Fedorov:2010sj}.

\subsection{Paramagnetic atoms: Cs and Tl}


The EDM of a paramagnetic system is most sensitive to the electron EDM $d_e$ and the strength $C_S$ of a nuclear-spin independent electron-nucleus coupling corresponding to a scalar nuclear current. The tensor nuclear current contribution is several orders of magnitude smaller, and the pseudoscalar contribution vanishes in the limit of zero velocity ({\it i.e} infinite nuclear mass).
\label{sec:CsTl}

The first direct atomic EDM experiment - measurement  of the frequency shift of the cesium atom EDM in an atomic beam with  a modulated electric field - was undertaken by ~\textcite{rf:Sandars1964} and collaborators.  There are many challenges to such a measurement that have led to the techniques applied to contemporary undertakings. First, the atomic beam, traveling at several-hundred m/s,  transited an apparatus of length less than a meter so that linewidths of several kHz were observed (Ramsey's separated oscillatory field technique was used). Second, due to the unpaired electron, the cesium atom has a large magnetic moment  that couples both to external magnetic fields and to the motional magnetic field $\vec B_m=\vec v\times \vec E/c^2$. The magnetic field produced by any  leakage currents  would change with the modulation of the electric field and could provide a false EDM signal. Misalignment of the applied magnetic and electric fields  also produces a false signal.  By determining the center of a resonance line to a precision better than the linewidth, {\it i.e.} line splitting  by more than 10,000, the frequency shift sensitivity was of order 0.1 Hz. The result is $d_{\rm Cs}=(2.2\pm 0.1)\times 10^{-19}$ \ecm~ with an electric field up to 60 kV/cm. The  error is statistical only, and the finite EDM signal is attributed to the motional effect due to a misalignment of 10 mrad~\cite{rf:Sandars1964}. Subsequent work using other alkali-metal species with lower $Z$ and less sensitivity to T-odd/P-odd interactions to monitor magnetic-field effects - now called a comagnetometer - led to the result $d_{\rm Cs}=(5.1\pm 4.4)\times 10^{-20}$ \ecm~\cite{rf:Carrico1968}. Shortly after that publication, ~\textcite{rf:Weisskopf1968} and collaborators presented a significantly improved result based on a longer interaction region, correspondingly narrower resonance lines and a sodium comagnetometer: $d_{\rm Cs}=(0.8\pm 1.8)\times 10^{-22}$ \ecm.

The cesium atomic beam EDM experiments were ultimately limited by the linewidths,  count-rate limitations, and by systematic errors due to motional magnetic field effects, though the atomic beam machines provided the capability to use other, lighter alkali-metal species, {\it i.e.} a comagnetometer, significantly reducing the motional-field systematic errors~\cite{rf:Weisskopf1968}. 
Another approach was the vapor cell experiment developed by Hunter and collaborators~\cite{Murthy:1989zz}. The confined atoms provided much narrower resonance linewidths  ($\approx$50 Hz), and also greatly mitigated motional field effects. Though a comagnetometer was not practical in the vapor cell, the leakage currents were directly measured and set the systematic uncertainty in the final result 
\be
d_{\rm Cs}=(-1.8\pm 6.7\ (\rm{stat}) \pm 1.8\ ({\rm sys}))\times 10^{-24}\ \text{\ecm}.
\ee

Sandar's group performed an atomic beam experiment to search for an electric dipole moment
in the $^3P_2$ metastable state of xenon with a comagnetometer beam of krypton~\cite{rf:Player1970,HPS71}.
 In the strong applied electric field,  the parity-allowed splittings are proportional to $m_J^2E^2$ and to the magnetic field component along  $\vec E$. The EDM signal would be a splitting linear in $\vec E$; however transitions that change the magnetic quantum  number $m_J$ are not practical due to the sensitivity of the $E^2$ term to a change of the magnitude of the electric field. Thus the $\Delta |m_J|=0$ transition $m_J=-1 \rightarrow m_J=+1$ was measured. With the xenon-krypton comparison, the difference of EDMs was found to be
$ |d_{\rm Xe}-d_{\rm Kr}|=(0.7\pm 2.2)\times 10^{-22}\ e{-\rm cm}$, where the errors are 90\% c.l.

~\textcite{rf:Commins1994}  developed a vertical counter-propagating  thallium atomic beam and subsequently added sodium beams as a comagnetometer~\cite{rf:Regan2002}. These experiments pioneered a new understanding of some of the most important systematic effects for EDM experiments, including those mitigated by the comagnetometer and the geometric phase effect~\cite{rf:ComminsAJPBerryPhase,rf:PendleburyGP,rf:Barabanov2006}. The most recent result can be interpreted as 
\be
d_{\rm Tl}=(-4.0\pm 4.3)\times 10^{-25}\ e{-\rm cm}.
\ee

\subsection*{Prospects for alkali-metal atoms}

Laser cooling and trapping of  cesium and francium offer promising new directions, and several approaches are being pursued.
Cesium atomic fountain clocks based on launching atoms from a laser cooled or trapped sample  have moved to the forefront of time-keeping. 
Narrow linewidths ($\tau\approx$ 1 s) are attained as the atoms move up and then down through a resonance region. While the $^{133}$Cs atomic frequency standard uses the $\Delta m_F=0$ transition, which is insensitive to small magnetic fields in first order, an EDM measurement must use $\Delta m_F\ge 1$. From equation~\ref{eq:EDMFreqEquation1},  with $T=1$ s, $\tau\gg T$  and $N=10^6$, the expected uncertainty  on $\omega$ is expected to be about $\delta_\omega\approx 10^{-3}$ Hz, which is consistent with observations of the Allan variance representing the short-term instability of cesium fountain clocks  ($ \sigma_\omega/\omega \approx 10^{-13}$ for $\omega=2\pi\times$ 9.2 GHz~\cite{rf:Weyers2009}). 
 For an EDM measurement with an electric field of 100 kV/cm, which may be feasible, 
  each 1 second shot would have a sensitivity of $6\times 10^{-24}$ \ecm, comparable to the sensitivities of both the cesium and thallium measurements. 
  Thus significant improvement is possible, and a demonstration experiment with about 1000 atoms per shot and $E=60$ kV/cm was reported as a measurement of $d_e$ by \textcite{rf:Amini2006}. The result can be interpreted  as  $d_{Cs}=(-0.57 \pm 1.6) \times 10^{-20}$ \ecm~ (the authors use $\eta_{e}=114$ for cesium).
The major limitation in this demonstration was the necessity to map out the entire resonance-line shape spectrum, which is the combination of transitions among the nine hyperfine sub-levels and inhomogeneities of the applied magnetic field in the resonance region.
  This subjected the measurement to slow magnetic field drifts that would need to be monitored or compensated. 
  If these problems are solved, the statistical sensitivity could be significantly improved with several orders of magnitude more atoms, higher electric field and duty-factor improvements. However
  the major systematic effect  due to $\vec v\times\vec E$ was about $2\times 10^{-22}$ \ecm~\cite{rf:Amini2006}. This could ultimately limit the sensitivity of a single species fountain measurement.
  
For francium, $\eta_{d_e}$, the ratio of atomic EDM to electron EDM from equation~\ref{eq:ParamagneticEDMs}, is in the range 900 to 1200~\cite{Ginges:2003qt}, and $k_{C_S}$ should be similarly enhanced for francium. Francium can be produced in significant quantities in isotope-separator rare-isotope production facilities, and $^{210}$Fr has been produced, laser cooled and trapped in a magneto-optical trap (MOT)~\cite{rf:Gomez2006}. The experiment has been moved to the isotope-separator facility (ISAC) at TRIUMF in Vancouver, Canada. A parallel effort is underway at Tohoku University Cyclotron and Radioisotope Center (CYRIC)~\cite{SakemiFrEDM1742-6596-302-1-012051,Kawamura-FrEDM}.
Francium isotopes have half-lives of 20 minutes (for $^{212}$Fr) or less, and any experiment would need to be ``on-line,'' that is the EDM apparatus would be at the site of the rare-isotope production facility. The applying the cesium-fountain approach to francium may lead the way to a future program  at a  the Facility for Rare Isotope Beams (FRIB) at Michigan State University.

The fountain concept allows linewidths on the order of 1 Hz, limited by the time for the cold atoms with vertical velocity of a few m/s to rise and fall about 1 meter. Another idea being pursued by D. Weiss and collaborators is to stop and cool alkali-metal atoms in optical molasses near the apogee of their trajectory and trap them in an optical lattice formed in a build-up cavity~\cite{rf:Fang2009}. 
Storage times in the lattice could be many seconds.  The lattice would be loaded with multiple launches, filling lattice sites that extend over 5-10 cm, and  $10^{8}$ or more  atoms could be used for the EDM measurement. After loading the lattice, the atoms would be optically pumped to maximum polarization and then the population transferred to the $m_F=0$ state by a series of microwave pulses. A large electric field ({\it e.g.} 150 kV/cm) would define the quantization axis in nominally zero magnetic field, and the energies would be  proportional to $m_F^2$ due to the parity-allowed interaction.  

In another planned innovation,  a Ramsey separated-field approach would be used with the free-precession interval  initiated by pulses that transfer atoms to a superposition of $m_F=F$ and $m_F=-F$ states and terminated by a set of pulses coherent with the initial  pulses. The relative populations transferred back to the $m_F=0$ state would be probed by optical fluorescence  that could be imaged with about 1 mm spatial resolution~\cite{rf:Zhu2013}.   With the large size of the lattice, the superposition of  stretched levels ($m_F=\pm F$) would amplify the sensitivity by a factor of $F$ relative to experiments that monitor $\Delta m_F=1$ transitions~\cite{rf:Xu1999}. In a measurement time $\tau=3$ s, and $N=2\times 10^8$, an EDM sensitivity of $6.5\times 10^{-26}$ \ecm~ for the cesium atom is expected. The optical lattice can also trap rubidium, which could be used as a comagnetometer. 


\subsection{Paramagnetic polar molecules: YbF, ThO and $^{180}$Hf$^{19}$F$^+$}
\label{sec:PolarMolecules}

Molecular-beam EDM experiments exploit several features of diatomic polar molecules, usually one light and one heavy atom;
 most significantly the strong interatomic electric field with characteristic strength of 10-100 GV/cm. The pioneering molecular-beam EDM approach of~\cite{rf:SandarsTlF} and~\cite{rf:RamseyTlF} used thallium-fluoride (TlF) which is {\it diamagnetic} and is discussed in the next section. A number of groups have followed the lead of~\cite{Hudson:2002az} and investigated paramagnetic molecules. The most sensitive measurements in paramagnetic systems are from YbF~\cite{Hudson:2011zz},  ThO~\cite{Baron:2013eja,Baron:2016obh} and HfF$^+$~\cite{Cairncross:2017fip}.  In addition, efforts using the molecular ion  ThF$^+$, which is similar in electronic structure, are underway by~\textcite{Loh1220} .

Experimentally, the internal electric dipole moment $\vec D$ is oriented parallel or antiparallel to a relatively modest applied electric field in the lab $\vec E_{\rm lab}=E_{\rm lab}\hat z$. In essence the molecule's electric polarizability is very large resulting in an effective molecular dipole moment $D_z\propto E_{\rm lab}$.   In the atomic beam, the average orientation of $D_z$ can be expressed as a polarization ${\mathcal P}_z=D_z/|\vec D|$, for example the dependence of ${\mathcal P}_z$ for YbF is shown in FIG.~\ref{fg:YbFPz}. For ThO and HfF$^+$, which are effectively fully electrically polarized by a relatively small laboratory electric field, the estimated effective internal electric fields are $E_{\rm eff}\approx$ 84 GV/cm and $E_{\rm eff}\approx$ 23 GV/cm, respectively, which can be found from Table~\ref{tb:paramagnetics} using $E_{\rm eff}=\hbar\alpha_{d_e}/e$.

\begin{figure}[ht]
 \includegraphics[width=3.5truein,angle=0]{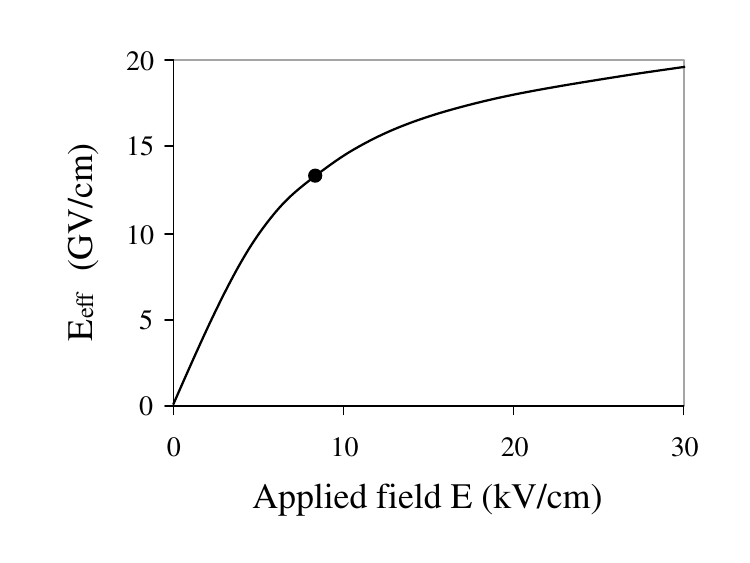}
 \caption{The effective electric field $E_{\rm eff}$ in YbF as a function of the applied electric field ($E_{\rm lab}$ in the text). The fully saturated $E_{\rm eff}$ corresponds to ${\mathcal P}_z=1$. Figure from~\textcite{Hudson:2002az}. 
 }
\label{fg:YbFPz}
\end{figure}

The electronic energy level structure of a $^\Sigma\Lambda_\Omega=^3\!\!\Delta_1$, {\it e.g.} $F=1$ state in the polar molecule ThO, is shown in FIG.~\ref{fg:MolecularStructure}. The quantum numbers labled by capital greek letters indicate angular momentum projected on the internuclear axis: ($\Sigma=3$ indicates the electron-spin triplet state, orbital angular momentum  $\Lambda=2$ is labeled by $\Delta$, and the total of spin and orbital angular momentum and rotation is $\Omega=1$.) The $^3\Delta_1$ state also results in a relatively small magnetic moment as the spin and orbital moments cancel. As shown, the laboratory electric field electrically polarizes the molecule, providing two directions of the internal electric field with different splittings due to the EDM, but the same magnetic field splitting. Thus the two orientations of the molecular dipole, which can be separately probed, for example by tuning a probe laser, provide an effective {\em internal} comagnetometer.

\begin{figure}[hb]
 \includegraphics[width=3.5truein,angle=0]{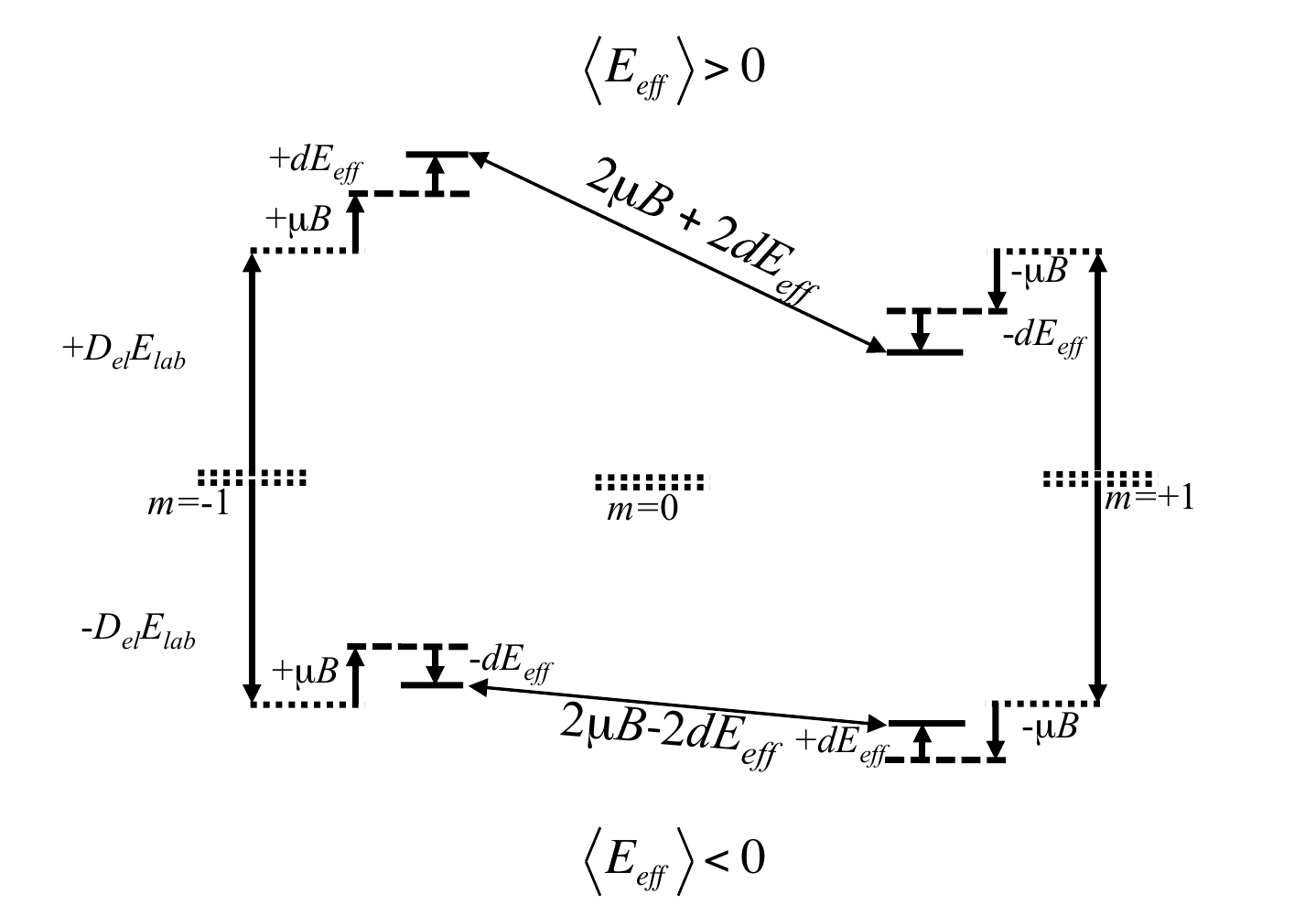}
\caption{Level structure of molecule in a  $^3\Delta_1$, $J=1$ state.   The laboratory electric field $E_{lab}$ splits the states by $\pm D_{el}E_{lab}$ into molecular dipole up and down. This results in  the two orientations of the effective internal field $E_{\rm eff}$, which is directed towards the lighter atom. The specific structure for the ThO experiment is shown; for HfF$^+$   $J=3/2$ and the stretched states ($m=\pm 3/2$) move up and down similarly. In a magnetic field $\vec B$ parallel or antiparallel to $\vec E_{lab}$, the $m=\pm 1$ states are further split by $\mu B$ as shown. The EDM $d$ further splits the $m=\pm 1$ states with $E_{eff}>0$, but reduces the splitting with $E_{\rm eff}<0$ antiparallel to $B$. }
\label{fg:MolecularStructure}
\end{figure}

The YbF experiment used ground-state $^2\Sigma_{1/2}^+$ molecules, where the $+$ indicates positive reflection-symmetry along a plane containing the internuclear axis~\cite{Hudson:2002az}. The resulting P-odd/T-odd  frequency shift is reported as
\begin{equation}
\omega^{EDM}({\rm YbF})=(5.3\pm 12.6\ {\rm (stat)}\pm 3.3\ {\rm (sys)})\ {\rm mrad/s},
\end{equation}
which can be interpreted as an electron EDM assuming $E_{\rm eff}=14.5$ GV/cm and $C_S=0$:
\begin{equation}
d_e({\rm YbF})=(-2.4\pm 5.7\ {\rm (stat)}\pm 1.5\ {\rm (sys)})\times 10^{-28}\ \text{\ecm}\ (C_S=0).
\end{equation}

The setup of the ACME ThO experiment is shown in FIG.~\ref{fg:ThOAcmeSetup} .The experiment used molecular-beam resonance methods based on the Ramsey separated-oscillatory-fields technique. For ThO, the $^3\Delta_1$ state is a metastable state originally populated from the ground state by the 943 nm optical pumping light.  In this case, $2D_{\rm el}E_{\rm lab}/h\approx$ 100 Mhz, and 1090 nm optical transitions  to an excited state with two opposite parity levels separated by 10 MHz are used to prepare and probe the orientations of a superposition of the $m=\pm1$ levels. If the state-preparation light is polarized along $\hat x$, then the initial electron spin is along $\hat y$. The EDM signal is a rotation around $\hat z$, which is detected by the component of the electron spin along $\hat x$ that reverses with the sign of $\vec E_{lab}\cdot \vec B$. The EDM frequency shift $\Delta\omega^{\rm EDM}$ is the rotation angle divided by $\tau\approx$1.1~ms, the transit time from pump to probe positions. ($\tau$ is determined from the magnetic precession angle $\phi^B=-\mu | B_z|\tau/\hbar$, where $\mu$ is the molecular magnetic moment, which is relatively small due to cancellation of spin and orbital effects.) A number of additional experimental parameters are changed to separate background and false-EDM signals including the molecular orientation ($E_{\rm eff}>0$ or $E_{\rm eff}<0$), the magnetic field  direction and magnitude, the electric field magnitude, the readout  laser polarization direction and various exaggerated imperfections. In all more than 40 parameters were varied.  Systematics were evaluated through a combination of anticipated effects informed by earlier experiments in YbF~\cite{Hudson:2011zz}  and PbO~\cite{Eckel:2013lsa}. The dominant systematic effects are generally the combination of two small effects, {\it e.g.} the AC stark shift caused by detuning the pump and probe lasers along with misalignments, gradients of the circular polarization and imperfect reversal of the electric field. The final result of the first-generation ThO experiment is a P-odd/T-odd precession frequency
\begin{equation}
\omega^{EDM}({\rm ThO})=(2.6\pm 4.8\ {\rm (stat)}\pm 3.2\ {\rm (sys)})\ {\rm mrad/s}.
\end{equation}

From Eqn.~\ref{PolarMoleculeDeltaOmega}, these  can be interpreted as the combination of contributions from the electron EDM $d_e$ and from a nuclear-spin independent (scalar) coupling labeled $C_S$; however adopting the ``sole-source'' approach (see Sec.~\ref{sec:TheoreticalInterpretation}), this can be interpreted as
\begin{eqnarray}
d_e({\rm ThO})&=&(-2.1\pm 4.5)\times 10^{-29}\ \text{\ecm}\quad (C_S=0),
\nonumber
\\
C_S({\rm ThO})&=&(-1.3 \pm 3.0)\times 10^{-9}\quad\quad\quad\quad (d_e=0).
\nonumber
\\
\end{eqnarray}
Further interpretation is provided in Sec.~\ref{sec:GlobalAnalysis}. 
This is considered a ``first-generation'' ThO effort by the experimenters, and upgrades to cold molecular beam promise improved statistical uncertainty.

Paramagnetic HfF$^+$molecular ions  in the metastable $^3\Delta_1$, $F=3/2$ state were confined in a radio frequency  trap to measure the P-odd/T-odd energy shifts by ~\textcite{Cairncross:2017fip}. (The specific isotopes were $^{180}$Hf and $^{19}$F.) 
The molecular ions were produced by laser ablation of Hf metal in a supersonic jet with a mixture of  argon and SF$_6$ gas, which produced neutral ground state HfF molecules. The cold argon gas from the supersonic jet cooled the rotational and vibrational degrees of freedom of the molecules. UV lasers ionized the HfF, and the ions were initially  trapped by an axial static field and a 50 kHz  radial (quadrupole) field and then confined by a uniform electric field rotating at 250kHz. The result was that the ions rotated in a circle of about 1 mm diameter.  An axial magnetic field gradient created the bias field in the rest frame of the ions. The Ramsey-style EDM measurement consisted of laser preparation of a polarized spin state  followed by a $\pi/2$ pulse, which created a superposition of $m_F=\pm3/2$, a free precession time of about 700 ms, and a second $\pi/2$ pulse. The final phase was read out by selective laser depopulation of alternating $m_F=\pm3/2$ levels followed by laser ionization and detection of the Hf$^+$ and background ions by a microchannel plate. Systematic effects are studied by observing frequency shifts in channels that are not sensitive to the P-odd/T-odd effects. The important systematic effects included non-ideal reversal of the rotating magnetic field combined with the different gyromagnetic ratios of the upper (labeled $E_{\rm eff}>0$) and lower (labeled $E_{\rm eff}<0$) pairs of states (doublets) due to Stark mixing with $J=2$ states, geometric phases, and background. The difference of frequencies for two flips that project the EDM - flipping the bias magnetic field and flipping $E_{\rm eff}$ by selecting the upper to lower pair of states -- was reported: 
\begin{equation}
\omega^{BD}({\rm HfF^+})=2\pi(0.1\pm 0.87\ {\rm (stat)}\pm 0.2\ {\rm (sys)})\ {\rm mrad/s}.
\end{equation}
Assuming $C_S=0$, the resulting sole-source electron EDM is
\begin{equation}
d_e({\rm HfF^+})=(0.9\pm 7.7\ {\rm (stat)}\pm 1.7\ {\rm (sys)})\times 10^{-29}\ {\rm \ecm}.
\end{equation}
The authors did not provide a soul-source limit on $C_S$; however \textcite{Skripnikov2017} has calculated   $\alpha_{C_S}\approx 2.0\times 10^6$ rad/s, which is used in the analysis presented in Sec.~\ref{sec:GlobalAnalysis}, which includes results from ThO and $^{180}$Hf$^{19}$F$^+$  to constrain $d_e$ and $C_S$ simultaneously.

A second generation ion trap that may confine ten times more HfF$^+$ ions in a larger volume combined with  improved electrode design is expected to provide an order of magnitude higher sensitivity~\cite{Cairncross:2017fip}. The JILA group also intends to perform an experiment on ThF$^+$ ($E_{\rm eff}\approx$ 36 GV/cm), 
for which the ground state is $^3\Delta_1$ providing for coherence times that are not limited by the lifetime of an excited state. It has also been pointed out that an experiment with the isotope $^{177}$Hf  (18.6\% abundance)  with nuclear spin $I=7/2$ would be sensitive to the P-odd/T-odd magnetic-quadruople moment~\cite{Skripnikov2017MQM}.  

\begin{figure}
 \includegraphics[width=3.5truein,angle=0]{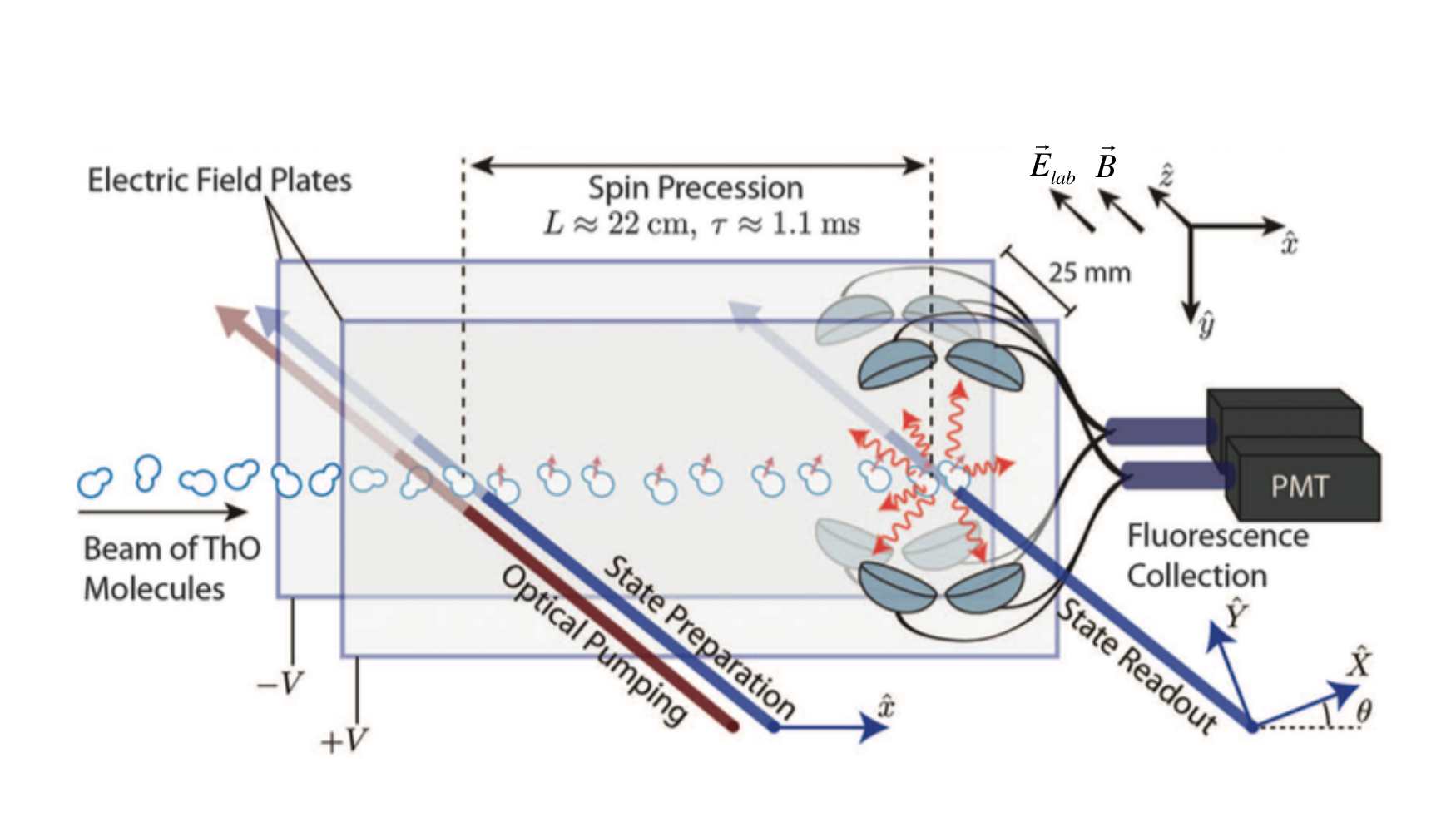}
  \caption{  (Color online) The experimental layout of the ACME ThO experiment from~\textcite{Baron:2013eja}. 
 }
\label{fg:ThOAcmeSetup}
\end{figure}

A recent proposal to study the orientation-dependent hyperfine structure of polar molecules in a rare-gas matrix, which is sensitive to the electron EDM has been presented by~\textcite{edm3-2}.
Another promising idea  is to store paramagnetic molecular ions or other particles in an electrostatic storage ring of a few meters diameter, used as a large ion trap.\footnote{A ring with in principle suitable parameters exists~\cite{MPIfurKernphysikHeidelberg}. } Such a configuration enables the storage of molecular ions of all possible configurations of states. 
In TaO, for example, the  ground state structure is $^\Sigma\Lambda_\Omega=^3\!\!\Delta_1$, and ions could be trapped electrostatically for several hours in bunches of  up to 10${^7}$  ions with kinetic energies of the order 100 keV. Preparation and readout of the molecular states relevant for EDM measurements would be done with lasers. 
Due to the sub-kHz angular frequency of the particles, the small molecular magnetic dipole moment, and the eddy-current and RF shielded environment provided by the vacuum housing of the storage ring, no compensation of the ambient magnetic fields is necessary. 
The long storage times in the ring allow for a large number of repetitions of the experiment for each 
configuration, and the large number of ions stored in the ring may enable  up to six orders of magnitude greater sensitivity to $d_e$. At this level, EDMs and Majorana neutrinos have model dependent connections, thus enabling a new path to access physics beyond the SM \cite{Archambault:2004td,Ng:1995cs}. Additionally, radium or radon ions could be stored, taking advantage of the octupole enhanced Schiff moment or nuclear EDM discussed below.  

\subsection{Solid-state systems}

The electron EDM can also be measured in special ferro-electric and paramagnetic solid-state systems with quasi-free electron spins that can be subjected to applied electric and magnetic fields. Advantages of such a system are
\begin{enumerate}[i.]
\item a high number density of unpaired electrons ($10^{22}$ cm$^{-3}$), providing signal amplification;
\item confinement of the electrons, mitigating such effects as motional fields; 
\item features of solid-state samples including collective effects, {\it e.g.} for ferro-electric systems,  a large electric field spin-polarizes the electrons resulting in a magnetization that reverses with the electric field;
\item minimal magnetic order to mitigate spurious magnetic effects. 
\end{enumerate}
A cryogenic experiment increases the electron polarization and provides for detection of the resulting magnetization  by SQUID magnetometers. 
The desired properties of an ideal material follows from consideration of the specific requirements of the EDM search~\cite{Ignatovich:1969tv,
Ignatovich:1969tv-ru,rf:Liu2004,rf:Shapiro1968,rf:Buhmann2002,rf:Sushkov2009,rf:Sushkov2010}. 

The polycrystal Gd$_3$Ga$_5$O$_{12}$ (gadolinium-gallium-garnet) provides seven unpaired atomic electrons, high resistivity ($10^{14}$ $\Omega$-m ) and high dielectric strength (1 GV/m). Enhancement of the electron EDM leads to an atomic EDM of Gd$^{3+}$  atoms in the lattice  $d_{Gd^{3+}}\approx 20d_e$, and the result $d_e=(-5.57\pm 7.98_{stat}\pm 0.12_{syst})\times 10^{-25}$ \ecm~ was reached with 5 days of data~\cite{PhysRevD.91.102004}. An experiment in the paramagnetic ferroelectric Eu$_{0.5}$Ba$_{0.5}$TiO$_{3}$ measured $d_e=(-1.07\pm 3.06_{stat}\pm 1.74_{syst})\times 10^{-25}$ \ecm. \cite{Eckel:2012aw}. Though this result is several orders of magnitude short of the sensitivity of paramagnetic molecules, improvements to magnetic noise and shielding 
can improve sensitivity. Other materials under consideration include SrTiO$_3$ doped with Eu$^{2+}$~\cite{PhysRevB.19.3593,PhysRevB.50.601}.
Another approach in paramagnetic ferroelectrics,  would detect the electric field produced by the electron EDM that would be magnetically aligned by polarized spins~\cite{rf:Heidenreich2005}.

\subsection{Diamagnetic atoms and molecules}
\label{sec:XeHg}

Diamagnetic atoms have the experimentally attractive feature that they can be contained in room-temperature cells with long polarization and spin-coherence lifetimes $T_1$ and $T_2^*$, because the nuclear spin is well shielded by the closed electron shell. Diamagnetic atoms can also be spin polarized using optical-pumping techniques, providing the largest possible signal-to-noise ratios and optimal statistical precision. Combined with techniques to carefully monitor and control systematic effects, measurements with $^{129}$Xe~\cite{rf:Vold1984}, with $^{129}$Xe/$^3$He~\cite{rf:Rosenberry2001} and the series of measurements with $^{199}$Hg~\cite{rf:Romalis2001,Griffith:2009zz,Graner:2016ses} are the most sensitive EDM measurements to date. The most recent $^{199}$Hg result stands alone in its sensitivity to various sources of CP violation~\cite{Graner:2016ses-erratum}. The diamagnetic molecule TlF was used in pioneering work by~\textcite{,rf:Hinds1980a} and~\textcite{rf:Wilkening1984}. The most recent and most precise TlF result was reported by~\textcite{rf:Cho1991}, and a new effort to greatly improve the sensitivity is underway~\cite{PhysRevA.95.062506}.

 
As discussed in Sec.~\ref{sec:EFTParameters} and in Eqn.~\ref{eq:DiamagneticAtoms}, the dominant contributions to the atomic EDM in diamagnetic atoms is the Schiff moment of the nucleus and the nuclear-spin-dependent electron-nucleus force with coefficient $C_T^{(0)}$. The Schiff moment  itself can arise from T-odd/P-odd NN interactions and from the EDMs of the individual nucleons  (both $^{129}$Xe and $^{199}$Hg have an unpaired neutron). However these sources can be related, depending on the nature of the P-odd/T-odd interactions as discussed in Sec.~\ref{sec:Theory}.

\noindent{\bf Xenon}
\label{sec:Xe}

Xenon is the heaviest stable noble gas, and $^{129}$Xe is a spin-1/2 isotope. Spin-1/2 atoms in cells have the advantage that only magnetic dipole interactions with external fields, with other atoms, and with the cell walls are allowed. This leads to longer spin-coherence times and narrow linewidths compared to atoms with nuclear spin $K>1/2$, which are subject, for example, to electric quadrupole interactions, in particular with the cell walls~\cite{rf:Wu1990,rf:Chupp1990}. Spin relaxation times of several 100's of  seconds and longer are observed for free-induction decay.  In natural xenon, the abundance of $^{129}$Xe is 26\%; however isotopically enriched gas is available. Polarization of $^{129}$Xe generally of greater than 10\%, and approaching 100\%, is possible using spin-exchange, mediated by the hyperfine interaction, with laser-optically-pumped alkali-metal vapor~\cite{rf:Zeng1985}. Spin exchange also makes it possible to use the alkali-metal vapor to monitor the free-precession of $^{129}$Xe polarization.

The first EDM measurement in $^{129}$Xe by Fortson and collaborators~\cite{rf:Vold1984} used spin exchange with laser-optically-pumped rubidium to polarize $^{129}$Xe in a stack of three cylindrical cells with electric fields of magnitude 3.2 to 4.9 kV/cm applied parallel and antiparallel to a uniform and well shielded 10$\mu$T  magnetic field. The stack of cells, treated as magnetometers, allows sums and differences of the free-precession frequencies to be used to determine the average magnetic field, and the average magnetic field gradient. A third combination of the three frequencies is the EDM signal. 
One potential systematic error for such a system was the effective magnetic field due to the hyperfine interaction, caused by any rubidium polarization projection along the electric field axis that somehow changed when the electric fields were changed. One successful approach was to ``quench'' the polarization of the two rubidium isotopes with resonance RF magnetic fields~\cite{OteizaDissertation}. Another concern was any change in the leakage currents that flowed across the cells due to the applied voltages that was different for different cells. Both effects were studied and found to be small compared to the statistical error of the measurement. The EDM of $^{129}$Xe was measured to be  $d_{\rm Xe}=(-0.3\pm 1.1)\times 10^{-26}$ \ecm, where the error is statistical only.

Another approach to measure the $^{129}$Xe EDM used  spin-exchange pumped noble-gas masers of $^{129}$Xe and $^{3}$He~\cite{rf:ChuppMaser1,Stoner:1996zz,Bear:1998zz}. Spin-exchange optical pumping is practical, in principle, for any odd~$A$ noble gas, and a population inversion can be pumped in multiple species with the same sign of the magnetic moment. The two species have very different sensitivity to the Schiff moment and to other P-odd/T-odd interactions, which are approximately proportional to $Z^2$, but similar sensitivity to magnetic field effects, particularly those produced by leakage currents that can change when the electric field is changed. Thus the $^3$He served as a comagnetometer occupying nearly the same volume as the $^{129}$Xe in a single measurement cell~\cite{rf:Chupp88}. 
The result reported by~\textcite{rf:Rosenberry2001} was
 \begin{equation}
d_A(^{129}{\rm Xe}) = (0.7\ \pm 3.3\ ({\rm stat})\ \pm 0.1\ ({\rm sys}))\times 10^{-27}\  e{-\rm cm}.
\end{equation}


Several experimental efforts to improve the $^{129}$Xe EDM sensitivity by 2-3 orders of magnitude are underway, including the active maser~\cite{rf:Yoshimi2002}, highly polarized liquid $^{129}$Xe detected with SQUID magnetometers~\cite{rf:RomalisLiqXe,LedbetterDissertation} and gas-phase experiments with $^3$He comagnetometry and SQUID-magnetometer detection~\cite{rf:FloHeXePaper,rf:MainzHeXePaper}. The SQUID-magnetometer experiments have demonstrated signal and noise that suggest one to three orders of magnitude improvement in sensitivity to the $^{129}$Xe EDM is possible in the near future.


%
%
%

\noindent{\bf Mercury}
\label{sec:Hg}

The $^{199}$Hg experiments undertaken by Fortson's group~\cite{rf:Romalis2001,Griffith:2009zz,Graner:2016ses} built on the ideas used in their $^{129}$Xe buffer-gas cell experiment~\cite{rf:Vold1984}. However there are two crucial differences with mercury: it is more chemically reactive, resulting in shorter coherence times, and it is heavier and thus generally more sensitive to sources of T and P violation. The most recent experiment~\cite{Graner:2016ses} used a stack of four cells sealed with sulfur-free 
 vacuum sealant and directly pumped and probed the $^{199}$Hg with a 254 nm laser~\cite{rf:Harber2000} as illustrated in FIG.~\ref{fg:HgSetup}. 
The outer two of the four cells have no electric field and the inner two have electric fields in opposite directions so that a difference of the free-precession frequencies for the two inner cells is an EDM signal. An EDM-like difference of the outer cell frequencies was attributed to spurious effects such as non-uniform leakage currents correlated with the electric field reversals and were therefore scaled and subtracted from the inner-cell frequency difference to determine the EDM frequency shift. The magnitudes of the leakage currents were also monitored directly and used to set a maximum E-field correlated frequency shift that contributed to the systematic error estimate. Other systematic error sources explored included effects of high-voltage sparks on the EDM signals and a number of possible correlations of experimentally monitored parameters ({\it e.g.} laser power and magnetic field fluctuations outside the magnetic shields). There were no apparent correlations, and the leakage current ($\pm$ 0.5 pA) was so small that only upper limits on the systematic errors could be estimated. The most recent result is~\cite{Graner:2016ses-erratum} 
\begin{equation}
d_A(^{199}{\rm Hg}) = (2.20\ \pm 2.75\ ({\rm stat}) \  \pm 1.48\ ({\rm sys})  )\times 10^{-30} e{-\rm cm}. 
\end{equation}

\begin{figure}
 \includegraphics[width=3.5truein,angle=0]{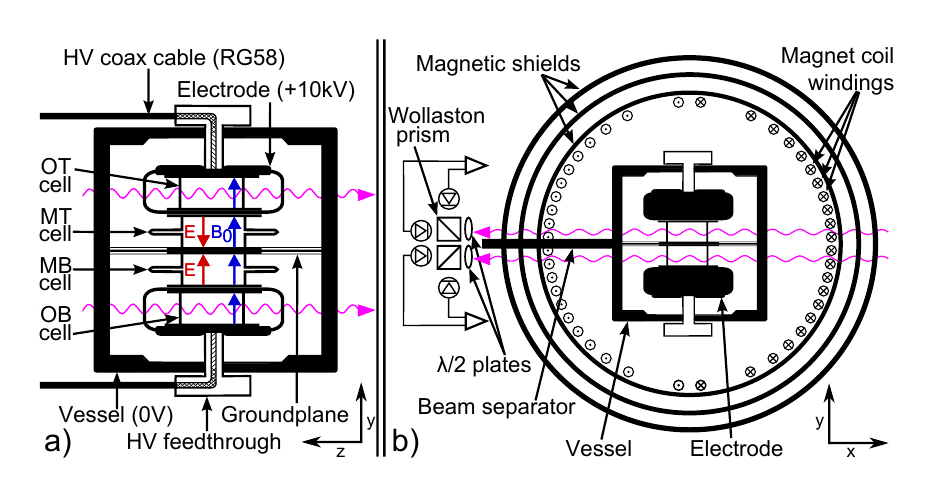}
  \caption{ (Color online) The experimental layout of the Seattle $^{199}$Hg experiment from~\textcite{Graner:2016ses}. 
 }
\label{fg:HgSetup}
\end{figure}

\noindent{\bf TlF}
\label{sec:TlF}

Molecular beam experiments using TlF were pursued by Sandars~\cite{rf:Harrison1969,rf:Harrison1969-erratum,rf:Hinds1980a}, by Ramsey~\cite{rf:Wilkening1984} and  by Hinds~\cite{rf:Schropp1987,rf:Cho1991}. For molecular beams, the  systematic errors associated with the $\vec v\times \vec E$ and leakage current effects are mitigated by using a relatively small applied electric field to align the intermolecular axis as is the case with polar molecules discussed in~\ref{sec:PolarMolecules}. This results in a large internal electric field at the thallium nucleus~\cite{rf:Coveney1983}.  The experiment is set up to detect  an alignment of a spin or angular momentum along the electric field by detecting precession around the internuclear axis, {\it i.e.} the frequency shift when the relative orientation of  applied electric and magnetic fields are reversed. 
When the average projection of the thallium nuclear spin on the internuclear axis is taken into account ($\langle\cos\theta_{\sigma\lambda}\rangle=0.524$), a frequency shift for full electric polarization is determined to be $d=(-0.13\pm 0.22)\times 10^{-3}$ Hz. With the applied electric field of 29.5 kV/cm, the most recent result~\cite{rf:Cho1991} is interpreted as a permanent dipole moment of the thallium molecule of  
\begin{equation}
d_{\rm TlF}=(-1.7\pm 2.9)\times 10^{-23}e\ {\rm cm}.
\end{equation}

For TlF, the electron spins form a singlet, but both stable  isotopes of thallium ($^{203}$Tl and $^{205}$Tl) have nuclear spin $J^\pi=1/2^+$, and the dipole distribution in the nucleus would be aligned with  the spin through T and P violation. This gives rise to the Schiff moment.
An alternative (and the original) interpretation is based on the observation that in the odd-A thallium isotopes, one proton  remains unpaired and can induce the molecular EDM through both the  Schiff moment (see Eqns.~(\ref{eq:dndp}-\ref{eq:etanetap})) and through magnetic interactions~\cite{rf:Coveney1983}. Separating these, the proton EDM would produce a magnetic contribution to a molecular EDM  of $d_{\rm TlF}^{p-mag}=0.13\ d_p$, and a contribution to the Schiff moment that would produce a molecular EDM estimated to be $d_{\rm TlF}^{p-vol}=0.46\ d_p$. The TlF molecular EDM can also arise from the electron EDM and from  P- and T-violating scalar and tensor electron-hadron interactions. However paramagnetic systems are more sensitive to $C_S^{(0,1)}$ and diamagnetic systems such as TlF are more sensitive to $C_T^{(0,1)}$. 
Thus this measurement has been interpreted as a (model dependent) measurement of the proton EDM: $d_p=(-3.7\pm 6.3)\times 10^{-23}$ \ecm.

A new effort is underway to measure CP violation in TlF using cooled molecules, which is based on work by~\textcite{PhysRevA.85.012511,PhysRevA.95.062506}. This would combine advantages of longer free-precession times and greater statistical power, which could provide several orders of magnitude greater sensitivity compared to the result of~\textcite{rf:Cho1991}.
 
\subsection{Octupole collectivity in diamagnetic systems}
\begin{figure}\vskip-1.7 truein
\hskip -1.1 truein
\includegraphics[width=4.4 truein,angle=0]{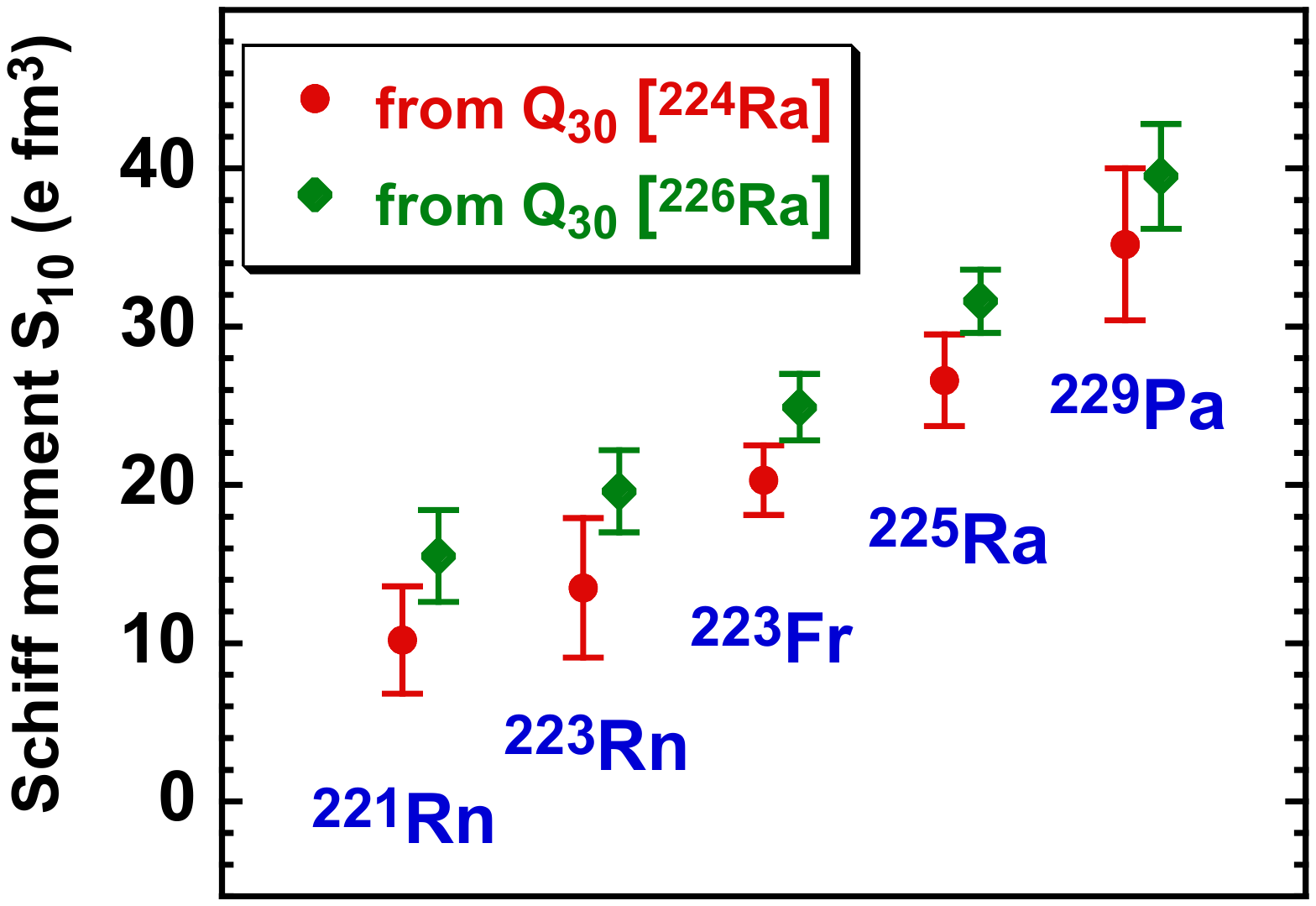}
\vskip -2 truein
  \caption{ (Color online) Estimates of intrinsic Schiff moments for octupole deformed nuclei from ~\textcite{Dobaczewski:2018nim}. Figure used with permission from the author. 
 }
\label{fg:OctSchiffDobaczewski}
\end{figure}

Recently, experimental efforts have focused on exploiting the enhanced Schiff moment in nuclei with closely spaced ground state parity doublets and strong E1 matrix elements, characteristic of  isotopes with nuclear octupole  collectivity. For  proton and neutron numbers in the range $Z$ or $N\approx 34,\ 56,\ 88$ and $N\approx 134$, nulceons near the Fermi surface populate states of opposite parity separated by total angular momentum 3$\hbar$, corresponding to reflection asymmetric states that lead to permanent octupole deformation or  octupole vibration. In either case the intrinsic dipole moment is polarized along the nuclear-spin by P-odd/T-odd interactions leading to the nuclear Schiff moment, which is enhanced by the electric polarizability of the nucleus, in analogy to polar molecules (see Sec.~\ref{sec:PolarMolecules}). In the two-state approximation, the resulting Schiff moment can be parameterized as
\begin{equation}
S\propto \eta e\frac{\beta_2\beta_3^2ZA^{2/3}r_0^3}{E_+-E_-},
\label{eq:SchiffMomentOctupole}
\end{equation} 
where $\eta$ represents the strength of the P-odd/T-odd interaction, $\beta_2$ and $\beta_3$ are the quadrupole and octupole deformation parameters and $E_\pm$  are the energies of the opposite-parity states~\cite{rf:Auerbach1997,rf:Auerbach1996,spevak95}. 
Note that the octupole deformation parameter enters $S$ quadratically, which means that both octupole vibrations and permanent deformation are equally effective~\cite{PhysRevC.78.014310}. Permanent deformation is indicative of (and indicated by) closely spaced parity doublets: $(E_+-E_-)\approx$ 50-100 keV. In $^{229}$Pa, the splitting was originally reported to be as small as 0.22 keV~\cite{Pa229AhmadPRL1982}, and evidence of strong octupole correlations support a ground-state parity doublet with $I=5/2$~\cite{Pa220AhmadPhysRevC.92.024313}, which was also predicted theoretically by~\textcite{CHASMAN19807}. In fact the evidence of the closely-spaced doublet in $^{229}$Pa provided motivation for the suggestion of enhanced T-nonconserving nuclear moments by~\textcite{Haxton:1983dq}, which followed suggestions in the earlier the work of~\textcite{rf:Feinberg1977}.

There is strong  evidence of octupole collectivity for nuclei with $A\approx 200-226$, including interleaved even/odd parity states in even-A nuclei~\cite{rf:NaturePaperRef7},   parity doublets in odd-A nuclei~\cite{rf:NaturePaperRef8} and enhanced electric-dipole (E1) transition moments~\cite{rf:NaturePaperRef9}.   The strongest direct evidence for octupole collectivity  has come from recent measurement of E3 strength using Coulomb excitation of radioactive beams of $^{220}$Rn and $^{224}$Ra at ISOLDE~\cite{rf:NaturePaper}. 
$\beta_2$ and $\beta_3$ for  $^{220}$Rn and $^{224}$Ra 
 are quite similar. However the larger moment $Q_3$ in $^{224/226}$Ra compared to  $^{220}$Rn 
 suggests that the deformation is permanent in $^{224}$Ra, while $^{220}$Rn is a vibrator~\cite{rf:NaturePaper}. Recently
 ~\textcite{Dobaczewski:2018nim} has estimated the intrinsic Schiff moments of nuclei in this region based on $Q_3$'s extracted from $^{224}$Ra~\cite{rf:NaturePaperRef9} and from $^{226}$Ra~\cite{rf:NaturePaperRef7}, which are shown in FIG.~\ref{fg:OctSchiffDobaczewski}. To estimate the observable Schiff moment further requires calculation of the P-odd/T-odd matrix elements arising from these intrinsic moments (see also~\cite{dobaczewski05}). 
 
 An intriguing possibility for future experimental efforts is to use a molecule with an ocutpole-enhanced nucleus, for example $^{225}$RaO~\cite{FlambaumRaO}. Though experimentally very challenging and potentially limited by production of appropriate molecules, this would take advantage of both the possible octupole-enhanced Schiff moment and the very large internal electric fields in the molecule. 
 
 


\bigskip
\noindent{\bf $^{225}$Ra}

\begin{figure}
 \includegraphics[width=3.25truein,angle=0]{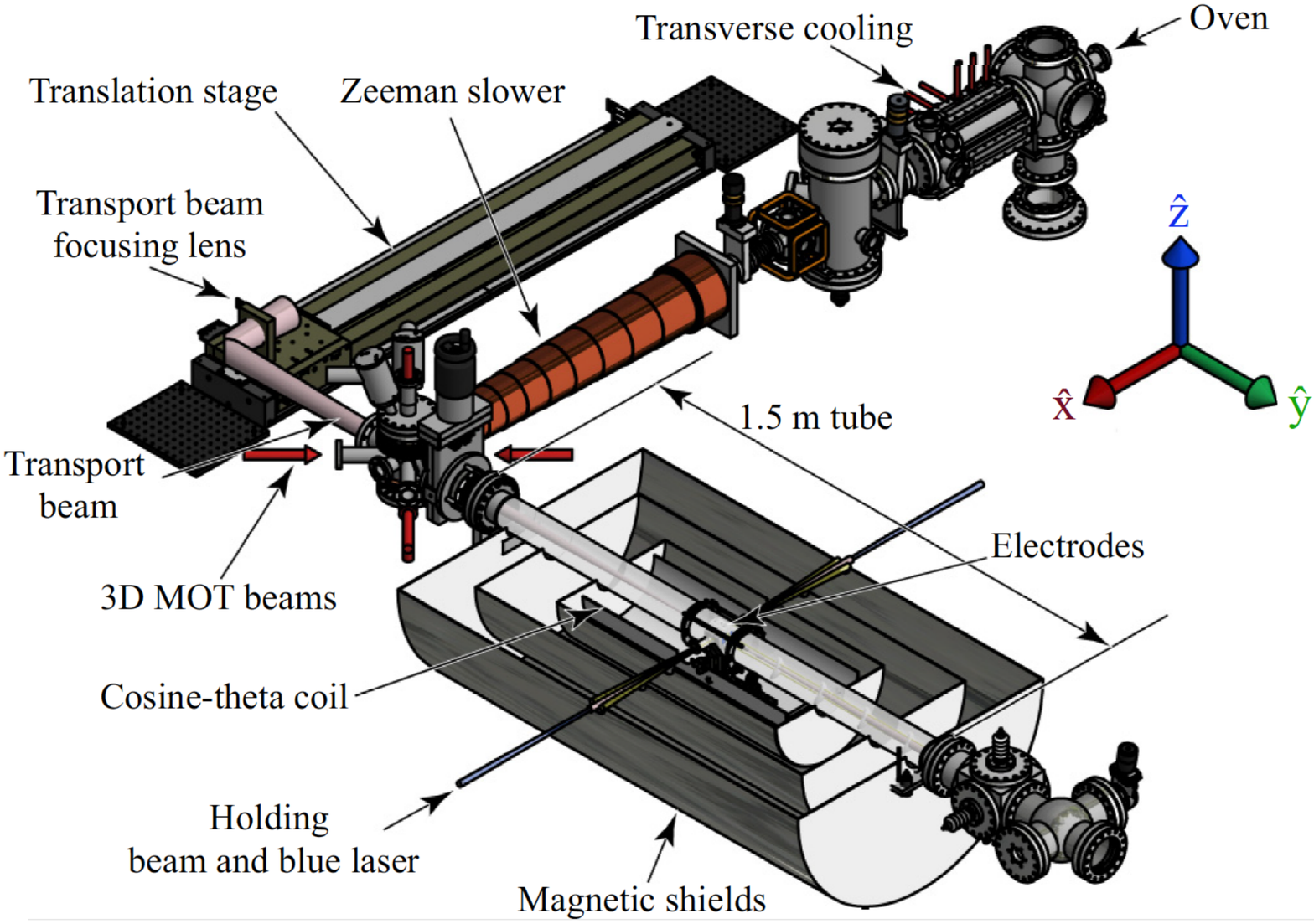}
  \caption{  (Color online) The experimental layout of the $^{225}$Ra experiment from~\textcite{rf:Parker2015,PhysRevC.94.025501}. 
}
\label{fg:RadiumSetUp}
\end{figure}

An ongoing effort at Argonne National Lab uses cold-atom techniques to measure the EDM of the $^{225}$Ra atom~\cite{rf:Parker2015,PhysRevC.94.025501}, which may have 2-3 orders of magnitude greater sensitivity to the P-odd/T-odd pion-nucleon couplings than $^{199}$Hg based on atomic~\cite{rf:DzubaPRAv60p02111a2002} and nuclear physics calculations~\cite{rf:Auerbach1996,rf:Auerbach1997,rf:EngelFriarHayes,rf:deJesus}. 

The apparatus is shown in FIG.~\ref{fg:RadiumSetUp}, and the atomic level structure of radium is shown in FIG.~\ref{fg:RadiumAtomLevels}. The key components of the apparatus were the radium source, the Zeeman slower loading the magneto-optical trap (MOT) and the optical-dipole trap (ODT). The radium was provided by sources of up to 9 mCi, which provide both $^{225}$Ra ($t_{1/2}$=14 d) and significantly greater quantities of $^{226}$Ra ($t_{1/2}$=1600 y), which has no spin or EDM, but which is useful for diagnostics and tuning the optical traps.
The Zeeman slower used the momentum transferred from photons in a counter-propagating laser beam which was kept close to the 714 nm intercombination-line resonance by the Zeeman shift in a spatially-varying magnetic field. The slowed atoms then entered the MOT operating on the same transition.
Atoms that leak to the $7s6d^3$ $^3D_1$ level were repumped to the ground state with a 1429 nm laser. The trapping efficiency of the  MOT was approximately $10^{-6}$, largely limited by the low scattering rate. Radium atoms were accumulated for about 40 s and cooled to $\approx 40$ $\mu$K with typically $10^5$ $^{226}$Ra atoms or $10^3$ $^{225}$Ra atoms trapped in the MOT. From the MOT, atoms were transferred with high efficiency (80\%) to the ODT, effected by a 40 W, 1550 nm, laser beam focused with an $f$= 2 m lens.  By translating the lens, the ODT-trapped atoms were moved one meter into a separate chamber within a cylindrical-multilayer magnetic shield, where a standing-wave 1550 nm ODT held the atoms for the EDM measurement. The ODT holding time-constant of $\approx40$ s was limited by collisions with residual gas atoms.  A pair of copper electrodes separated by 2.3 mm provide an electric field of 67 kV/cm.

\begin{figure*}
 \includegraphics[width=6.5truein,angle=0]{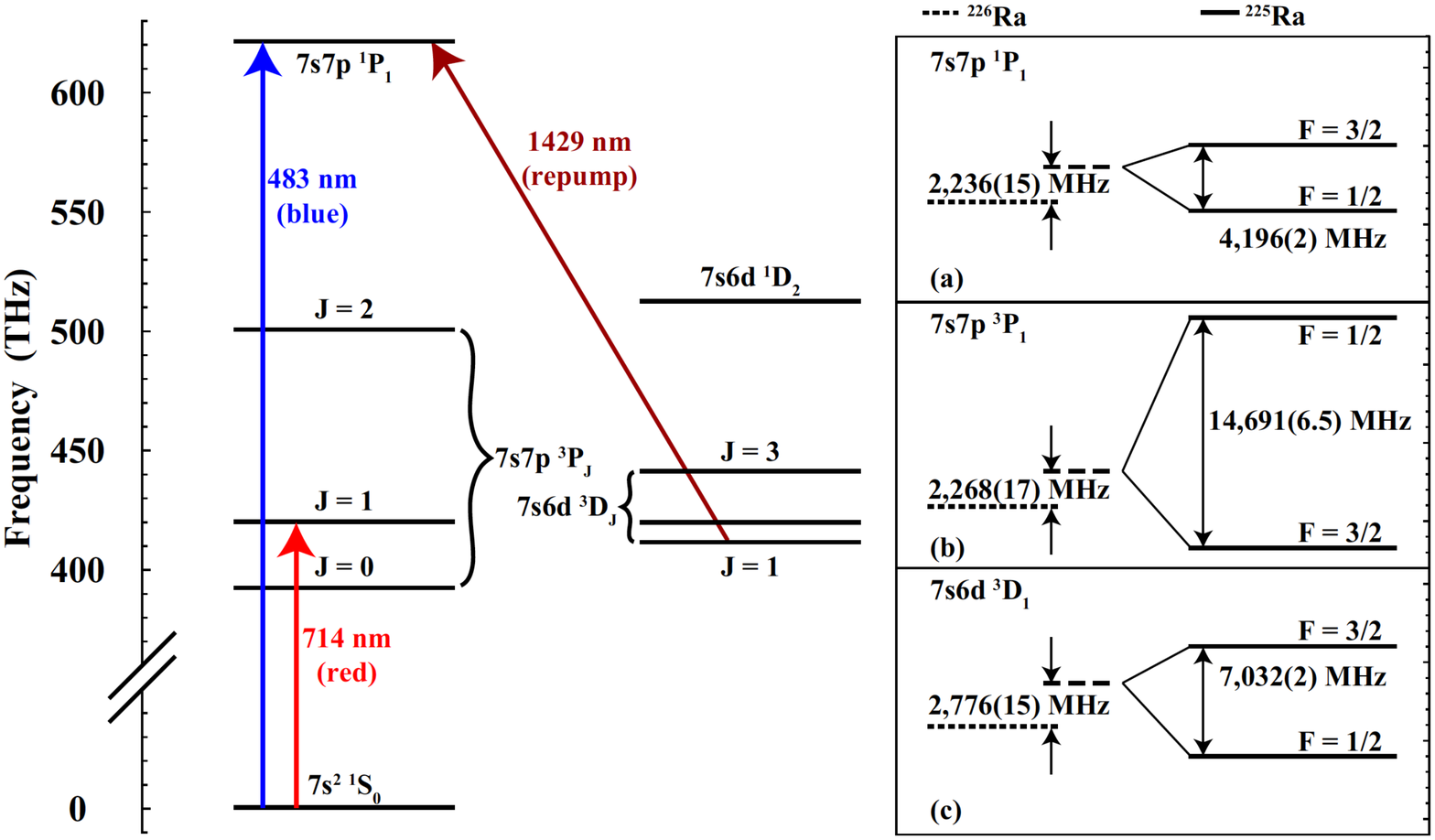}
 \vskip -0.35 truein
  \caption{  (Color online) Left: Radium energy-level diagram. Right: isotope shifts for $^{225}$Ra ($I=1/2$) and $^{226}$Ra ($I=0$) relative to the isotopic average; panels (a). (b) and  (c) show the specific levels of interest. Figure from~\textcite{rf:Parker2015,PhysRevC.94.025501}.
}
\label{fg:RadiumAtomLevels}
\end{figure*}

The EDM measurement was based on a 100 second  cycle consisting of 60 seconds to cool, trap and transfer the atoms into the ODT and two approximately 20 second free precession periods to extract the EDM signal. 
The Ramsey separated-oscillatory-field measurement consisted of state preparation with a circularly polarized laser beam (483 nm) followed by a nuclear-spin precession period of $(20+\delta)$ seconds and measurement of atoms of the opposite polarization detected by absorption of the laser light imaged onto a CCD camera. By varying $\delta$, the change in accumulated phase over the free-precession period, shown in FIG.~\ref{fg:RadiumEDMData}, was converted to an EDM induced frequency shift.

A number of systematic effects were considered, the most important of which were Stark-shift related $E^2$  effects, correlations with drifts in the magnetic field between subsequent  electric field flips, correlations with the ODT laser power, and Stark interference.
\begin{figure}
 \includegraphics[width=3.25truein,angle=0]{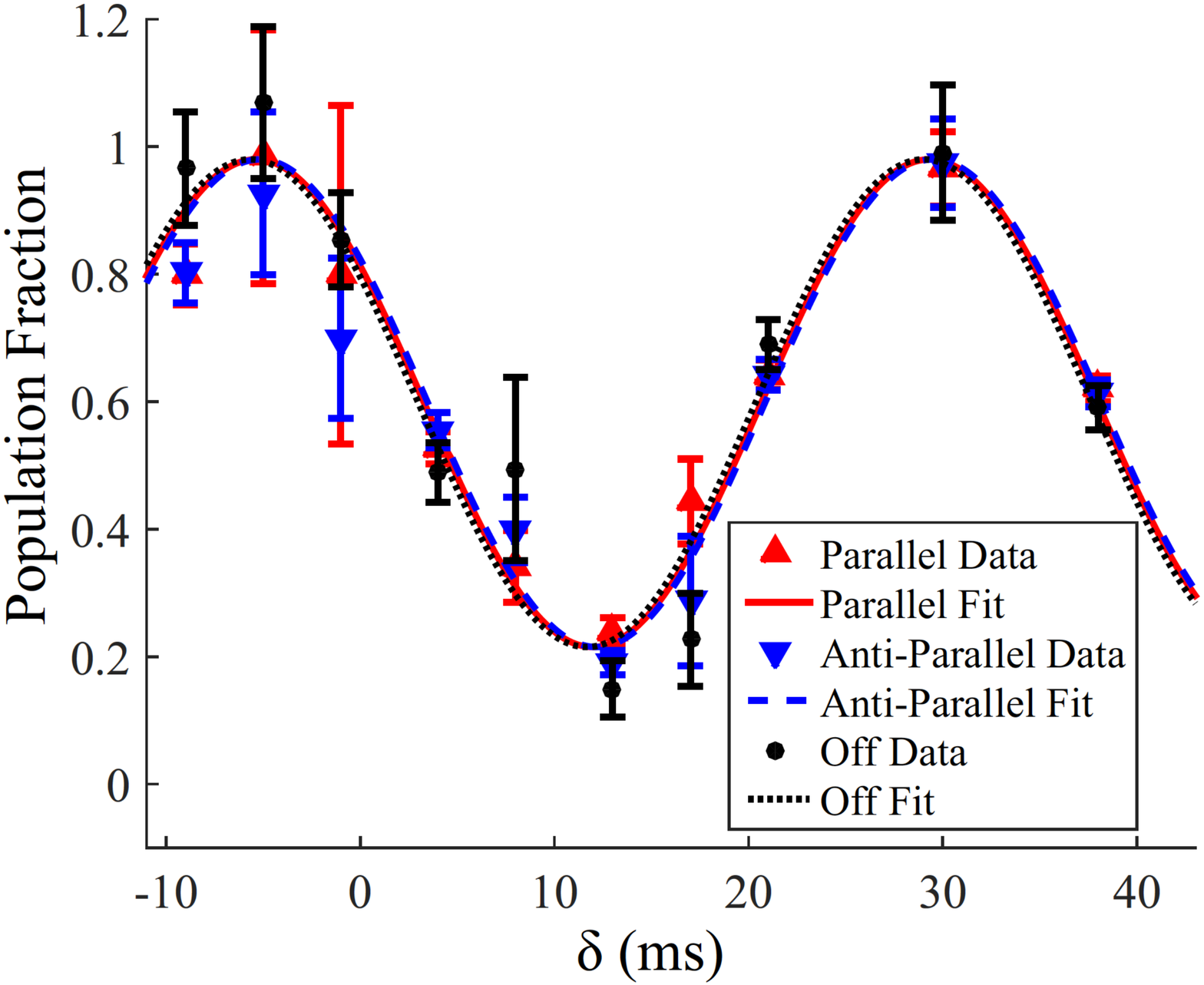}
  \caption{ (Color online) Phase-shift data showing the fraction of spin-down $^{225}$Ra nuclei after the nuclear-spin precession cycle of $20+\delta$ seconds. Figure from~\textcite{rf:Parker2015,PhysRevC.94.025501}.
}
\label{fg:RadiumEDMData}
\end{figure}

The most recent  $^{225}$Ra EDM result is
\begin{equation}
d(^{225}{\rm Ra})=(4\pm 6\ {\rm (stat)}\pm 0.2\ {\rm(sys)})\times 10^{-24}\ {\rm \ecm}.
\end{equation}
This was  a 36-fold improvement over the first run and corresponds to an upper limit of $1.4\times 10^{-23}$ \ecm~ (95\%)

A disadvantage of the imaging technique for probing the spin-state population is that only a small fraction of the atoms absorb the probe light so that the contrast is low and the statistical error is dominated by photon counting statistics and not the number of $^{225}$Ra atoms. This can be mitigated by essentially counting atoms using the STIRAP technique~\cite{rf:STIRAP}
Estimated production rates at FRIB, based on fragmentation models for light targets, may provide significantly  more $^{225}$Ra for future experiments.

\bigskip 
\noindent{\bf $^{221/223}$Rn}

Radon isotopes present the possibility of using techniques developed for the $^{129}$Xe EDM measurements and exploiting potential enhancements  due to octupole collectivity in $^{221/223}$Rn, which have half-lives on the order of 20-30 minutes. A program to develop an on-line EDM experiment at TRIUMF has been underway, and the prospects of producing and harvesting significantly greater quantities of atoms from the beam dump  at FRIB is very promising~\cite{Pen201462}. 
The on-line EDM measurement consists of the following elements:
1.)  on-line collection and transfer of radon isotopes, 
2.) optical pumping polarization by spin exchange (SEOP),
3.) polarized gas transport  to the EDM cell and gas recovery,
4.) EDM cells and high voltage,
5.) magnetic field, magnetic shielding and monitoring,
6.) spin-precession monitoring.


For collection and transfer, a 40 kV rare-gas isotope beam was incident on a tantalum foil~\cite{rf:Warner} for typically two half-lives and then transferred to a LN$_2$-cooled cold-finger, from which the gas was then released by rapidly warming the cold finger and pushed into the SEOP cell by a piston of N$_2$ gas. The N$_2$ gas also served as a buffer gas for noble gas polarization and as an insulating gas for the high voltage applied to the EDM cells.  Nearly 100\% collection efficiency was demonstrated~\cite{rf:NussWarren,rf:Warner}. SEOP produced $^{209}$Rn and $^{221}$Rn polarization of  $\approx 10\%$~\cite{rf:Tardiff2008,rf:Kitano1988} and spin relaxation times ($T_1\approx 15\ s$)~\cite{rf:Tardiff2008} were measured.
Each EDM measurement is limited by the spin-coherence time to about one minute, whereas the half-life of the Rn isotopes of interest is about 30 min. The Rn/N$_2$ mixture can be recycled into the polarizer cell by a circulating system.
A  measurement cycle will be initiated by a $\pi/2$  pulse after which the nuclear spins  freely precess in the $x$-$y$ plane ($\vec E$ and $\vec B$ are along $\hat z$). The radon-isotope free precession can be monitored by the gamma-ray anisotropy technique or by two-photon magnetometry~\cite{ref:xenon_twophoton}. 

Estimates of the $^{221/223}$Rn  production rates are based on measured rates at ISOLDE and TRIUMF. At  ISOLDE with a 1.6 $\mu$A 1.4 GeV proton beam incident on a thick uranium-carbide target, $1.4\times 10^7$ $^{220}$Rn-s$^{-1}$ were delivered to the low-energy end of the REX-ISOLDE accelerator~\cite{rf:NaturePaper}. With a 10 $\mu$A proton beam, $10^{8}$ $^{221}$Rn-s$^{-1}$ and $2\times 10^{7}$ $^{223}$Rn-s$^{-1}$ are expected. This would provide about $3\times 10^{10}$ nuclei for each 1-hour collection cycle. Estimated production rates at FRIB, based on fragmentation models on light targets {\it i.e.} the water in the FRIB beam dump, are up to 100$\times$ greater; however developing schemes for extraction of a large fraction of this remains a challenge. 


Radon spin precession can be measured by gamma-ray anisotropy or by direct optical detection. The gamma-anisotropy is a $P_2(\cos\theta)$ distribution of photons emitted after a polarized nucleus decays to the excited states of the daughter. For a precessing-polarized sample, the detection rate for photons in a detector at a specific azimuthal position is modulated at twice the precession frequency ($2\omega$). The statistical power of the gamma-ray anisotropy technique for the Radon-EDM measurement is limited by the intrinsic photo-peak count rate limit for typical high-purity germanium gamma-ray detectors.  The projection of the nuclear spin is monitored by detecting the transmission or fluorescence of circularly polarized light or by the optical rotation of linearly polarized light. For radon, the single and two-photon transitions correspond to wavelengths of 178 nm and 257 nm respectively, and the two-photon magnetometry techniques discussed in Sec.~\ref{sec:Magnetometry} can be applied. 
In the transmission/fluorescence case, photons and angular momentum are absorbed by the atoms and the measurement is destructive in the sense that the initial state of an atom is changed and thus affects the spin-coherence time $\tau$. Therefore the intensity of the light is adjusted to optimize the measurement. It is also possible to use the Ramsey separated oscillatory field technique to allow the spins to ``precess in the dark.'' 
With a fluorescence collection efficiency of 10\%,  a photon-statistics-limited EDM sensitivity of $3\times 10^{-26}$~\ecm~ could be achieved in one day assuming a 50\% duty cycle, $E=10$ kV/cm and $\tau=15$ sec.  
With anticipated FRIB production rates, a sensitivity of $3\times 10^{-28}$ \ecm~ in 100~days is a reasonable running-time scenario for an  off-line experiment using isotopes harvested from the beam dump~\cite{Pen201462}. 
The octupole enhancement also provides for different isotopes of the same atomic species as comagnetometers, {\it e.g.} $^{211}$Rn and $^{209}$Rn.






\subsection{Storage ring EDMs}
\label{sec:StorageRingEDMs}

Over the last two decades, EDM measurement techniques using storage rings have been developed using inspiration from the muon ``g-2'' experiment at Brookhaven National Laboratory~\cite{Bennett:2006fi}. 
The concept, introduced by~\cite{Farley04}, is based on the evolution of the momentum and spin of a charged particle  in the presence of magnetic and electric fields.

For a relativistic particle of charge $q$ and rest mass $m$, the equation of motion in the lab frame is
\begin{equation}
\frac{d\vec{p}}{dt} = q \left ( \vec{v} \times \vec{B} + \vec{E} \right).
\end{equation}
where $\vec{B}$ and $\vec{E}$ are the static and uniform magnetic and electric fields in the lab frame, $\gamma=(1-v^2/c^2)^{-1/2}$ is the Lorentz factor,   
$t=\gamma\tau$ is the time measured in the lab frame, and $\tau$ is the proper time in the particle rest frame. 
The acceleration in the lab frame is \cite{rf:Jackson1998pp564565}
\begin{eqnarray}
\vec{a} = \frac{d\vec{v}}{dt} & = & \frac{ q}{m \gamma} \left [ \vec{v} \times \vec{B} + \vec{E}   - \left (\frac{\vec{v} \cdot \vec{E}}{c^2} \right ) \vec{v}\right ] \nonumber\\
& = &  \vec{\omega}_v \times  \vec{v} +\frac{ q}{m \gamma}\frac{1}{\gamma^2-1}\left ( \frac{\vec{v} \cdot \vec{E}}{c^2} \right ) \vec{v},\nonumber\\
\end{eqnarray}
where the rotation of the velocity in the lab frame is 
\begin{equation}
\vec{\omega}_v = \frac{q}{m\gamma}\left [ \frac{\gamma^2}{\gamma^2-1} \left ( \frac{\vec{v}\times\vec{E}}{c^2} \right ) - \vec{B}\right ].
\end{equation} 

The torque on the particle's spin $\vec{s}$ is 
\begin{equation}
\frac{d\vec{s}}{d\tau} = \gamma \left[ \frac{d\vec{s}}{dt} \right ] = \vec{\mu} \times \vec{B}' + \vec{d}\times \vec{E}' ,
\end{equation}
where the magnetic moment $\vec \mu$ and the EDM $\vec d$ are
 \be
 \vec{\mu} = g (q/2m) \vec{s}  \hskip 0.5 truein  \vec{d} = \eta (q/2mc) \vec{s},
 \ee
and  fields coupling to $\vec\mu$ and $\vec d$ in the particle rest frame are 
\begin{eqnarray}
\vec{B}' & = & \gamma \left ( \vec{B} - \frac{\vec{v} \times \vec{E}}{c^2} \right ) - \frac{\gamma^2}{\gamma+1}\frac{\vec{v}\cdot (\vec{v} \cdot \vec{B})}{c^2} \nonumber \\
\vec{E}' & = & \gamma \  \  \left (\vec{E} + \vec{v} \times \vec{B} \right ) - \frac{\gamma^2}{\gamma+1} \frac{\vec{v}\cdot (\vec{v} \cdot \vec{E})}{c^2} .\nonumber \\
\end{eqnarray}
For $v\approx c$,  $\vec{v} \times \hat{B} \approx 3000\ \mathrm{kV/cm/T}$. For the Brookhaven muon $g-2$ experiment, $B=1.45$ T, and the motional electric field is about two orders of magnitude larger than a typical laboratory electric field. In the lab frame, accounting for the Thomas precession \cite{doi:10.1080/14786440108564170}, the equation of motion for the spin is
\begin{equation}
	 \frac{d\vec{s}}{dt}  =  \left [ \frac{d\vec{s}}{dt} \right ]_\mathrm{rest} + \frac{\gamma^2}{\gamma+1}   \frac{\vec{s} \times(\vec{v} \times \vec{a})}{c^2}  
\end{equation}
Combining the preceding equations, the evolution of the spin in the lab frame is
\begin{eqnarray}
\frac{d\vec{s}}{dt}  &=&   \vec{\omega}_s\times \vec{s} \nonumber\\
         &=&   \big (\frac{q}{m\gamma}\big )\ \ \vec s\ \times\  \left [  \vec{B}  -   \frac{\gamma}{(\gamma + 1)} \frac{\vec{v} \times \vec{E}}{c^2} \right ]\nonumber\\
	 &+&\  \big (\frac{aq}{m}\big )\ \ \vec s\ \times\  \left [  \vec{B} -  \frac{\vec{v} \times \vec{E}}{c^2}	 - \frac{\gamma}{\gamma+1} \frac{\vec{v}(\vec{v}\cdot\vec{B})}{c^2}  \right ]  \nonumber \\ 
	 &+&\big ( \frac{\eta q}{2mc} \big )  \ \vec{s}\  \times\ \left [ \vec{E} + \vec{v} \times \vec{B} -\frac{\gamma}{\gamma+1} \frac{\vec{v}(\vec{v}\cdot\vec{E})}{c^2}  \right ] . \nonumber\\
	 \nonumber \\ 
\end{eqnarray}
The first two terms are the non-covariant form of the Bargmann-Michel-Telegedi or BMT equation~\cite{PhysRevLett.2.435} and give the torque on the spin in the lab frame due to the magnetic moment, where $a$ is the magnetic moment anomaly. The third term in square brackets, proportional to $\eta$, is due to the EDM. 

To illustrate the principle of the EDM measurement, consider the rotation of the spin with respect to the velocity in the lab frame $\vec \omega_s-\vec\omega_v$. This can be separated into two terms: $\vec \omega_a$ and $\vec\omega_d$, where
\begin{eqnarray}
	\vec{\omega}_a   = & -&\ \frac{a q}{m} \left [ \vec{B} 
	+\!\! \left ( \frac{1}{a(\gamma^2 - 1) }-1\right )\! \frac{\vec{v} \times \vec{E}}{c^2}\! -\! \frac{\gamma}{\gamma+1}  \frac{\vec{v}(\vec{v}\cdot\vec{B})}{c^2} \right ] \nonumber\\
\vec{\omega}_d	 = & -&\! \frac{d}{\hbar J} \left [\vec{v} \times \vec{B} + \vec{E} -\frac{\gamma}{\gamma+1} \frac{\vec{v}\cdot\vec{E}}{c^2}  \vec{v} \right ]  
= -\frac{d \vec{E}'}{\hbar J \gamma}.
	\end{eqnarray}
The magnetic-moment anomaly 
for leptons is
\begin{equation}
a = \frac{g-2}{2} = \left ( \frac{\mu}{\mu_B} \right ) \left ( \frac{e}{q} \right ) \left ( \frac{m}{m_e} \right )- 1
\end{equation}
and for bare nuclei is~\cite{Khriplovich98}
\begin{equation}
a = \frac{g-2}{2} = \frac{1}{2J}\left ( \frac{\mu}{\mu_N} \right ) \left ( \frac{A}{Z} \right ) - 1= \frac{\kappa}{Z},
\end{equation}
where $\mu_{N(B)}$ is the nuclear (Bohr) magneton, $Z=q/e$, $e$ is the elementary charge, $A=m/m_p$, $m_e$ is the mass of the electron, $m_{p}$ is the mass of the proton, and $\kappa$ is the customary anomalous magnetic moment.

The EDM experiments envision the charged particle of a carefully chosen energy with initially only longitudinal spin polarization, {\it i.e.}  $\vec s$ parallel to $\vec p$, injected into a storage ring.
In the presence of appropriately chosen static laboratory electric and magnetic fields, the spin polarization of the particle will slowly develop a spin component linearly proportional to $\eta$, which is  transverse to its velocity,  {\it i.e.} pointing out of the storage ring plane. 
The direction of the particle's spin polarization vector can be determined, for example  in the case of the muon, by measuring the polarization-dependent decay asymmetry or, for stable nuclei, by spin-dependent elastic scattering.
The statistical uncertainty follows Eqn.~\ref{eq:EDMSigmaEquation1} for a total of $T/\tau$  EDM measurements of duration $\tau$ by substituting  for $E$, the electric field in the particle rest frame  $E^\prime/\gamma$: 
\be
\sigma_d=\frac{\gamma\hbar J}{2E^\prime P A}\frac{1}{\sqrt{N T\tau}},
\ee
where  the particle polarization is  $P$ and the experimental analyzing power is $A$, which are both $\le 1$, and $N$ is the number of particles detected for each measurement.
	
The key insight of storage ring EDM  measurements  is choosing the electric and magnetic fields as well as the particle's momentum such that the $\omega_a$ is  suppressed~\cite{Farley04}.  This 
 is accomplished by first making $\hat E$, $\hat B$ and the velocity all mutually orthogonal and then either choosing a radial electric field that cancels $\omega_a$ ($E_r=aBv\gamma^2/[1-av^2\gamma^2/c^2]$) or by setting  $\vec B\approx 0$ and storing particles with momentum $p\approx mc/\sqrt{a}$, {\it i.e.} purely electric confinement. A critical challenge is to minimize undesired radial magnetic fields due to misalignments and fringe fields, which would result in the transverse polarization due to the normal ``g-2'' spin precession and would mimic the EDM signal. Injection of simultaneous counter-propagating beams into the storage ring has been proposed to control these effects~\cite{Anastassopoulos:2015ura}. Unlike the EDM signal, the effect of the radial field depends on the  beam propagation direction, thus providing a way to disentangle the two sources of transverse polarization.
Other false effects, for example due to the non-orthogonality of the electric and magnetic fields, would be cancelled by summing over detectors separated by $180^\circ$ in azimuth around the ring.

Generic proposals have been made to search for EDMs with unstable charged ions using the beta-decay asymmetry for polarimetry similar to the muon $g-2$ concept~\cite{Khriplovich98,Khriplovich00HI,Khriplovich00NPA} and with
highly charged ions \cite{HCI-PR}. Current efforts to develop storage ring EDM experiments for muons, protons, deuterons, and $^{3}\mathrm{He}$  nuclei are summarized in 
Table~\ref{tab:srpar}.

\begin{table*}
\begin{tabular}{|c|c|c|c|c|c|c|c|c|c|c|}
\hline\hline
particle & $J$ & $a$ & $|\vec{p}|$ & $\gamma$ & $|\vec{B}|$ & $|\vec{E}|$ & $|\vec{E}'|/\gamma$ & $R$  & $\sigma^\mathrm{goal}_d$  & Ref. \\
(units) &  &  & $(\mathrm{GeV}/c)$ &  & $(\mathrm{T})$ & $(\mathrm{kV/cm})$ & $(\mathrm{kV/cm})$ & $(\mathrm{m})$  & (\!\ecm) &  \\
\hline
{$\mu^\pm$} & {$1/2$} & {$+0.00117$} & $3.094$ & $29.3$ & $1.45$ & $0$ & $4300$ & $7.11$ & $10^{-21}$ & E989 \\
 &  &  & $0.3$ & $3.0$ & $3.0$ & $0$ & $8500$ & $0.333$ & $10^{-21}$ & E34 \\
 &  &  & $0.5$ & $5.0$ & $0.25$ & $22$ & $760$ & $7$ & $10^{-24}$ & srEDM \\
 &  &  & $0.125$ & $1.57$ & $1$ & $6.7$ & $2300$ & $0.42$ & $10^{-24}$ & PSI \\ 
\hline
{$p^+$} & {$1/2$} & {$+1.79285$} & $0.7007$ & $1.248$ & $0$ & $80$ & $80$ & $52.3$ & $10^{-29}$ & srEDM \\
 &  &  & $0.7007$ & $1.248$ & $0$ & $140$ & $140$ & $30$ & $10^{-29}$ & JEDI \\
\hline
{$d^+$} & {$1$} & {$-0.14299$} & $1.0$ & $1.13$ & $0.5$ & $120$ & $580$ & $8.4$ & $10^{-29}$ & srEDM \\
 & &  &  $1.000$& $1.13$ & $0.135$ & $33$ & $160$ & $30$ &  $10^{-29}$ & JEDI \\
\hline
$^3\mathrm{He^{++}}$ & $1/2$ & $-4.18415$ & $1.211$ & $1.09$ & $0.042$ & $140$ & $89$ & $30$ &  $10^{-29}$ & JEDI \\
\hline\hline
\end{tabular}
\caption{\label{tab:srpar}Relevant parameters for proposed storage ring EDM searches.  The present muon EDM limit is ${1.8}\times 10^{-19}$\ecm~and the indirect limit on the proton EDM derived from the atomic EDM limit of $^{199}\mathrm{Hg}$ is
${2}\times 10^{-25}$\ecm. The magnetic moment anomaly is calculated using values for the unshielded magnetic moments of the particles from CODATA 2014 \cite{RevModPhys.88.035009}.
The sign convention for  positively charged particles is such that the magnetic field is vertical and the particles are circulating clockwise.
References are 
E989: Muon $g\!-\!2$ experiment at Fermilab \cite{Gorringe201573};
E34: Muon $g\!-\!2$ experiment at JPARC \cite{Gorringe201573};
srEDM: Muon EDM at JPARC \cite{KANDA2014212},  ``All-Electric'' Proton EDM at Brookhaven \cite{Anastassopoulos:2015ura}, Deuteron EDM at JPARC \cite{Morse2011}; 
PSI: Compact Muon EDM  \cite{compactMU10};
JEDI: ``All-In-One'' Proton, Deuteron, and Helion EDM at COSY \cite{JEDI}.}
\end{table*}  
 
  
\textcite{Farley04} presented the first storage ring proposal for a dedicated EDM measurement, which focused on a muon EDM. Their proposal suggested technically feasible values for $\vec{E}$ and $\beta=v/c$  to make $\vec{B}_a$ equal to zero. 
The current muon EDM limit $|d_\mu|\le 1.8\times 10^{-19}$ \ecm~ is derived from ancillary measurements of the muon decay asymmetry taken during a precision measurement of the muon anomalous magnetic moment \cite{Bennett:2008dy}.
The sensitivity of this measurement was limited by the fact that the apparatus was designed to be maximally sensitive to $\omega_a$.
For dedicated muon EDM experiments, under development at JPARC \cite{KANDA2014212} and PSI \cite{compactMU10}, $\vec{E}$ and $\gamma$ are chosen to make  $\omega_a=0$. 
The spin coherence time $\tau$ in this case is limited by the muon lifetime in the lab frame ($\gamma\times$2.2~$\mu$s).
An alternative muon EDM approach using lower energy muons and a smaller and more compact storage ring is being developed at  PSI.
A proposal for injecting muons into such a compact storage ring as well as an evaluation of the systematic effects due specifically to the lower muon energy is presented by \textcite{compactMU10}.

The two main differences between between an experiment designed for muons and one designed for light nuclei
are the need for more careful control of the beam properties to preserve the spin coherence and, of course, a different spin polarimetry scheme. 
A spread in the beam position and momentum smears the cancellation of the ``g-2'' spin precession which would, after many cycles, result in decoherence of the beam.
Since the muon spin coherence time is limited by the finite muon lifetime, this is not as critical for the muon EDM experiment.
For the case of a proton EDM search,  choosing $\vec{B}=0$ and   $\beta=1/\sqrt{a+1}$  supresses the $\vec{\beta}\times\vec{E}$ term~\cite{Anastassopoulos:2015ura}. This requires effective magnetic shielding, such as that discussed in sec.~\ref{sec:MagneticShielding}. 
The electric storage ring with bending radius $R=(m/e)/(E\sqrt(a(a+1))$  is generally only possible for particles with positive magnetic moment anomalies ($a > 0$).
With $E=10^6$ V/m, a  bending radius of $R\approx10$ m is required for protons.
Progress has been made in describing the challenging problem of orbital and spin dynamics inside electrostatic rings \cite{Mane08,Mane12,Mane14a,HS14,Mane15a,Metodiev15, Mane14b,Mane14c,Mane15b,Mane15c}, developing simulation code for electrostatic rings \cite{Talman15a,Talman15b}, and calculating the fringe fields for different plate geometries \cite{Metodiev14}.
To achieve sensitivity of 10$^{-29}$ \ecm,  impractically small residual magnetic fields would be required, thus two counter propagating beams within the same storage ring are envisioned, for which a vertical 
separation would develop in the presence of a radial magnetic field. 
After several cycles around the ring, this vertical separation would be large enough to measure using SQUID magnetometers as precision beam position monitors (BPMs).
 The development of an electric storage ring experiment dedicated to measurement of the proton EDM is being pursued by the Storage Ring EDM collaboration srEDM~\cite{Rathmann:2013rqa}.


A magnetic storage ring could also be used to measure the $J=1$ deuteron EDM using a similar technique. The deuteron polarization would be analyzed by the asymmetry in elastic scattering from a carbon target~\cite{Brantjes12}.
The goal for the deuteron EDM experiment is to maintain the spin coherence for at least as long as the vacuum-limited ion storage time which is about $10^3$ seconds for a vacuum of $10^{-10}\ \mathrm{Torr}$, which has been demonstrated at COSY~\cite{COSYDSpinLifetimePhysRevLett.117.054801}. The theory of spin evolution for a $J=1$ particle  in electromagnetic fields has been developed by \textcite{Silenko15}. 


The J\"{u}lich Electric Dipole moment Investigations (JEDI) collaboration in  Germany is undertaking precursor experiments while developing  long term plans  to measure the EDMs of the proton, deuteron, and $^{3}\mathrm{He}$ using an ``all-in-one'' electirc and magnetic storage ring~\cite{Rathmann:2013rqa}.
An intermediate step is  direct measurement of the proton and deuteron EDMs with lower statistical sensitivity using the presently available magnetostatic Cooler Synchrotron (COSY) storage ring with some modifications.
The main challenge is to introduce beam-line elements that prevent the spin precession due to the magnetic moment anomaly from washing out the  torque on the spin generated by the presence of an EDM. 
One suggestion is to synchronize the EDM torque to the magnetic moment anomaly spin precession \cite{Orlov06}, however the approach being developed for COSY by the JEDI collaboration is to partially ``freeze'' or lock the spin to the momentum using a beam element called a ``magic''  RF Wien filter \cite{Morse13}.
If the parameters of the Wien filter are carefully chosen,  one component of the particle's spin does not undergo the usual magnetic moment anomaly spin precession, which would allow the EDM torque to build up a transverse polarization.

Spin polarimetry is critical for both measuring the EDM signal as well as for diagnosing and improving the spin coherence time.
Significant progress has been made towards controlling systematics related to spin polarimetry for deuterons.
Results indicate that precision polarimetry for both deuterons and protons is feasible at the ppm level, which is required for a $10^{-29}\ e \cdot \mathrm{cm}$ EDM sensitivity.
Preliminary efforts to measure and improve spin coherence times of deuterons using the COSY storage ring have also been reported \cite{Benati12etal,Benati13etal,Bagdasarian14etal}.
High precision ($10^{-10}$) control and monitoring of the spin motion of deuterons at COSY has also been demonstrated \cite{Eversmann15}. Plans are also underway to develop an ion source and polarimetry for $^{3}\mathrm{He}$ by the JEDI collaboration.
Although significant effort is still required to perform storage ring EDM experiments, they would provide the most direct and clean measurements of the EDM of light ions and muons and would improve the limits on their EDMs by several orders of magnitude.



\section{Interpretations of current and prospective experiments}
\label{sec:Interpretation}
\label{sec:GlobalAnalysis}

In general there are many possible contributions to the EDM of any system accessible to experiment, for example the neutron EDM may arise due to a number of sources including short range, {\it e.g.} quark EDMs, and long range pion-nucleon couplings characterized by $\gpiz$ and $\gpio$. One approach to putting EDM results in context has been to use the upper limit from an experiment to set limits on individual phenomenological parameters by making use of theoretical calculations that establish the dependence on the individual parameters. This is the conventional approach, and is based on the reasoning that if the measured EDM is small then either all the contributions to the EDM (all the $\alpha_i C_i$) are small as well or large contributions must effectively cancel, that is have opposite signs and similar magnitudes. 
 While such a cancellation would be fortuitous, it may be ``required" in the sense that any underlying source of CP violation generally contributes CP violation in more than one way. Take, for example, a Left-Right Symmetric model, which contributes to both $\gpio$ and the short-range part of the neutron EDM, ${\bar d}_n^\mathrm{sr}$ as given in equations~\ref{eq:LRSM2},~\ref{eq:LRSM3} and \ref{eq:dnfull}. A cancellation would require a value of $\sin\xi$ less than $2\times 10^{-6}$. Thus in this model, either the phase $\alpha$  is very small or the mixing angle is very small, or both.

\subsection{Sole source}

Sole-source limits on the low-energy parameters are presented in Table~\ref{tb:SoleSource} along with the system that sets the limits. The most conservative upper limit is derived using the smallest $|\alpha_{ij}|$ from the ranges presented in Tables~\ref{tb:paramagnetics} and \ref{tb:diamagnetics}. The sole source short-range neutron contribution assumes $\gpbz=\gpbo=0$. For the short-range proton contribution, the model of~\textcite{rf:SandarsTlF} is used for TlF and from~\textcite{dmitriev03} for $^{199}$Hg. The combination of light quark EDMs $d_d-1/3d_u$ is derived from the limit on $d_n$. The parameter $\bar \theta$ and the combination of CEDMs $\tilde d_d-\tilde d_u$ are derived from the sole-source limits on  $\gpbz$ and $\gpbo$, respectively.

\begin{table}[ht]
\begin{centering}
\begin{tabular}{|c|c|c|}
\hline\hline
LE Parameter &  system & 95\% u.l. \\
\hline
$d_e$ &  ThO & $9.2\times 10^{-29}$ \ecm\\
\hline
$C_S$ &   ThO & $8.6\times 10^{-9}$ \\
\hline
$C_T$ &   $^{199}$Hg &  $3.6\times 10^{-10}$\\
\hline
$\gpbz$ &  $^{199}$Hg & $3.8\times 10^{-12}$\\
\hline
$\gpbo$ &  $^{199}$Hg & $3.8\times 10^{-13}$\\ 
\hline
$\bar g_\pi^{(2)}$ &  $^{199}$Hg & $2.6\times 10^{-11}$\\ 
\hline
$\bar d_n^{sr}$ & neutron & $3.3\times 10^{-26}$ \ecm\\
\hline
$\bar d_p^{sr}$ & TlF & $8.7\times 10^{-23}$ \ecm\\
\hline
$\bar d_p^{sr}$ & $^{199}$Hg & $2.0\times 10^{-25}$ \ecm\\
\hline\hline
\multicolumn{3}{|c|}{Other parameters}  \\
\hline
$d_d$ & $\approx 3/4 d_n$ & $2.5\times 10^{-26}$ \ecm \\
\hline
$\bar\theta$ & $\approx \gpbz/(0.015)$ & $2.5\times 10^{-10}$\\
\hline
$\tilde d_d-\tilde d_u$ & $5\times 10^{-15}\gpbo$ \ecm & $2\times 10^{-27}$ \ecm  \\
\hline\hline
\end{tabular}
\caption{Sole-source limits (95\% c.l.) on the absolute value of the low energy (LE) parameters presented in Sec.~\ref{sec:LowEnergyParameters}  for several experimental systems assuming a single contribution to the EDM or, for molecules, the P-odd/T-odd observable. The lower part of the table presents limits on other parameters derived from the best limits on the low energy parameters. \label{tb:SoleSource}}
\end{centering}
\end{table}
\subsection{Global analysis}

A global analysis of EDM results has been introduced by~\textcite{Chupp:2014gka} and is updated here. In this approach simultaneous limits are set on six low-energy parameters: $d_e$, $C_S$, $C_T$, $\gpbz$, $\gpbo$ and the short-range component of the neutron EDM $d_n^{sr}$. New results from HfF$^+$, $^{199}$Hg and $^{225}$Ra along with clarifications of the isospin dependence of $C_T$ are included in the analysis presented below.

\subsection*{Paramagnetic systems: limits on $d_e$ and $C_S$}

Results are listed in Table~\ref{tb:EDMResults} for paramagnetic systems  Cs, Tl, YbF, ThO and HfF$^+$. Following~\textcite{rf:Dzuba2011,rf:Dzuba2011-erratum} we take the electron EDM result reported by each author to be the combination
 \begin{equation}
 d^\mathrm{exp}_j= d_e +\bigl(\frac{\alpha_{C_S}}{ \alpha_{d_e}}\bigr)_j C_S.
 \label{eq:deEquation}
 \end{equation} 
 The $\alpha_{C_S}/ \alpha_{d_e}$ are listed in Table~\ref{tb:paramagnetics}. 
As pointed out by \textcite{rf:Dzuba2011,rf:Dzuba2011-erratum}, though there is a significant range of $\alpha_{d_e}$ and $\alpha_{C_S}$ from different authors for several cases, there is much less dispersion in the ratio ${\alpha_{C_S}/\alpha_{d_e}}$.

In  Figure~\ref{fig:Paramagnetics}, we plot $d_e$ vs $C_S$ for the   $d^{exp}_\mathrm{para}$ for ThO and HfF$^+$ along with 68\% and 95\% confidence-level contours for $\chi^2$ on the $d_e$-$C_S$ space, where
\begin{equation}
\chi^2=\sum_{i} \frac{\bigl[ d^\mathrm{exp}_i-d_e -\bigl(\frac {\alpha_{C_S}}{ \alpha_{d_e}}\bigr)_i C_S\bigr]^2}{\sigma_i^2}. 
\end{equation}
Here $i$ sums over Cs, Tl, YbF, ThO and HfF$^+$, but only ThO and HfF$^+$ have significant impact. The range of $\bigl(\frac {\alpha_{C_S}}{ \alpha_{d_e}}\bigr)_j$ expressed in Table~\ref{tb:paramagnetics}, about 10\%, is accommodated by adding in quadrature to the total experimental uncertainty for each system. 
The resulting constraints from all paramagnetic systems on $d_e$ and $C_S$ at 68\%\ {\rm c.l.} are
\begin{equation}
d_e=(0.8\pm 4.2)\times 10^{-28}\ {\rm e\ cm} \quad C_S=(-0.9\pm 3.7)\times 10^{-8}.
\label{eq:deGlobal68}
\end{equation}
The upper limits at 95\% confidence level are
\begin{equation}
|d_e|<8.4\times 10^{-28}\ {\rm e\ cm} \quad |C_S|<7.5\times 10^{-8}\quad (95\%\ {\rm c.l.}).
\end{equation}

\begin{figure}[ht]
\centerline{\includegraphics[width=3.5in]{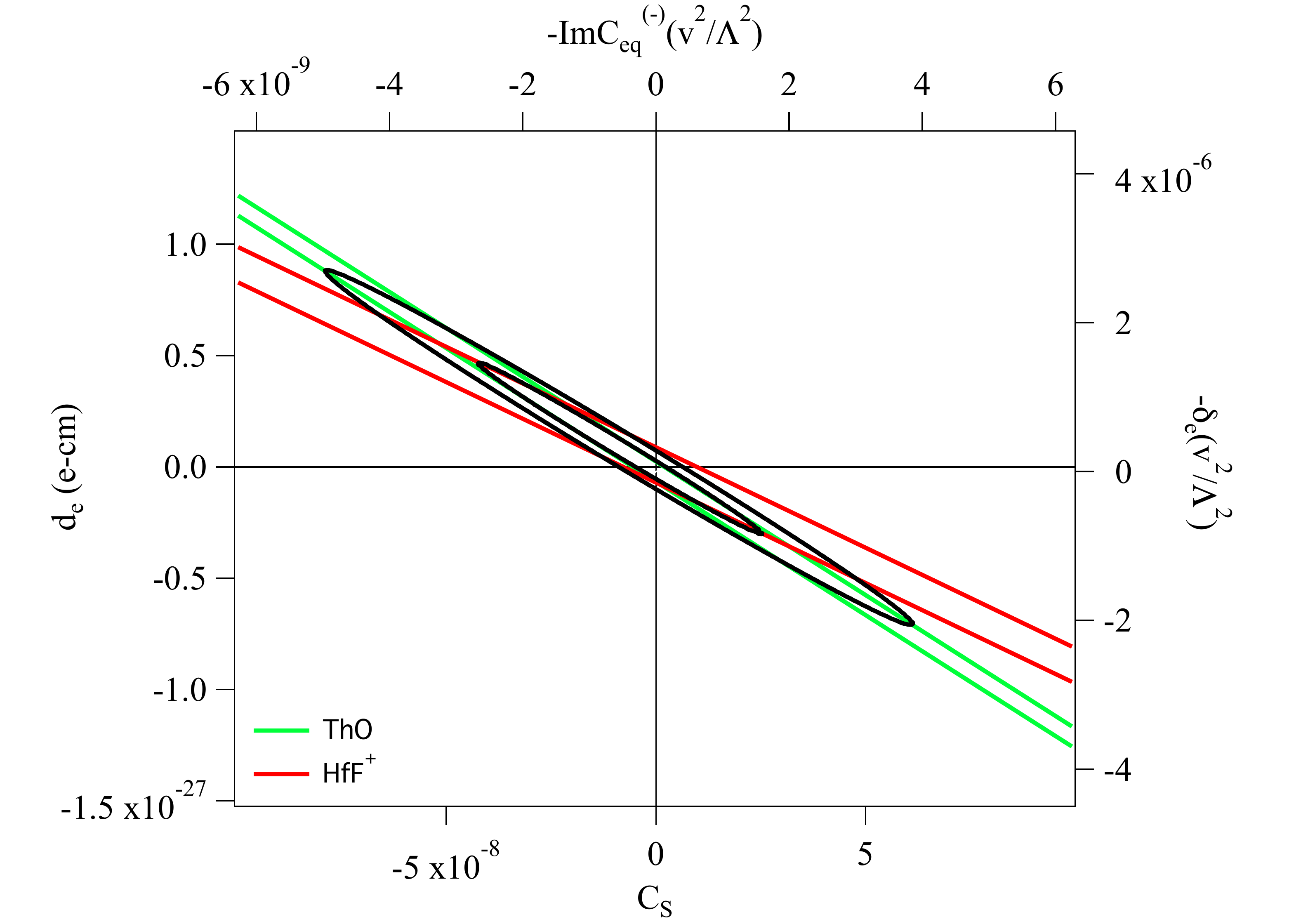}}
\caption{(Color online) Electron EDM $d_e$ as a function of $C_S$ from the experimental results in ThO and HfF$^+$ with 1$\sigma$ experimental error bars. Also shown are 68\% and 95\% $\chi^2$ contours for all paramagnetic systems including Cs, Tl, YbF.  The top and right axes show the corresponding dimensionless Wilson coefficients $\delta_e$ and $\mathrm{Im}\, C_{eq}^{(-)}$ normalized to the squared scale ratio $(v/\Lambda)^2$.}
\label{fig:Paramagnetics}
\end{figure}

Corresponding 95\% c.l. constraints on $\delta_e(v/\Lambda)^2$ and $\mathrm{Im}\, C_{eq}^{(-)} (v/\Lambda)^2$, obtained from those for $d_e$ and $C_S$ by dividing by $-3.2\times 10^{-22}$ e\,cm and $-12.7$, respectively are
\be
|\delta_e(v/\Lambda)^2|<2.6\times 10^{-6}\quad \mathrm{Im}\, C_{eq}^{(-)} (v/\Lambda)^2<5.9\times 10^{-9}.
\ee
The $^{199}$Hg result is not included in the constraints of Eq~\ref{eq:deGlobal68}, but has been used to  constrain $d_e$ and $C_S$, for example by~\textcite{Chupp:2014gka,Jung:2013mg,Fleig:2018bsf}. Particularly notable is that the limits on the scalar quark-pseudoscalar electron interaction may place a lower bound on the mass scale $\Lambda$ of roughly $10^3$ TeV or more. This has been applied in the context of leptoquark models by~\textcite{Fuyuto:2018scm}.


\subsection*{Hadronic parameters and $C_T$}

Since the introduction of our global analysis, there have been advances in hadronic and atomic theory along with three significant experimental developments in the diamagnetic/hadronic systems: 
\begin{enumerate}[i.]
\item
the four-times more sensitive result for $^{199}$Hg~\cite{Graner:2016ses-erratum}
\item  reanalysis of the neutron-EDM  which increased the uncertainty and moved the centroid by about 1/4\ $\sigma$~\cite{Afach:2015sja}, 
\item results from the octupole deformed $^{225}$Ra~\cite{PhysRevC.94.025501} 
\end{enumerate}
There are experimental results in five systems and four parameters $d_n^{sr}$, $C_T$,  $\gpbz$ and $\gpbo$, which are fully contstrained once $d_e$ and $C_S$ are fixed from the paramagnetic-systems results. In order to provide estimates of the allowed ranges of the  four parameters,  $\chi^2$ is defined as
 \begin{equation}
 \chi^2({\bf C_j})=\sum_i \frac{(d_i^\mathrm{exp}-d_i)^2}{\sigma_{d_i^\mathrm{exp}}^2},
 \end{equation}
 where $d_i$ have the form given in equation~\ref{eq:d_i}.
The four parameters ${\bf C_j}$ are varied to determine $\chi^2$ contours for a specific set of $\alpha_{ij}$. For 68\% confidence and four parameters, $(\chi^2-\chi^2_{min})<4.7$. 
The  $\alpha_{ij}$ are varied over the ranges presented in Table~\ref{tb:diamagnetics} to reflect the hadronic-theory uncertainties.
Estimates of the constraints are presented as ranges in Table~\ref{tab:gpiCTdn}, which has been updated from~\textcite{Chupp:2014gka}. The significant improvement in limits on $C_T$ is largely due to the change in sensitivity estimates ($\alpha_{i C_T^{0}}$) due to the recent calculations of the tensor form factors (see Eqn.~\ref{eq:CTs})\cite{Bhattacharya:2016zcn,Bhattacharya:2015wna}. Limits on $\gpiz$ also improve by about 50\% while limits on $\gpio$ and $\bar d_n^{sr}$ are  about 50\% less stringent.


Since $\gpiz$ and $\gpio$ also contribute to the neutron EDM,  the short-range neutron contribution $\bar d_n^{sr}$ is notably much less constrained than the experimental limit on the neutron EDM itself.  The anticipated improved sensitivity in the next few years for the diamagnetic systems $^{199}$Hg, $^{129}$Xe, $^{225}$Ra and TlF will provide tighter constraints on $\gpiz$ and $\gpio$; however the constraints do have significant correlations. The correlations of pairs of parameters are illustrated  in FIG.~\ref{fg:dnsrvsgpis}, which shows the 68\% contour on plots of allowed values of $\bar d_n^{sr}$ vs $\gpiz$, $\gpio$ and $C_T^{(0)}$ as well $\gpio$ vs $\gpiz$.

\begin{figure*}[h]
\centerline{\includegraphics[width=4truein]{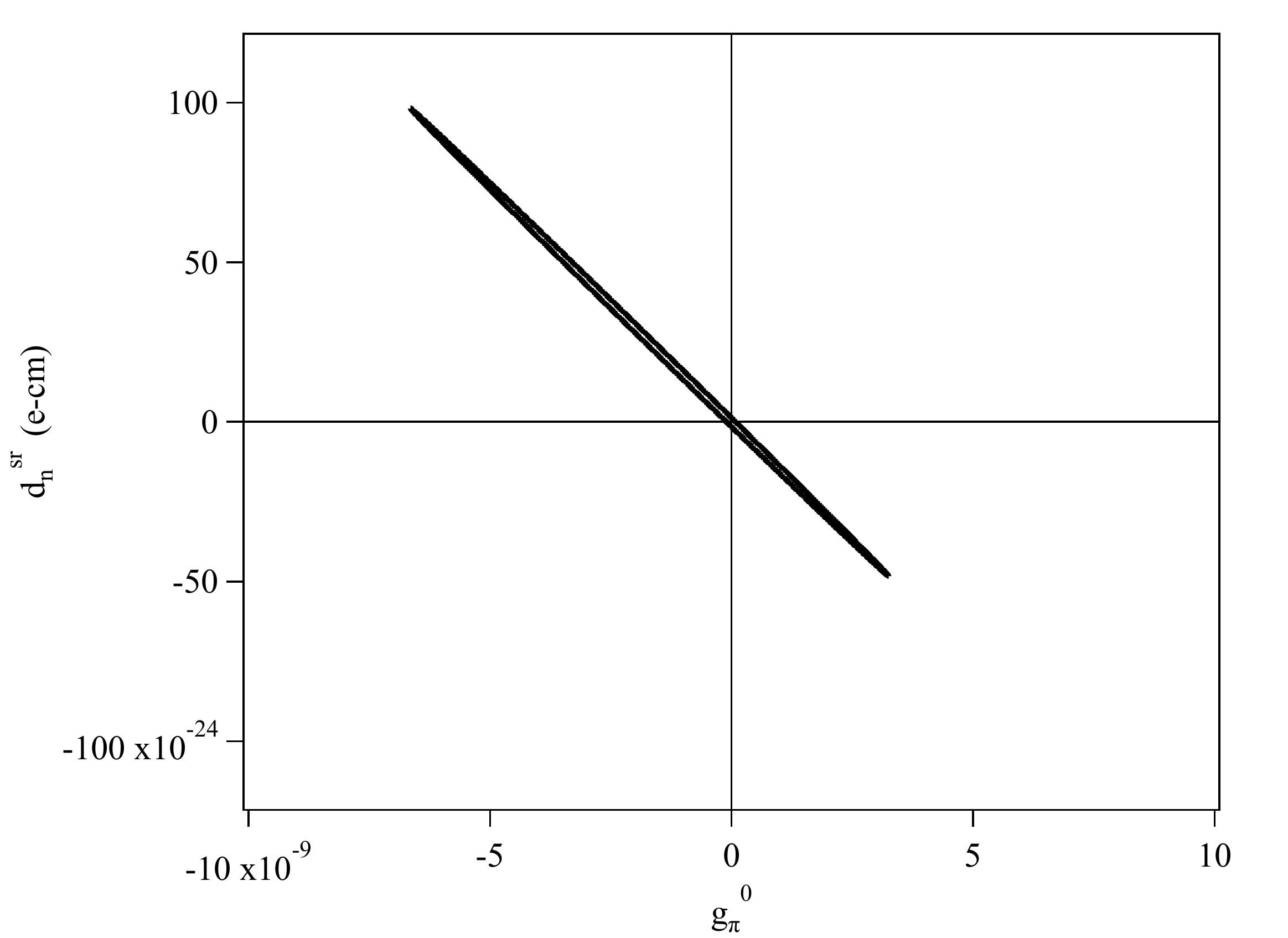}\includegraphics[width=4truein]{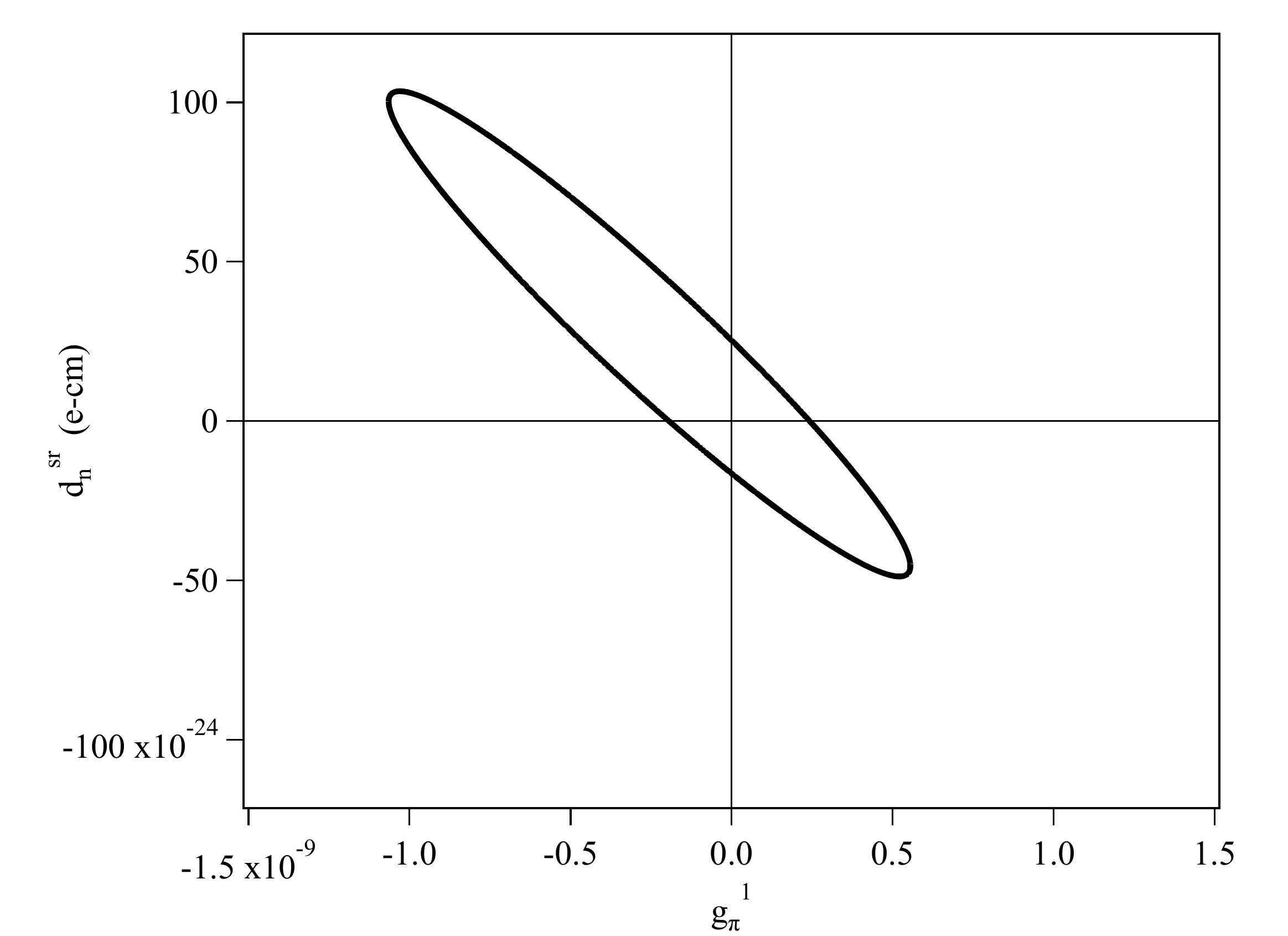}}
\centerline{\includegraphics[width=4truein]{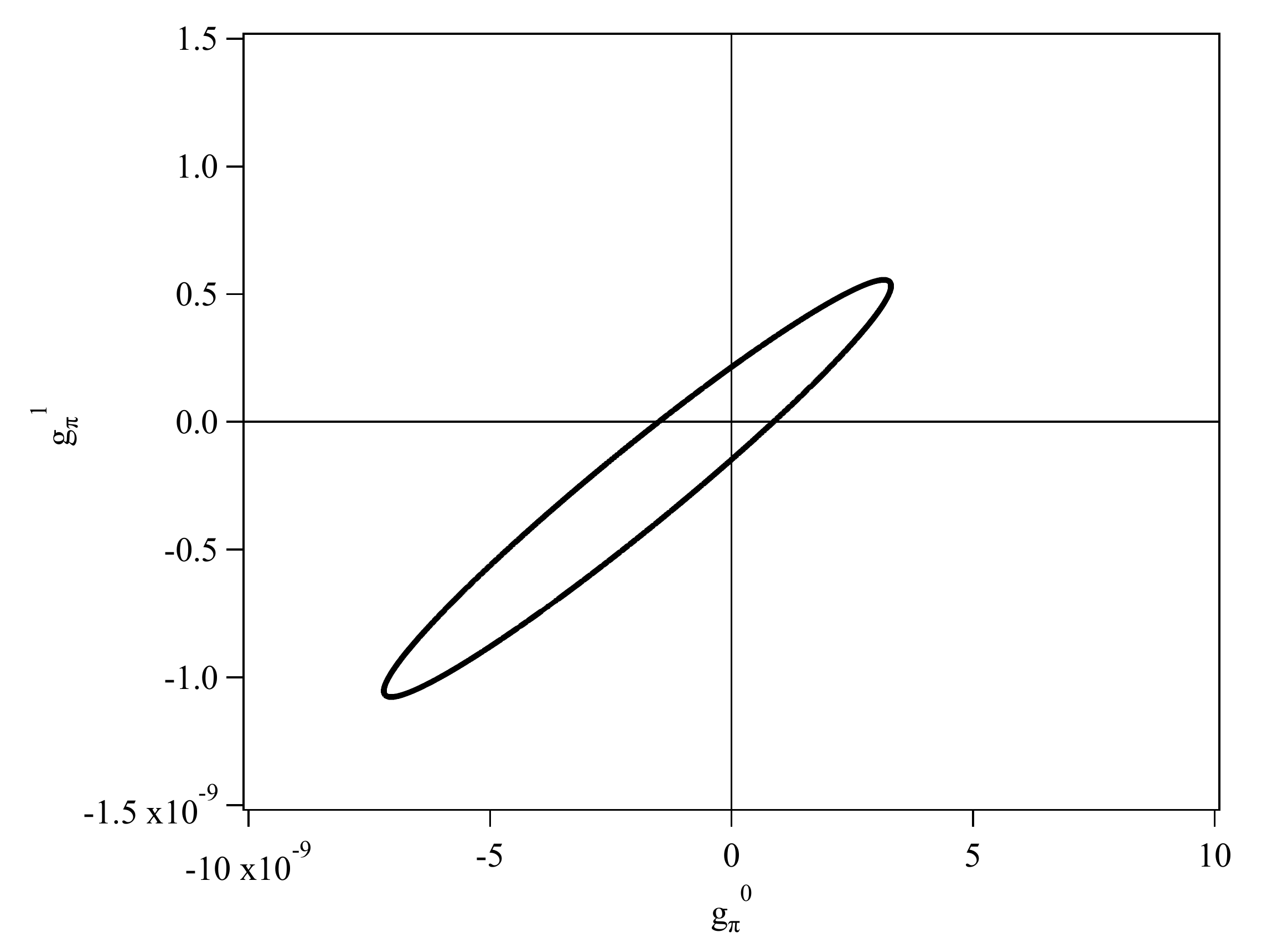}\includegraphics[width=4truein]{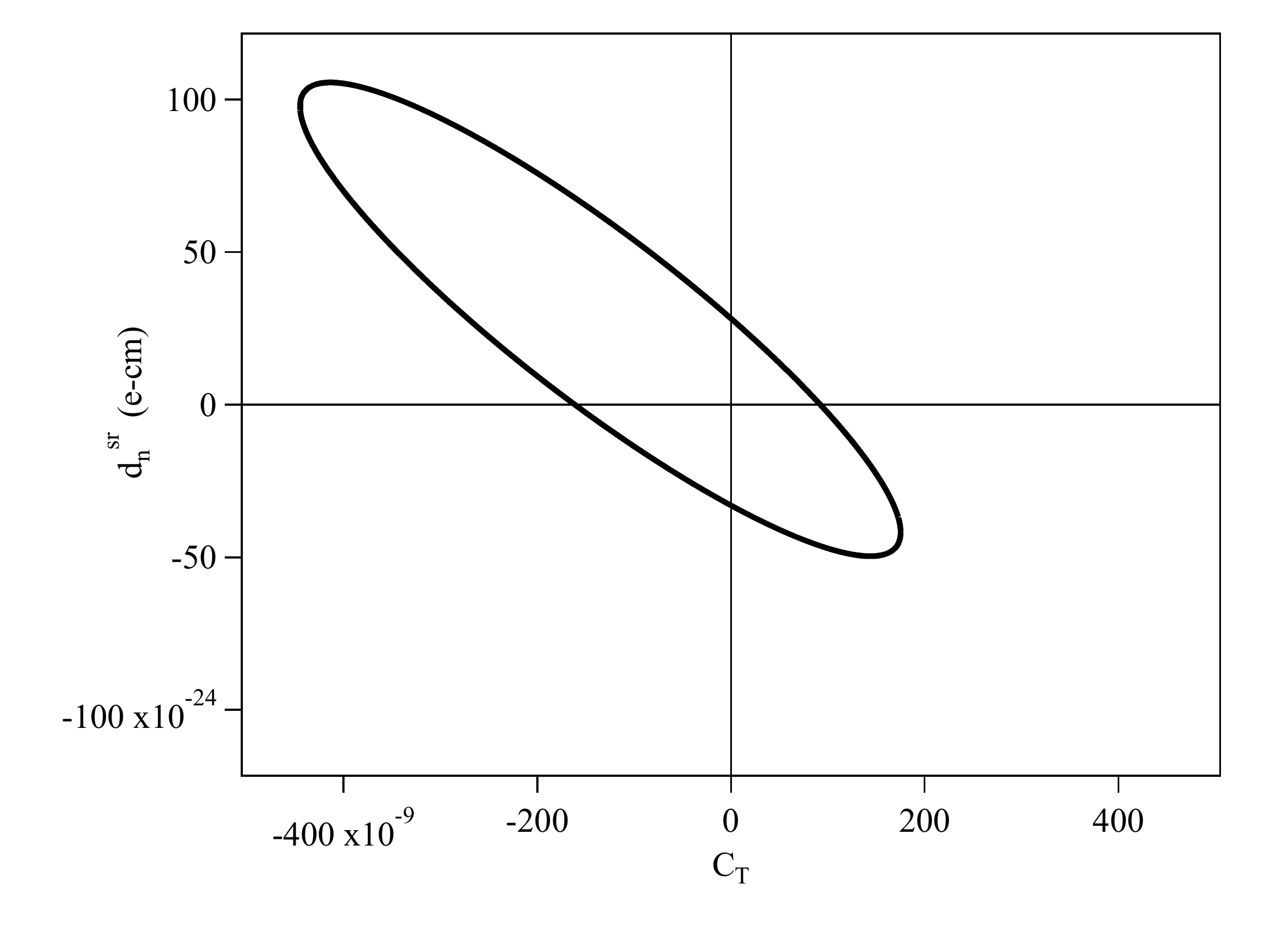}}
\caption{Combinations of hadronic parameters allowed by experimental results for the best values for $\alpha_{ij}$ in Table~\ref{tb:diamagnetics} with $\alpha_{{\rm Hg},\gpbo}=1.6\times 10^{-17}$ and $\alpha_{{\rm Ra},\bar d_n^{sr}}=-8\times 10^{-4}$. The allowed values at 68\% c.l. are contained within the ellipses for each pair of parameters.}
\label{fg:dnsrvsgpis}
\end{figure*}

\begin{table*}[h]
\begin{tabular}{||c|c|c|c|c||}
\hline\hline
&&&&\\
&${\bar d}_n^\mathrm{sr}$ (e\,cm) &$\gpbz$ & $\gpbo$& $C_T^{(0)}$  \\
\hline
Range from best values &&&&\\
with $\alpha_{g_\pi^1}(\mathrm{Hg})=+1.6\times 10^{-17}$  & $(-4.8$-$9.8)\times 10^{-23}$ &$(-6.6$-$3.2)\times 10^{-9}$ & $(-1.0$-$0.5)\times 10^{-9}$& $(-3.5$-$1.6)\times 10^{-7}$ \\
\hline
Range from best values &&&&\\
with $\alpha_{g_\pi^1}(\mathrm{Hg})=0$ & $(-4.3$-$3.4)\times 10^{-23}$ &$(-2.3$-$2.9)\times 10^{-9}$ & $(-0.6$-$1.3)\times 10^{-9}$& $(-3.2$-$4.0)\times 10^{-7}$\\
\hline
Range from best values &&&&\\
with $\alpha_{g_\pi^1}(\mathrm{Hg})=-4.9\times 10^{-17}$  & $(-9.3$-$2.6)\times 10^{-23}$ &$(-1.8$-$6.3)\times 10^{-9}$ & $(-1.2$-$0.4)\times 10^{-9}$& $(-11$-$3.8)\times 10^{-7}$\\
\hline
Range from full variation of $\alpha_{ij}$  &   $(-12$-$12)\times 10^{-23}$ &$(-7.9$-$7.8)\times 10^{-9}$ & $(-1.3$-$1.1)\times 10^{-9}$& $(-6.6$-$4.6)\times 10^{-7}$\\
\hline
Upper limits (95\% c.l.) & $2.4\times 10^{-22}$ & $1.5\times 10^{-8}$  & $2.4\times 10^{-9}$ & $1.1\times 10^{-6}$ \\
\hline\hline
\end{tabular}
\caption{\label{tab:gpiCTdn} Revised values and ranges for coefficients for diamagnetic systems and the neutron. The first three rows give the 68\% c.l. range allowed by experiment combined with the best values of the coefficients $\alpha_{ij}$ covering the reasonable range of $\alpha_{{\rm Hg},\gpio}$ with  $\alpha_{{\rm Ra},\bar d_n^{sr}}=-8\times 10^{-4}$; the fourth row gives ranges of coefficients for the entire reasonable ranges of the coefficients $\alpha_{ij}$ given in Table~\ref{tb:diamagnetics}, and the bottom row presents the 95\% c.l. upper limits on the coefficients for the full reasonable ranges of the coefficients.}
\end{table*}


In this global analysis approach, the constraints on each parameter depend on all experiments, the sensitivity of the EDM results, and the range of theoretical uncertainties of the $\alpha_{ij}$  given in Table~\ref{tb:diamagnetics}. To illustrate the dependence of the four dominant hadronic parameters on the experimental results  this we choose four of five experiments: $^{199}$Hg, $^{129}$Xe, $^{225}$Ra, and the neutron. The inverse of the matrix $\alpha_{ij}$ from Eqn.~\ref{eq:d_i} is
\begin{eqnarray}
& &\mbox{\footnotesize $
\left[ 
\begin{array}{c}
\bar d_n^{\rm sr} \\ \gpbz  \\ \gpbo\\ C_T^{(0)}
\end{array}
\right] $} = \nonumber\\
& & 
\mbox{\footnotesize $
\begin{bmatrix}
5.2 &\ \, 4.7\times 10^{4}&\ \, 9.5\times 10^{3} & 21\\
-2.8\times 10^{14} &-3.1\times 10^{18}&-6.3\times 10^{17}&-1.4\times 10^{15}\\
-7.0\times 10^{13} &-7.7\times 10^{17}&-1.6\times 10^{17}&-4.8\times 10^{14}\\
\ \ 1.9\times 10^{16} &\ \  1.4\times 10^{19}&\ \  3.6\times 10^{19}&\ \  8.4\times 10^{16}\\
\end{bmatrix}
\left[
\begin{array}{c}
 d_n \\ d_{\rm Xe}  \\d_{\rm Hg}\\ d_{\rm  Ra} 
\end{array}
\right]
$},
\nonumber\\
\end{eqnarray}
for the best values from Table~\ref{tb:diamagnetics} with  $\alpha_{{\rm Hg},\gpbo}=1.6\times 10^{-17}$ and $\alpha_{{\rm Ra},\bar d_n^{sr}}=-8\times 10^{-4}$ . For example
\be
 {\bar d_n^{sr}}=5.2 {d_n} + 4.7\times 10^{4} {d_{\rm Xe}} + 9.5\times 10^{3}{d_{\rm Hg}}+21    {d_{\rm Ra}}\nonumber \\
\ee
This combined with the results from  Table~\ref{tb:EDMResults}   shows that the $^{129}$Xe and $^{225}$Ra results have comparable contributions to the constraints and that improving each by a factor of about 500 would make their impact similar to that of $^{199}$Hg in the context of this global analysis.

\section{Summary and Conclusions}

We live in exciting times for EDMs. The observation and explanations of the baryon asymmetry call for BSM sources of CP violation that produce EDMs  that may be discovered in the next generation of experiments in a variety of systems. Experiment has marched forward with greater sensitivity recently achieved for the neutron and $^{199}$Hg, the tremendous advance in complexity and sensitivity for ThO and HfF$^+$ polar-molecule experiments sensitive to the electron EDM, and with new techniques providing results from the octupole deformed $^{225}$Ra. The next generation of experiments on the neutron will take advantage of new ideas and techniques incorporated into UCN sources and EDM techniques at a number of laboratories around the world; and new approaches to magnetic shielding and magnetometry/comagnetometry along with deeper understanding of systematic effects will be essential to achieving the next step in sensitivity in all systems. Storage rings and rare isotopes are expected to be new approaches that move forward in the coming years.

Interpretations of EDM limits and eventually finite results continue to advance with more quantitative connections to baryogenesis and clarification of effective-field theory approaches that connect fundamental quantum field theory to low-energy parameters relevant to the structure of nucleons, nuclei, atoms and molecules. The theory of EDMs brings together theoretical approaches at each of these scales, however the nucleus is a particularly difficult system for calculations and introduces the largest uncertainties in connecting experiment to theory. 
The best experimental result - in $^{199}$Hg -  is challenged by significant nuclear theory uncertainties.
 With the increasing interest in EDMs due to their role in connecting cosmology, particle physics and nuclear/atomic and molecular physics, the motivations for tackling these problems in hadronic theory become stronger. 

Even in light of current uncertainties, interpretation of EDM results from the sole-source perspective or in the context of a global analysis show that CP violating parameters are surprisingly small. In the case of the QCD parameter $\bar\theta$ this leads to the strong-CP problem and its potential solution via the axion hypothesis, which may also provide an explanation of non-baryonic dark matter. In a generic approach to CP violation consistent with current limits, combined with an assumption that the phases are of order unity, the mass scale probed is tens of TeV or greater, emphasizing the complementarity of EDMs and the LHC as well as future higher-energy colliders. In the context of models that introduce new phases, such as SUSY variants and Left-Right Symmetric Models, either the phases appear to be far less than naturally  expected or the mass scale of CP violation is quite large, which introduces challenges with the connection to Electroweak Baryogenesis. 

The definitive observation of an EDM in any system will be a tremendous achievement, but a single system alone may not clarify the questions arising in the connections to fundamental theory and to cosmology, for example separating weak and strong CP-violation. We therefore conclude by calling for efforts in several systems - paramagnetic systems most sensitive to the electron EDM and electron-spin-dependent CP violating interactions as well as diamagnetic atoms/molecules, nucleons and nuclear systems where hadronic CP violation is dominant. We also call for advanced theory efforts, in particular nuclear theory, which must improve to sharpen interpretation of EDM results in all systems.

\begin{acknowledgments}

The authors would like to thank everyone in the community who has provided input and advice as well as encouragement. In particular we are grateful to Martin Burghoff, Will Cairncross, Vincenzo Cirigliano, Skyler Degenkolb,  Matthew Dietrich, Peter Geltenbort, Takeyasu Ito, Martin Jung, Zheng-Tian Lu, Kent Leung, Kia Boon Ng, Natasha Sachdeva, Z Sun,  Yan Zhou, and Oliver Zimmer. The authors also acknowledge the Excellence Cluster Universe, MIAPP, the Munich Institute for Astronomy and Astrophysics and MITP, the Mainz Institute for Theoretical Physics for hosting and providing
 the opportunity to collaborate.  TC has been supported by US Department of Energy  grant No. DE-FG0204ER41331; PF is supported by the Deutsche Forschungsgemeinschaft (DFG) Priority Program SPP 1491 ÒPrecision Experiments with Cold and Ultra-Cold NeutronsÓ ; MJRM is supported by US Department of Energy Grant DE-SC0011095; JS is supported by Michigan State Univeristy.
 \end{acknowledgments}
\bibliography{rmp-handc}


\end{document}